\documentstyle[12pt,pennames,cite,epsf,amssymb]{book}
\global\def\E#1{{\cal M}\!\left\{#1\right\}}

\def\text#1{\hbox{#1}}
\def\GEN{{\cal G}}
\def\aver#1{\langle#1\rangle}
\def\FMM#1#2{\aver{{#1}^{[#2]}} }
\def\FT{\tilde{F}}
\def\XVEC#1#2{#1_1,\ldots,{#1}_{#2}}

\def\DXX#1#2#3{\,\prod_{#3=1}^{#2}\rd{#1}_{#3}}
\def\FACT#1#2{{#1}_m({#1}_m-1)\ldots({#1}_m-#2+1)}

\def\ifmath#1{\relax\ifmmode #1\else $#1$\fi}%
\def\mb{\mbox}
\def\ra{\ifmath{{\mathrm{a}}}}
\def\rb{\ifmath{{\mathrm{b}}}}
\def\rc{\ifmath{{\mathrm{c}}}}
\def\rd{\ifmath{{\mathrm{d}}}}
\def\rD{\ifmath{{\mathrm{D}}}}
\def\re{\ifmath{{\mathrm{e}}}}
\def\rE{\ifmath{{\mathrm{E}}}}
\def\rf{\ifmath{{\mathrm{f}}}}
\def\rF{\ifmath{{\mathrm{F}}}}
\def\rg{\ifmath{{\mathrm{g}}}}
\def\rh{\ifmath{{\mathrm{h}}}}
\def\rH{\ifmath{{\mathrm{H}}}}
\def\rL{\ifmath{{\mathrm{L}}}}
\def\rN{\ifmath{{\mathrm{N}}}}
\def\rq{\ifmath{{\mathrm{q}}}}
\def\rR{\ifmath{{\mathrm{R}}}}
\def\rs{\ifmath{{\mathrm{s}}}}
\def\rS{\ifmath{{\mathrm{S}}}}
\def\rT{\ifmath{{\mathrm{T}}}}
\def\rV{\ifmath{{\mathrm{V}}}}
\def\rX{\ifmath{{\mathrm{X}}}}
\def\rZ{\ifmath{{\mathrm{Z}}}}
\def\sbar{\overline{\rs}}

\font\dik=cmbxsl10
\def\vec#1{{\mbox{\dik #1}}}

\def\cm{{\mathrm{cm}}}
\def\cut{{\mathrm{cut}}}
\def\dyn{{\mathrm{dyn}}}
\def\excl{{\mathrm{excl}}}
\def\min{{\mathrm{min}}}
\def\max{{\mathrm{max}}}
\def\all{{\mathrm{all}}}
\def\inv{{\mathrm{inv}}}
\def\incl{{\mathrm{incl}}}
\def\inel{{\mathrm{inel}}}
\def\perm{{\mathrm{perm}}}
\def\QCD{{\mathrm{QCD}}}
\def\st{{\mathrm{st}}}
\def\th{{\mathrm{th}}}
\def\const{{\mathrm{const}}}
\def\PPC{\ifmath{\mathrm{PPC}}}
\def\PPCA{\ifmath{\mathrm{PPCA}}}

\newcommand{\beqa}{\begin{eqnarray}}
\newcommand{\eeqa}{\end{eqnarray}}
\newcommand{\beqan}{\begin{eqnarray*}}
\newcommand{\eeqan}{\end{eqnarray*}}
\newcommand{\beq}{\begin{equation}}
\newcommand{\eeq}{\end{equation}  }

\def\a{\alpha}
\def\b{\beta}
\def\g{\gamma}
\def\G{\Gamma}
\def\d{\delta}
\def\D{\Delta}
\def\e{\epsilon}
\def\ve{\varepsilon}
\def\z{\zeta}
\def\h{\eta}
\def\k{\kappa}
\def\la{\lambda}
\def\La{\Lambda}
\def\m{\mu}
\def\n{\nu}

\def\O{\Theta}

\def\p{\pi}
\def\P{\Pi}
\def\r{\rho}
\def\s{\sigma}
\def\S{\Sigma}

\def\t{\tau}

\def\f{\phi}
\def\vf{\varphi}
\def\x{\chi}

\def\w{\omega}
\def\W{\Omega}
\def\E{\re^+\re^-}

\def\cent{\centerline}

\def\ol{\overline}
\def\hs{\hskip}

\def\vs{\vskip}

\def\\{\hfill\break}
\def\ni{\noindent}

\def\i{\item}

\def\bu{\bullet}

\def\ls{\leftskip}

\def\ran{\rangle}
\def\lan{\langle}

\def\hf{\hfill}

\def\rf{\hrulefill}

\def\ds{\displaystyle}

\def\simgr{\ ^>\hs-2.5mm_\sim\ }
\def\simkl{\ ^<\hs-2.5mm_\sim\ }
\def\00{^o\hs-0.8truemm/\hs-0.8truemm_{oo}}
\def\12{^1\hs-0.8truemm/\hs-0.8truemm_2}
\def\blok{\sqcup\hs-3mm\sqcap}

\newcommand{\ZF}[3]{Z. Phys. {\bf C{#1}} ({#2}) {#3}}
\newcommand{\PL}[3]{Phys. Lett. {\bf {#1}} ({#2}) {#3}}
\newcommand{\MPL}[3]{Mod. Phys. Lett. {\bf {#1}} ({#2}) {#3}}
\newcommand{\PR}[3]{Phys. Rep. {\bf {#1}} ({#2}) {#3}}
\newcommand{\NP}[3]{Nucl. Phys. {\bf {#1}} ({#2}) {#3}}
\newcommand{\PRL}[3]{Phys. Rev. Lett. {\bf {#1}} ({#2}) {#3}}
\newcommand{\PRV}[3]{Phys. Rev. {\bf {#1}} ({#2}) {#3}}
\newcommand{\SJNP}[3]{Sov. J. of Nucl. Phys. {\bf {#1}} ({#2}) {#3}}
\newcommand{\APP}[3]{Acta Phys. Pol. {\bf {#1}} ({#2}) {#3}}

\renewcommand{\ZF}[3]{Z. Phys. {\bf C{#1}} ({#2}) {#3}}
\renewcommand{\PL}[3]{Phys. Lett. {\bf {#1}} ({#2}) {#3}}
\renewcommand{\MPL}[3]{Mod. Phys. Lett. {\bf {#1}} ({#2}) {#3}}
\renewcommand{\PR}[3]{Phys. Rep. {\bf {#1}} ({#2}) {#3}}
\renewcommand{\NP}[3]{Nucl. Phys. {\bf {#1}} ({#2}) {#3}}
\renewcommand{\PRL}[3]{Phys. Rev. Lett. {\bf {#1}} ({#2}) {#3}}
\renewcommand{\PRV}[3]{Phys. Rev. {\bf {#1}} ({#2}) {#3}}
\renewcommand{\SJNP}[3]{Sov. J. of Nucl. Phys. {\bf {#1}} ({#2}) {#3}}
\renewcommand{\APP}[3]{Acta Phys. Pol. {\bf {#1}} ({#2}) {#3}}

\newcommand{\WSCP}{World Scientific, Singapore}

\newcommand{\FESTHOVE}{Festschrift L.~Van Hove,
Eds.~A.~Giovannini and W. Kittel (\WSCP, 1990)}
\newcommand{\MPGOA}{\rm Proc. Xth Int. Symp. on Multiparticle Dynamics, \
Goa 1979, Eds.~S.N. Ganguli, P.K. Malhotra and A. Subramanian (Tata Inst.)}

\newcommand{\MPLUND}{\rm Proc. XV Int. Symp. on Multiparticle Dynamics,
Lund, Sweden 1984, Eds.~G.~Gustafson and C.~Peterson  (\WSCP, 1984)}

\newcommand{\MPTASHKENT}{%
\rm Proc. XVIII Int. Symp. on Multiparticle Dynamics, Tashkent, USSR,  1987,
Eds.~I.~Dremin and K.~Gulamov (\WSCP, 1988)}
\newcommand{\MPARLES}{%
\rm Proc. 19th Int. Symp. on Multiparticle Dynamics, Arles, 1988,
Eds.~D.~Schiff and J.~Tran~Thanh~Van (Editions Fronti\`eres, France
and \WSCP, 1988)}
\newcommand{\MPHOLMECKE}{%
\rm Proc. XX Int. Symp. on Multiparticle Dynamics, Gut~Holmecke, Germany,
1990,
Eds.~R.~Baier and D.~Wegener  (\WSCP, 1991)}
\newcommand{\MPWUHAN}{\rm Proc. XXI Int. Symp. on Multiparticle Dynamics,
Wuhan,
China, 1991, Eds.~Y.F. Wu and L.S. Liu (\WSCP, 1992)}
\newcommand{\MPSANT}{%
\rm Proc. XXII Int. Symp. on Multiparticle Dynamics, Santiago de Compostela,
Spain, 1992, Ed.~A.~Pajares  (\WSCP, 1993)}
\newcommand{\RINGBERG}{%
\rm Proc. Ringberg Workshop on Multiparticle Production, Ringberg Castle,
Germany 1991, Eds.~R.C.~Hwa, W.~Ochs and N.~Schmitz  (\WSCP, 1992)\ }
\newcommand{\MARBURG}{%
\rm Proc. Int. Workshop on Correlations and Multiparticle Production,
Marburg,
Eds.~M.~Pl\"umer,  S.~Raha and R.M.~Weiner   (\WSCP, 1991)}
\newcommand{\SANTAFE}{%
\rm Proc. Santa F\'e Workshop Intermittency in High Energy Collisions, 1990,
Eds.~F.~Cooper, R.C.~Hwa and I.~Sarcevic (\WSCP, 1991)}

\begin{document}

\bibliographystyle{unsrt}
\thispagestyle{empty}
{}~
\vs-2truecm
\hf HEN-362 (1993)

\hf IIHE-93.01

\hf FIAN/TD-09/93

\hf update July 1995
\vs 1cm

\begin{center}
{\large
SCALING LAWS FOR DENSITY CORRELATIONS AND \\
FLUCTUATIONS IN MULTIPARTICLE DYNAMICS
}
\end{center}

\vspace{2cm}
\ni
\cent{E.A.~De~Wolf\footnotemark,  I.M.~Dremin\footnotemark,
W.~Kittel\footnotemark}

\addtocounter{footnote}{-2}\footnotetext{\ Dept. of Physics, Universitaire
Instelling Antwerp, B-2610 Wilrijk and Inter-University Institute for
High Energies, Universities of Brussels, B-1050 Brussels,  Belgium}
\addtocounter{footnote}{1}\footnotetext{\ P.N. Lebedev Institute of Physics,
Acad. of Sciences of the Russia, 117~924, Moscow, Russia}
\addtocounter{footnote}{1}\footnotetext{\ University of Nijmegen/NIKHEF,
NL-6525 ED Nijmegen, The Netherlands}

\vs 4cm
\cent{\sl Abstract}
\vs 2mm
Experimental data are presented on particle correlations and
fluctuations in various high-energy multiparticle
collisions, with special  emphasis on evidence
for scaling-law evolution
in small phase-space domains.
The notions of intermittency and fractality as related to
the above findings are described.
Phenomenological and theoretical work on the subject is reviewed.

\newpage
\setcounter{page}{2}
\tableofcontents

\newpage
{\ls=6truecm\tolerance=10000\ni
Of the achieved triumph pangs and tricks \\
Are just tightly stretched bow-strings.

\hf B. Pasternak
\par}

\vs 4mm
\chapter{Introduction}

Recent years have witnessed a remarkably intense experimental and
theoretical activity in search of scale-invariance
and  fractality in multihadron production processes,
for short also called ``intermittency''.
These investigations cover all types of reactions
ranging from e$^+$e$^-$ annihilation to nucleus-nucleus collisions,
up to the highest attainable energies.
The creation of soft hadrons in these processes, a major
fraction of the total cross section, relates to the
strong-coupling long-distance regime of Quantum Chromodynamics (QCD),
at present one of the least explored sectors in the whole of
high-energy particle physics.

A primary motivation is the expectation that
scale-invariance or self-similarity,
analogous to that
often encountered in complex non-linear systems, might open new
avenues ultimately leading towards  deeper insight into  long-distance
properties of QCD and the unsolved problem of colour confinement.

History  shows that studies of fluctuations have
often triggered significant advances in physics.
In the present context, it was the observation of
``unusually large'' particle density fluctuations, reminiscent
of intermittency spikes in spatio-temporal turbulence, which prompted
the pioneering suggestion to investigate the pattern of
multiplicity fluctuations in multihadron events for ever decreasing
domains of phase-space.
Scale-invariance or fractality  would manifest itself in
power-law behaviour for scaled factorial moments
of the multiplicity distribution in such domains.
It is important to stress here that, in practice, one deals with the
problem of evolution of particle number distributions for ever smaller
bins and intermittent behaviour implies that, for small phase space bins,
the distributions become wider in a specific way. The same problem can be
stated as an increasing role of correlations within a small phase space
volume.

Through a multitude of increasingly sophisticated
experimental studies of factorial moments,
much new information has been gathered in a surprisingly short time.
This work indeed confirms approximate power behaviour down to the
experimentally possible resolution,
especially when carried out in two- and three-dimensional phase space.

The proposal to look for intermittency also has triggered
a thorough revival of interest in the old subject of particle correlations,
by experimentalists and theorists alike. The need
for greater sensitivity in measurements of correlation
functions has directly inspired important work on refined
analysis techniques. A promising and long overdue systematic
approach to correlation phenomena of various sorts,
including Bose-Einstein interferometry, is finally emerging.

The large body of experimental observations
now available  is calling for satisfactory explanation and, indeed,
theoretical ideas of all sorts abound.

The level of theoretical understanding is  quite different
for the various types of collision processes.
In e$^+$e$^-$ annihilation, parton cascade models based on
leading-log QCD have met considerable success and
good overall description of multiplicity fluctuations is claimed.
Closer inspection, nevertheless, reveals potentially serious
deviations from the data, thus requiring further study.

For other processes, in particular hadron initiated collisions,
models are faced with large and partly unexpected obstacles.
This may be a reflection of
insufficient knowledge of the reaction dynamics, although
present evidence points to hadronization as the main culprit.

Within the framework of perturbative QCD, results
of considerable interest on the
emergence of power behaviour and multifractality  have  been obtained.
However, these are asymptotic in nature and most likely quite
unrelated to present-day experiment. Being
related to the mechanism of confinement, not surprisingly,
the role of hadronization remains unclear.

Random  self-similar multiplicative branching models have inspired
much of the original work on intermittency.
Among many scale-invariant physical systems,
the  cascade process is a particularly natural candidate
for the description of  strong  fluctuations
self-similar over a wide range of scales.
It finds support in the cascade nature, not only  of
perturbative QCD, but also of  the subsequent hadronization.
However, further work is needed to help understand the details of the
process.

Alternatively, ``classic'' and extensively studied possibilities
are scale-invariant
systems at the critical point of a high-order phase transition.
This subject has attracted particular attention in view of potential
application to quark-gluon plasma formation in heavy-ion collisions.

This paper  contains a review of the present status of work on
intermittency and correlations as performed over the last years.
In Chapter~2 we introduce the necessary formalism and collect
useful results and relations  widely scattered in the literature.
Chapter~3 describes experimental data
on correlations in various experiments and discusses predictions
of popular models.
Chapter~4 is devoted to
data and models on the subject of particle fluctuations and
the search for power laws.
Chapter~5 gives an overview of the many theoretical ideas
related to the problem of multiplicity scaling and fractality.
Conclusions are summarized in Chapter~6.

\chapter{Formalism}
\section{Definitions and notation}
In this section, we compile and
summarize definitions  and various relations
among the physical quantities used in the sequel of this paper.
No originality is claimed in the presentation of this material.
It merely serves the purpose of fixing the notation and assembling
a number of results scattered throughout the literature.

\subsection{Exclusive and inclusive densities}\\
We start by considering a collision between particles a and b yielding
exactly $n$ particles  in a sub-volume $\Omega$ of the total phase space
$\Omega_{\mbox{\small tot}}$.
Let the single symbol $y$ represent the kinematical variables needed to
specify the position of each particle in this space (for example, $y$ can be
the c.m. rapidity\footnote{\ The rapidity $y$ is defined as
$y={1\over 2} \ln [(E+p_\rL)/(E-p_\rL)]$, with $E$ the energy and $p_\rL$ the
longitudinal component of momentum vector $\vec p$ along a given direction
(beam-particles, jet-axis, etc.); pseudo-rapidity is defined as
$\h={1\over 2}\ln [(p+p_\rL)/(p-p_\rL)]$}
variable of each particle and $\Omega$ an interval of length
$\delta y$).  The distribution of points in $\Omega$ can be characterized
by continuous probability densities $P_n(y_1,\ldots,y_n)$; $n=1,2,\ldots$.
For  simplicity, we assume all  final-state particles to be of the same type.
In this case, the {\bf exclusive} distributions $P_n(y_1,\ldots,y_n)$
can be taken fully symmetric in
$y_1,\ldots,y_n$; they describe the distribution in $\Omega$ when the
multiplicity is exactly $n$.

The corresponding {\bf inclusive} distributions are given for
$n=1,2,\ldots$ by:
\begin{eqnarray}
\rho_n(y_1,\ldots,y_n)&=&P_n(y_1,\ldots,y_n)\nonumber\\
&&\mbox{}+\sum_{m=1}^{\infty}\frac{1}{m!} \int_\Omega
P_{n+m}(y_1,\ldots,y_n,{y'}_1,\ldots,{y'}_m)\,\DXX{y'}{m}{i}.
\label{dr:1}
\end{eqnarray}
The inverse formula is
\begin{eqnarray}
P_n(y_1,\ldots,y_n)&=&\rho_n(y_1,\ldots,y_n)\nonumber\\
&&\mbox{}+\sum_{m=1}^{\infty}(-1)^m \frac{1}{m!} \int_\Omega
\rho_{n+m}(\XVEC{y}{n},\XVEC{y'}{m})\,\DXX{y'}{m}{i}\ .
\label{dr:2}
\end{eqnarray}
$\rho_n(y_1,\ldots,y_n)$ is  the probability density  for $n$ points to be at
$\XVEC{y}{n}$, irrespective of the presence and location of any further
points. The probability $P_0$ of multiplicity zero is given by
\begin{equation}
P_0=1-\sum_{n=1}^{\infty}\frac{1}{n!} \int_\Omega
P_{n}(\XVEC{y}{n})\,\DXX{y}{n}{i}\ .
\label{dr:3}
\end{equation}
This suggests to define $\rho_0=1$ in (\ref{dr:1}).
It is often convenient to summarize the above results with the help of
the generating functional\footnote{\ The technique of generating
functions has been known since Euler's time and was used for functionals by
N.N.~Bogoliubov
in statistical mechanics already in
1946~\cite{bogolubov}; see also~\cite{Brown}}
\begin{equation}
\GEN^{\excl}\left[z(y)\right]\equiv P_0+\sum_{n=1}^\infty\frac{1}{n!}
\int_\Omega\ P_n(\XVEC{y}{n})\,
z(y_1)\ldots z(y_n)\,\DXX{y}{n}{i}\ ,
\label{dr:4}
\end{equation}
where $z(y)$ is an arbitrary function of $y$ in $\Omega$.
The substitution
\begin{equation}
z(y)=1+u(y)
\label{dr:5}
\end{equation}
gives through (\ref{dr:1}) the alternative expansion
\begin{equation}
\GEN^{\incl}\left[u(y)\right]=1+\sum_{n=1}^\infty\frac{1}{n!}
\int_\Omega\,\rho_n(\XVEC{y}{n})\,u(y_1)\ldots u(y_n)\DXX{y}{n}{i}
\label{dr:6}
\end{equation}
and the relation
\begin{equation}
\GEN^{\incl}\left[z(y)\right]=\GEN^{\excl}\left[z(y)+1\right].
\label{dr:7}
\end{equation}

{}From (\ref{dr:4}) and (\ref{dr:7}) one recovers by functional
differentiation:
\begin{equation}
P_n(\XVEC{y}{n})=\left.
\frac{\partial^n \GEN^{\excl}\left[z(y)\right]}{%
\partial z(y_1)\ldots\partial z(y_n)} \right|_{z=0},
\label{dr:8}
\end{equation}
and
\begin{equation}
\rho_n(\XVEC{y}{n})=\left.
\frac{\partial^n \GEN^{\incl}\left[u(y)\right]}{%
\partial u(y_1)\ldots\partial u(y_n)} \right|_{u=0}.
\label{dr:9}
\end{equation}
To the  set of inclusive number-densities $\rho_n$ corresponds a sequence of
inclusive differential cross sections:
\begin{equation}
\frac{1}{\sigma_{\inel}}\,\rd\sigma=\rho_1(y)\,\rd y,
\label{dr:10}
\end{equation}

\begin{equation}
\frac{1}{\sigma_{\inel}} \rd^2\sigma=\rho_2(y_1,y_2)\,\rd y_1 \rd y_2.
\label{dr:10a}
\end{equation}
Integration over an interval $\W$ in $y$ yields
\begin{eqnarray}
 &~&\int_\W \r_1(y) \rd y = \lan n\ran \nonumber \\
 &~&\int_\W \int_\W \r_2(y_1,y_2)\rd y_1\rd y_2 =
\lan n(n-1)\ran \nonumber \\
 &~&\int_\W \rd y_1 \dots \int_\W \rd y_q \r_q (y_1,\dots,y_q) =
\lan n(n-1)\dots (n-q+1)\ran \ ,
\end{eqnarray}
where the angular brackets imply the average over the event ensemble.

\subsection{Cumulant correlation functions}\\
The inclusive $q$-particle densities $\rho_q(\XVEC{y}{q})$ in general
contain ``trivial'' contributions from lower-order densities. Under
certain conditions, it
is, therefore, advantageous to consider a new sequence of functions
$C_q(\XVEC{y}{q})$ as those statistical quantities which vanish whenever one
of their arguments becomes statistically independent of the others.
It is well known that the quantities with such  properties are the
correlation functions--also called (factorial) cumulant functions--or, in
integrated form,
Thiele's semi-invariants~\cite{thiele}. A formal proof of this property was
given by Kubo~\cite{kubo} (see also Chang et al.~\cite{Chang:69}).
The cumulant correlation functions are defined as in the cluster expansion
familiar from statistical mechanics via the sequence
{}~\cite{kahn:uhlenbeck,huang,Mue71}:
\begin{eqnarray}
\rho_1(1)& =& C_1(1),\\
\rho_2(1,2)& =& C_1(1)C_1(2) +C_2(1,2),\\
\rho_3(1,2,3)& =& C_1(1)C_1(2)C_1(3)
+C_1(1)C_2(2,3)
+C_1(2)C_2(1,3)
+\nonumber\\
& &\mbox{}
+C_1(3)C_2(1,2)+C_3(1,2,3);
\end{eqnarray}
and, in general, by
\begin{eqnarray}
\rho_m(1,\ldots,m) &=& \sum_{{\{l_i\}}_m}\sum_{\text{perm.}}
\underbrace{\left[C_1()\cdots C_1()\right]}_{l_1\,\text{factors}}
\underbrace{\left[C_2(,)\cdots C_2(,)\right]}_{l_2\,\text{factors}}
 \cdots\nonumber\\
& & \cdots \underbrace{\left[C_m(,\ldots,)\cdots C_m(,\ldots,)
\right]}_{l_m\,\text{factors}}.
\label{a:4}
\end{eqnarray}
Here, $l_i$ is either zero or a positive integer and the sets of integers
$\{l_i\}_m$ satisfy the condition
\begin{equation}
\sum_{i=1}^m i\, l_i=m.\label{a:5}
\end{equation}
The arguments in the $C_i$ functions are to be filled by the $m$ possible
momenta in any order. The sum over permutations is a sum over all
distinct ways of filling these arguments. For any given factor product there
are precisely~\cite{huang}
\begin{equation}
\frac{m!}{
\left[(1!)^{l_1} (2!)^{l_2}\cdots(m!)^{l_m}\right] {l_1!}{l_2!}\cdots{l_m!}}
\label{a:6}
\end{equation}
terms.
The complete set of relations is contained in the  functional identity:
\begin{equation}
\GEN^{\incl}\left[u(y)\right]=\exp{\left\{g\left[u(y)\right]\right\}}\ ,
\label{dr:11}
\end{equation}
where
\begin{equation}
g\left[u(y)\right]=\int\rho_1(y) u(y)\,\rd y +\sum_{q=2}^\infty\frac{1}{q!}
\int_\Omega\,C_q(\XVEC{y}{q})\,u(y_1)\ldots u(y_q)\DXX{y}{q}{i}.
\label{dr:12}
\end{equation}
It follows that
\begin{equation}
C_q(\XVEC{y}{q})=
\left.\frac{\partial^q g\left[u(y)\right]}{%
\partial u(y_1)\ldots\partial u(y_n)}\right|_{u=0}.
\label{dr:13}
\end{equation}

The relations (\ref{a:4}) may be inverted with the result:
\begin{eqnarray}
C_2(1,2)&=&\rho_2(1,2) -\rho_1(1)\rho_1(2)\ ,\nonumber\\
C_3(1,2,3)&=&\rho_3(1,2,3)
-\sum_{(3)}\rho_1(1)\rho_2(2,3)+2\rho_1(1)\rho_1(2)\rho_1(3)\ ,\nonumber\\
C_4(1,2,3,4)&=&\rho_4(1,2,3,4)
-\sum_{(4)}\rho_1(1)\rho_3(1,2,3)
-\sum_{(3)}\rho_2(1,2)\rho_2(3,4)\nonumber\\
&&\mbox{} +2\sum_{(6)}\rho_1(1)\rho_1(2)\rho_2(3,4)-6\rho_1(1)
\rho_1(2)\rho_1(3)\rho_1(4).
\label{a:4b}
\end{eqnarray}
In the above relations  we have abbreviated $C_q(\XVEC{y}{q})$ to
 $C_q(1,2,\ldots,q)$; the summations indicate that all possible permutations
have to be taken (the number under the
summation sign indicates the number of  terms).
Expressions for higher orders can be derived from the related formulae given
in~\cite{kendall}.

It is often convenient to divide  the functions
$\rho_q$ and $C_q$ by the product of one-particle densities. This leads to
the  definition of the  normalized inclusive densities and correlations:
\begin{eqnarray}
r_q(\XVEC{y}{q}) &=& \rho_q(\XVEC{y}{q})/\rho_1(y_1)\ldots
\rho_1(y_q),\label{3.8}\\
K_q(y_1,\ldots,y_q)& =& C_q(y_1,\ldots,y_q)/\rho_1(y_1)\ldots
\rho_1(y_q).\label{3.9}
\end{eqnarray}

{}From expression (\ref{dr:11}) it can be deduced that, at finite energy, an
infinite number of $C_q$ will be non-vanishing: The densities
$\rho_q$ vanish for $q>N$, where $N$ is the maximal number of particles
in $\Omega$  allowed e.g. by energy-momentum conservation. As a consequence,
the functional $\GEN$ is a ``polynomial'' in $u(y)$. This in turn requires
the exponent in (\ref{dr:11}) to be an ``infinite series'' in $u(y)$.
In other words, the higher-order correlation functions must cancel
the lower-order ones that contribute to a vanishing  density function.
Phenomenologically, this implies that it is meaningful to use
correlation functions $C_q$ only if the number of correlated particles
in the considered phase-space domain $\Omega$ is considerably smaller
than the average multiplicity in that region~\cite{Brown}. These
conditions are not always fulfilled in present-day experiments for very
small phase-space cells, with the exception of perhaps $AA$-collisions.

\subsection{Correlations for particles of different species}\\
The generating functional technique of Sect.~2.1.1 can be extended to
the general situation where several different species of particles are
distinguished. This will not be pursued here and we refer to the literature
for details~\cite{Brown,koba,webber:72,eggers:phd}.
Considering two particle species a and b, the two-particle rapidity
correlation function is of the form:
\begin{equation}
C_2^{\ra\rb}(y_1,y_2)= \rho_2^{\ra\rb}(y_1,y_2) -f \rho_1^\ra(y_1)
\rho_1^\rb(y_2),
\label{dr:ex1}
\end{equation}
with
\begin{equation}
\rho^{\ra}_1(y_1)=\frac{1}{\sigma_{\inel}}\,\frac{d\sigma^\ra}{\rd y_1}\,;
\quad
\rho^{\ra\rb}_2(y_1,y_2)=\frac{1}{\sigma_{\inel}}\,\frac{\rd\sigma^{\ra\rb}}
{\rd y_1\rd y_2}.
\label{dr:ex2}
\end{equation}
Here, $y_1$ and $y_2$ are the c.m. rapidities, $\s_{\inel}$ the inelastic
cross section and a, b represent particle properties, e.g. charge.

The normalization conditions are:
\begin{equation}
\int \r^\ra_1(y_1) \rd y_1 = \lan n_\ra \ran\;;  \int\hs-2mm\int
\r^{\ra\rb}_2(y_1, y_2) \rd y_1 \rd y_2 = \lan n_\ra(n_\rb-\d^{\ra\rb})\ran\ ,
\label{dr:2.3}
\end{equation}
\begin{equation}
\int\hs-2mm \int C^{\ra\rb}_2(y_1, y_2) \rd y_1 \rd y_2 = \lan n_\ra
(n_\rb-\d^{\ra\rb})\ran - f\lan n_\ra\ran \lan n_\rb\ran \ \ ,
\label{dr:2.4}
\end{equation}
where $\d^{\ra\rb}=0$ for the case when a and b are particles of different
species and $\d^{\ra\rb}=1$ for identical particles, and $n_\ra$ and $n_\rb$
are the corresponding particle multiplicities.

Most experiments use
\begin{equation}
f=1\ \ ,
\label{dr:2.5a}
\end{equation}
so that the integral over the correlation function (equal to the ratio
$\bar n^2/k$ of
the negative binomial parameters~\cite{GiovHove86}) vanishes for the case of
a Poissonian multiplicity distribution.
Other experiments use
\begin{equation}
f = \frac{\lan n_\ra(n_\rb-\d^{\ra\rb})\ran}{\lan n_\ra\ran\lan n_\rb\ran}
\label{dr:2.5b}
\end{equation}
to obtain a vanishing integral also for a non-Poissonian multiplicity
distribution.

To be able to compare the various experiments, we use both definitions and
denote the correlation function $C_2^{\ra\rb}(y_1,y_2)$ when following
definition (\ref{dr:2.5a}) and $C_2^{'\ra\rb}(y_1,y_2)$
when following definition (\ref{dr:2.5b}).
We, furthermore,
use a reduced form of definition (\ref{dr:2.5b}),
\begin{equation}
\tilde C^{\ra\rb}_2(y_1,y_2)=C'^{\ra\rb}_2(y_1,y_2)/\lan n_\ra(n_\rb-
\d^{\ra\rb})\ran.
\label{dr:2.6}
\end{equation}

\ni
The corresponding normalized correlation functions
\begin{equation}
K^{\ra\rb}_2 (y_1,y_2) = {C^{\ra\rb}_2(y_1,y_2)\over f \r^\ra_1(y_1)
\r^\rb_1(y_2)}
\label{dr:2.7}
\end{equation}
follow the relations
\begin{equation}
{K_2}' = {1\over f} (K_2+1) - 1\ \ ,
\label{dr:2.8}
\end{equation}
and $\tilde K_2$ is defined as $\tilde K_2={K_2}'$.
These are more appropriate than $C_2$ when comparisons have to be performed
at different average multiplicity and are less sensitive to acceptance
problems.

The correlation functions defined by expressions
(\ref{dr:ex1})-(\ref{dr:2.8}),  contain a
pseudo-correl\-ation due to the summation of events with different charge
multiplicity $n$ and different
semi-inclusive single-particle densities $\rho^{(n)}_1$.

The relation between inclusive and semi-inclusive correlation functions has
been carefully analyzed in~\cite{Carr90}. Let $\s_n$ be the topological
cross section and
\begin{equation}
P_n = \s_n/\S\s_n\ \ .
\end{equation}
The semi-inclusive rapidity single- and two-particle densities
for particles a and b are defined as
\begin{equation}
\r^{(n)}_1(y)={1\over \s_n}{\rd\s^a_n\over \rd y}\ \ {\rm and}\ \
\r^{(n)}_2(y_1,y_2)={1\over \s_n} {\rd\s^{ab}_n\over \rd y_1\rd y_2}\ .
\label{dr:2.12}
\end{equation}
The inclusive correlation function $C_2(y_1,y_2)$ can then be written as
\begin{equation}
C_2(y_1,y_2) = C_\rS (y_1,y_2) + C_\rL(y_1,y_2)\ ,
\label{dr2.9}
\end{equation}
where
\begin{equation}
C_\rS(y_1,y_2) = \S P_n C_2^{(n)} (y_1,y_2)  \label{cs37}
\end{equation}
\begin{equation}
C_\rL(y_1,y_2) = \S P_n \D\r^{(n)}(y_1)\D\r^{(n)}(y_2)
\end{equation}
with $C^{(n)}_2(y_1,y_2) = \r_2^{(n)}(y_1,y_2) - \r_1^{(n)}(y_1)
\r_1^{(n)}(y_2)$ and $\D\r^{(n)}(y)=\r_1^{(n)}(y)-\r_1(y)$. In
{}~(\ref{cs37}) $C_\rS$ is the average of the semi-inclusive correlation
functions (often misleadingly denoted as ``short-range") and is more
sensitive to dynamical correlations. The term $C_\rL$ (misleadingly called
``long-range") arises from mixing different topological single-particle
densities.

A normalized form of $C_\rS$ can be defined as
\begin{equation}
K_{\rS} (y_1,y_2) = {C_{\rS}(y_1,y_2)\over \sum_nP_n\r^{(n)}_1(y_1)
\r^{(n)}_1(y_2)}={\sum_nP_n\r^{(n)}_2(y_1,y_2)\over
\sum_nP_n\r^{(n)}_1(y_1)\r^{(n)}_1(y_2)} - 1
\ \ .\label{dr:2.13}
\end{equation}
$C'_{\rS}$ and $\tilde C_{\rS}$ and their normalized forms $K'_{\rS}$ and
$\tilde K_{\rS}$ are defined accordingly, with the averages $\lan n\ran$
and $\lan n_\ra(n_\rb-\d^{\ra\rb})\ran$ replaced by $n$ and
$n_\ra(n_\rb-\d^{\ra\rb})$,
respectively.

Analogous expressions may be derived for  three-particle correlations.  They
are discussed in Sect.~3.4.

\subsection{Factorial and cumulant moments}\\
When the parametric function $z(y)$ is replaced by a constant $z$, the
generating functionals reduce to the generating function for
the multiplicity distribution.
Indeed, the probability $P_n$ for producing  $n$ particles is
given by
\begin{equation}
P_n=\sigma_n^{\excl}/\sigma_{\inel}
\label{dr:14}
\end{equation}
and we have
\begin{eqnarray}
G(z)&=&\sum_{n=0}^\infty P_n(1+z)^n=\GEN^{\excl}[z+1]=
\GEN^{\incl}[z]\\
&=& 1 +\sum_{q=1}^\infty \frac{z^q}{q!}\int_\Omega\rho_q(\XVEC{y}{q})\,
\rd y_1\ldots \rd y_q\\
&=& 1 +\sum_{q=1}^\infty \frac{z^q}{q!}\;\tilde{F}_q\ .
\label{dr:15}
\end{eqnarray}
The $\FT_q$ are the unnormalized factorial (or binomial) moments
\begin{eqnarray}
\tilde F_q \ \equiv \ \FMM{n}{q}&\equiv& \aver{n(n-1)\ldots(n-q+1)}\nonumber\\
&=& \int_\Omega \rd y_1\ldots\int_\Omega \rd y_q\;
\rho_q(\XVEC{y}{q})\nonumber\\
&=& \sum_n P_n\, n(n-1)\ldots(n-q+1).
\label{dr:16}
\end{eqnarray}
This relation can (formally) be inverted. If $P_n=0$ for
$n>N$ then an approximation for $P_n$ is given by:
\begin{equation}
P_n=\frac{1}{n!}\sum_{j=0}^{N-n} (-1)^j \frac{\tilde{F}_{j+n}}{j!}
\quad (n=0,1,\ldots N),
\label{dr:28b}
\end{equation}
and $P_n$ is included between any two successive values obtained by
terminating the sum at $j=s$ and $j=s+1$, respectively.

In (\ref{dr:16}) $n$ denotes the multiplicity in $\Omega$ and the average is
taken over
the ensemble of events. All the integrals are taken over the
same volume $\Omega$ such that $y_i\in\Omega$  $\forall i\in\{1,\ldots,q\}$.
Using  the correlation-function cluster decomposition, one further has
\begin{equation}
\log G(z)=\aver{n}z +\sum_{q=2}^{\infty} \frac{z^q}{q!}\; f_q.
\label{dr:17}
\end{equation}
The $f_q$ are the unnormalized factorial cumulants, also known as Mueller
moments~\cite{Mue71}
\begin{equation}
f_q=
\int_\Omega \rd y_1\ldots\int_\Omega \rd y_q\, C_q(\XVEC{y}{q}),
\label{dr:18}
\end{equation}
the integrations being performed as in (\ref{dr:16}).
The quantities $\tilde{F}_q$ and $f_q$ are easily found if $G(z)$ is
known:
\begin{eqnarray}
\tilde{F}_q&=& \left.\frac{\rd^q G(z)}{\rd z^q}\right|_{z=0}\ \ \ ,\\
f_q        &=& \left.\frac{\rd^q \log{G(z)}}{\rd z^q}\right|_{z=0}\\
\noalign{and}
P_q&=& \frac{1}{q!}\left.\frac{\rd^q G(z)}{\rd z^q}\right|_{z=-1}.
\end{eqnarray}
Using Cauchy's theorem, this can also be written as
\begin{equation}
P_n=\frac{1}{2\pi i}\oint\frac{G(z)}{(1+z)^{n+1}}\,\rd z,
\label{dr:33a}
\end{equation}
where the integral is on a circle enclosing $z=-1$.
Equation~(\ref{dr:33a}) is sometimes useful in deriving   asymptotic
expressions for $P_n$ in terms of factorial moments or
cumulants~\cite{weisberger,Mue71}.

As a simple example, we consider  the Poisson distribution
$$P_n = \re^{-\aver{n}}\frac{\aver{n}^n}{n !}\ \ \ ,$$
for which
\begin{equation}
G(z)=\sum_0^\infty P_n\,(1+z)^n=\exp{\{\aver{n}z\}}\ \ \ ,
\label{dr:19}
\end{equation}
showing that $f_q\equiv0$ for $q>1$.
In that case one has:
\begin{equation}
\tilde{F}_q=\aver{n(n-1)\ldots(n-q+1)}=\aver{n}^q.
\label{dr:ex3}
\end{equation}

The expressions  of density functions in terms of  cumulant correlation
functions, and the reverse relations,
are duplicated for their integrated counterparts. They
follow directly from the equations:
\begin{eqnarray}
1+\sum_{q=1}^{\infty} \frac{z^q}{q!}\,\tilde{F}_q&=&\exp{\{\aver{n}z +
\sum_{q=2}^\infty \frac{z^q}{q!}f_q  \}}\label{dr:37}\\ \noalign{or\hfill}
\log\left(1+\sum_{q=1}^{\infty} \frac{z^q}{q!}\,\tilde{F}_q\right)&=&
\aver{n}z+\sum_{q=2}^{\infty} \frac{z^q}{q!} f_q\label{dr:38}
\end{eqnarray}
by expanding either the exponential  in (\ref{dr:37}) or the logarithm
in (\ref{dr:38})  and equating the
coefficients of the same power of $z$. One finds~\cite{kendall}:

\begin{eqnarray}
\tilde{F}_1&=& f_1\ \ \ , \nonumber\\
\tilde{F}_2&=& f_2+f_1^2\ \ \ ,\nonumber\\
\tilde{F}_3&=& f_3+ 3f_2f_1+f_1^3\ \ \ ,\nonumber\\
\tilde{F}_4&=& f_4 + 4f_3f_1 + 3f_2^2 +6f_2f_1^2 +f_1^4\ \ \ ,\nonumber\\
\tilde{F}_5&=& f_5 + 5f_4f_1 +10 f_3f_2 +10f_3f_1^2 +15f_2^2 f_1 +10f_2f_1^3
  +f_1^5\ ;
\label{dr:37a}
\end{eqnarray}
and in general:
\begin{equation}
{\tilde{F}_q} =  q! \sum_{{\{l_i\}}_q} \prod_{j=1}^q
\left(\frac{{f_j} }{j!}\right)^{l_j} \frac{1}{l_j!}\ , \label{dr:39}
\end{equation}
with the summation as in (\ref{a:4}) and
$\sum_{l=1}^q i\, l_i=q$.

The latter formula can also be written as:
\begin{equation}
\tilde{F}_q=\sum_{l=0}^{q-1} \left(\!\begin{array}{c}
q-1\\ l \end{array} \!\right)\,f_{q-l}\,\tilde{F}_l \ ,
\label{dr:20}
\end{equation}
(with $\tilde{F}_0\equiv1$, $f_0\equiv0$)
and  is well-suited for computer calculation.
An equivalent relation was derived in~\cite{Mue71}.
The (ordinary) moments:
\begin{equation}
\mu_q=\aver{n^q}=\sum_{n=0}^{\infty} n^q\,P_n\ \ ,
\label{dr:39a}
\end{equation}
may be derived from the moment generating function
\begin{equation}
M(z)=\sum_{n=0}^\infty \re^{\textstyle nz}P_n\ \ ,
\label{dr:40}
\end{equation}
since
\begin{equation}
\mu_q=\left.\frac{\rd^qM(z)}{\rd z^q}\right|_{z=0}.
\label{dr:41}
\end{equation}
We note the useful relations
\begin{eqnarray}
M(z)&=&G\left(\re^{\textstyle z}-1\right)\ \ ,\\
G(z)&=&M\left(\log(1+z)\right)\ \ .
\label{dr:41a}
\end{eqnarray}

Moments and factorial moments are related to each other by  series
expansions.
{}From the identities~\cite{abramowitz}:
\begin{eqnarray}
n(n-1)\ldots(n-q+1)&=&\sum_{m=0}^{q}S_q^{(m)}\,n^m\ \ ,\\
n^q&=&\sum_{m=0}^q{\cal S}_q^{(m)}\,n(n-1)\ldots(n-m+1)\ \ ,
\label{dr:42}
\end{eqnarray}
where $S_q^{(m)}$ and ${\cal S}_q^{(m)}$ are Stirling numbers of the first
and second kind, respectively, follows directly:
\begin{eqnarray}
\tilde{F_q}&=&\sum_{m=0}^{q}S_q^{(m)}\;\mu_m\ \ ,\\
\mu_q&=&\sum_{m=0}^q{\cal S}_q^{(m)}\;\tilde{F}_m\ \ .
\label{dr:43}
\end{eqnarray}

Cumulants  $\kappa_q$ can be defined in terms of the moments $\mu_q$ in
the standard way~\cite{cramer,kendall}. They obey relations identical to
(\ref{dr:37a}). The cumulants are integrals of the type (\ref{dr:18}) of
differential quantities known as density moments. These are discussed in
{}~\cite{weingarten,koba:weingarten}. Relations expressing central moments
in terms of factorial moments via non-central Stirling numbers are
derived in~\cite{Kout82}.

\subsection{Cell-averaged factorial moments and cumulants;\\
 generalized moments}\\
In practical work, with limited statistics,
 it is almost always necessary to perform
averages over more than  a single phase-space cell.  Let $\Omega_m$
be such a cell (e.g. a single rapidity interval of size $\delta y$) and
divide the phase-space volume into $M$ non-overlapping cells $\Omega_m$ of
size $\d\Omega$, independent of $m$.
Let $n_m$ be the number of particles in cell $\Omega_m$.
Different cell-averaged moments may be considered, depending
on the type of averaging.

Normalized factorial moments~\cite{bialas1,bialas2},
which have become known as {\it  vertical moments},
are defined as\footnote{\ Here and in the following we consider rapidity space
for definiteness}
\begin{eqnarray}
F^\rV_q(\delta y)&\equiv&
\frac{1}{M}\;\sum_{m=1}^M \frac{\aver{\FACT{n}{q}}}{\aver{n_m}^q}
\label{dr:44}\\
&\equiv & \frac{1}{M} \sum^M_{m=1}
\frac{\int_{\delta y} \rho_q(y_1,\ldots,y_q) \prod^q_{i=1} \rd y_i}
{\left(\int_{\d y} \r(y)dy\right)^q}\ \ , \nonumber \\
&= & \frac{1}{M(\d y)^q} \sum^M_{m=1} \int_{\d y} \frac{\r_q(y_1,\dots,y_q)
\prod^q_{i=1} \rd y_i}{\left(\bar\r_m\right)^q}. \label{dr:45}
\end{eqnarray}
The full rapidity interval $\D Y$ is divided into $M$ equal bins:
$\D Y=M\delta y$; each $y_i$ is within the $\delta y$-range and
$\aver{n_m}\equiv\overline{\rho}_m\delta y \equiv \int_{\d y} \r_1(q)\rd y$.

One may also define normalized {\it  horizontal moments}
by
\begin{equation}
F^\rH_q(\delta y)\equiv \frac{1}{M}\;\sum_{m=1}^M
\frac{\aver{\FACT{n}{q}}}{\aver{\overline{n}_m}^q}\ \ .
\label{dr:46}
\end{equation}
with $\overline{n}_m=\sum_m n_m/M;\quad \aver{\overline{n}_m}=\aver{n}/M;
\quad n=\sum_mn_m$.

Horizontal and vertical moments are equal if $M=1$. Vertical moments
are normalized locally and thus sensitive only to fluctuations within each
cell but not to the overall shape of the single-particle density.
Horizontal moments are sensitive to the shape of the single-particle density
in $y$ and further depend on the correlations between cells.
To eliminate the effect of a  non-flat rapidity distribution,
it was suggested to either introduce correction factors~\cite{Fial89}
or use ``cumulative'' variables which transform
an arbitrary distribution into a uniform one~\cite{Ochs,BiGa90}.

Likewise, cell-averaged normalized factorial cumulant moments may be
defined as
\begin{equation}
K_q(\delta y) = \frac{1}{M(\delta y)^q}
 \sum^M_{m=1} \int\limits_{\d y} \prod_i\rd y_i
\frac{C_q(y_1,\ldots,y_q)}{(\bar\r_{m})^q}\ \ \ . \label{dr:47}
\end{equation}
They are related~\cite{CaES91} to the factorial moments by\footnote{\ The
higher-order relations can be found in~\cite{CaES91}}
\begin{eqnarray}
F_2&=&1+K_2\ \ \ ,\nonumber\\
F_3&=&1+3K_2+K_3\ \ \ ,\nonumber\\
F_4&=&1+6K_2+3\overline{K^2_2}+4K_3+K_4 \,.\label{dr:48}
\end{eqnarray}

In $F_4$ and higher-order moments, ``bar averages" appear. They are defined
as $\ol{AB}\equiv\sum\limits_m A_mB_m/M$.

Besides factorial and cumulant moments, other measures of multiplicity
fluctuations have been proposed. In particular,
$G$-moments~\cite{Feder88}---known in statistics as
frequency moments~\cite{kendall}---were
extensively used to investigate
whether multiparticle processes possess
(multi)fractal properties~\cite{Hwa90,Hwa90-91}.
$G$-moments are defined as
\beq
G_q=\sum^{M}_{m=1}\hs-1mm ' p^q_m , \hs2truecm p_m=n_m/n, \hs2truecm
n=\sum^M_{m=1}n_m\ \ .
\label{gghwa}
\eeq
Also here, $n_m$ is the number of particles in bin $m$,
the absolute frequency;  $n$ is the total
multiplicity in an initial interval
and $M$ is the number of bins at ``resolution'' $M$.
Bins with zero content (``empty bins'')
are excluded in the sum, so that $q$ can cover the whole
spectrum of real numbers. For $q$ negative, $G_q$ is sensitive to ``holes"
in the rapidity distribution of a single  event.
Note that $p_m$ in (\ref{gghwa}) is not a probability but
a relative frequency or ``empirical measure'' in
modern terminology. For small $n$, $G$-moments are very sensitive to
statistical fluctuations (``noise''),  especially  for large M.
This seriously limits their potential.
In attempts to  reduce this noise-sensitivity,  modified definitions  have
been proposed in~\cite{Flor91}.

\subsection{Multivariate distributions}\\
The  univariate factorial moments $\tilde{F}_q$
characterize multiplicity fluctuations in a single phase-space cell and thus
reflect only local properties. More information is contained in the
correlations between fluctuations (within the same event) in two or more
cells. This has led to consider multivariate factorial moments. For
non-overlapping cells, the 2-fold factorial moments, also called
{\it correlators}, are defined as:
\begin{equation}
\tilde{F}_{pq}=\aver{n^{[p]}_m {n}^{[q]}_{m'}},
\label{f4:1}
\end{equation}
where $n_m$ ($n_{m'}$) is the number of particles  in cell $m$ (cell $m'$). A
normalized version of the two-fold {correlator}  is discussed in
{}~\cite{bialas1} and  defined as:
\begin{equation}
F_{pq}=\frac{\lan n^{[p]}_m n^{[q]}_{m'}\ran}{\tilde{F}_p \tilde{F}_q}\ .
\label{f4:2}
\end{equation}
For reasons of statistics, these quantities are usually averaged over many
pairs of cells, keeping  the ``distance'' ($D$) between the cells
constant\footnote{\ In one-dimensional rapidity space, $D$ is
defined as the distance between the centers of two rapidity intervals; in
multidimensional phase space  a proper metric must first be defined}.
This averaging procedure requires the same precautions regarding
stationarity of single particle densities as for their single-cell
equivalents.

Multi-fold factorial moments are a familiar tool in radio- and
radar physics and in quantum optics~\cite{saleh}. There, they  relate to
simultaneous measurement of photo-electron counts
detected in,  say $M$, time-intervals,
or in $M$ space points,  leading to a joint probability distribution
$P_M(n_1,n_2,\ldots,n_M)$. The importance of
multi-fold moments derives from the fact that, e.g. in the
simplest case of two cells, $\tilde{F}_{11}=\aver{n_m n_{m'}}$ is directly
related to the auto-correlation function of the radiation field and obeys,
for small cells, the Siegert-relation~\cite{saleh}, whatever the
statistical properties of the field.
The higher-order moments are sensitive to higher-order correlations
and to the phase of the field.

Factorial moments and factorial correlators are intimately related
quantities. In terms of inclusive densities one has:
\begin{equation}
 \tilde{F}_{pq}=
  \int_{\Omega_1}\,\rd y_1\ldots \rd y_p \int_{\Omega_2} \rd y_{p+1}\ldots
\rd y_{p+q}\, \rho_{p+q}(y_1,\ldots,y_p;y_{p+1},\ldots,y_{p+q}),\label{f4:3}
\end{equation}
where $\rho_{p+q}$ is the inclusive density of order $p+q$.
The integrations are performed over two  arbitrary (possibly overlapping)
phase-space cells
$\Omega_1$ and $\Omega_2$, separated by a ``distance'' $D$.

It should be noted that
the definition (\ref{f4:3}) is more general than (\ref{f4:1}).
For $\Omega_1=\Omega_2$  or $D=0$, (\ref{f4:3}) reduces to the
 correct definition of $\tilde{F}_2$
whereas  (\ref{f4:1}) is, in this case,
equal to  $\aver{n^2}$ and misses the so-called
``shot-noise'' term $-\aver{n}$.

Factorial moments and factorial correlators of the same order
are thus seen to differ only in the choice of the integration domains.
Note that for $p\neq q$, definition (\ref{f4:3}) is not symmetric
in $p$ and $q$ and a symmetrized version is often used in experimental work:
\begin{equation}
\tilde{F}^{(s)}_{pq}=(\tilde{F}_{pq}+\tilde{F}_{qp})/2\,.
\label{dr:49}
\end{equation}
{}From (\ref{f4:3}) follows that $F_{11}$ is directly derivable from measured
two-particle correlation functions or from appropriate analytical
parametrizations. Higher-order correlators involve higher-order
density-functions which, in general,  are unknown.

We now turn to a discussion of multivariate factorial cumulants.
For $M$ non-overlapping cells, we introduce the $M$-variate multiplicity
distribution $P_M(\XVEC{n}{M})$ and the corresponding moment- and
factorial-moment generating functions:
\begin{eqnarray}
\hs-7mm M(\XVEC{z}{M})&=&\sum_{n_1=0}^\infty\sum_{n_2=0}^\infty\cdots
\sum_{n_M=0}^\infty \;e^{z_1n_1+\cdots+ z_Mn_M}\,P_M(\XVEC{n}{M})\ \ ,
\label{dr:50}\\
\hs-7mm G(\XVEC{z}{M})&=&\sum_{n_1=0}^\infty\sum_{n_2=0}^\infty\cdots
\sum_{n_M=0}^\infty
\;(1+z_1)^{n_1}\cdots(1+z_M)^{n_M}\,P_M(\XVEC{n}{M})\ \ ,\label{dr:51}
\end{eqnarray}
from which the $M$-variate moments are easily obtained by differentiation:
\begin{eqnarray}
\hs-7mm \mu_{q_1\dots q_M}=\aver{n_1^{q_1}\dots n_M^{q_M}}&=&
\left.\left(
\frac{\partial}{\partial z_1}\right)^{q_1}\cdots\left(
\frac{\partial}{\partial z_M}\right)^{q_M}\; M(\XVEC{z}{M})
\right|_{z_1=\cdots z_M=0} , \label{dr:52}\\
\hs-7mm \tilde{F}_{q_1\dots q_M}=\aver{n_1^{[q_1]}\dots n_M^{[q_M]}}&=&
\left. \left( \frac{\partial}{\partial z_1}\right)^{q_1}\cdots\left(
\frac{\partial}{\partial z_M}\right)^{q_M}\; G(\XVEC{z}{M})
\right|_{z_1=\cdots z_M=0} . \label{dr:53}
\end{eqnarray}

The multivariate (ordinary) cumulants $\kappa_{q_1\dots q_M}$ and multivariate
factorial cumulants $f_{q_1\dots q_M}$ are likewise
obtained by replacing $M(.)$ and $G(.)$ in (\ref{dr:52}) and
(\ref{dr:53}) by their respective natural  logarithms~\cite{Cantrell:70}.
The same expressions serve to extend the relations between univariate
moments and cumulants to their multivariate counterparts.

For $M=2$ and non-overlapping cells, one has the identity
[cfr.~(\ref{dr:37})]:
\begin{equation}
\sum_{l=0}^{\infty}\sum_{m=0}^{\infty} \frac{(z_1)^l(z_2)^m}{l!\,m!}
\tilde{F}_{lm}=
\exp{( \sum_{l=0}^{\infty} \sum_{m=0}^{\infty}
\frac{(z_1)^l(z_2)^m}{l!\,m!} f_{lm})},\label{s1:14}
\end{equation}
where $\tilde{F}_{00}\equiv1$ and $f_{00}$ is defined equal to zero.
It follows that\footnote{\ See also~\cite{eggers:correlators}}

\begin{eqnarray}
 \tilde{F}_{11} & = & f_{11}+f_{01} f_{10}, \label{eq:b7} \\
\tilde{F}_{12} & = &
f_{12}
+f_{01} f_{20}
+2 f_{10} f_{11}
+f_{01} f_{10}^2, \label{eq:b8} \\
\tilde{F}_{13} & = &
f_{13}
+f_{01} f_{30}
+3 f_{11} f_{20}
+3 f_{01} f_{10} f_{20}
+3 f_{10} f_{12}
+3 f_{10}^2 f_{11}
+f_{01} f_{10}^3, \label{eq:b9} \\
\tilde{F}_{2 2} & = &
f_{2 2}
+2 f_{10} f_{2 1}
+f_{02} f_{20}
+f_{01}^2 f_{20}
+2 f_{01} f_{12}
+2 f_{11}^2
+4 f_{01} f_{10} f_{11}
\nonumber\\& &\mbox{}%
+f_{02} f_{10}^2
+f_{01}^2 f_{10}^2\ \ . \label{eq:b10}
\end{eqnarray}

Similarly, expanding the logarithm in
\begin{equation}
\sum_{l=0}^{\infty}\sum_{m=0}^{\infty} \frac{(-z_1)^l(-z_2)^m}{l!\,m!}
f_{lm}=
\log{\left(\sum_{l=0}^{\infty} \sum_{m=0}^{\infty}
\frac{(-z_1)^l(-z_2)^m}{l!\,m!} \tilde{F}_{lm}\right)},\label{s1:14bis}
\end{equation}
in powers of $s$ and $t$
and identifying coefficients, the reverse relations follow:
\begin{eqnarray}
\hs-7mm f_{11}& = &
      \tilde{F}_{11}-\tilde{F}_{01}\tilde{F}_{10}\ \ ,\\
\hs-7mm f_{12}& = &
\tilde{F}_{12}
 -\tilde{F}_{01}\tilde{F}_{20}
-2\tilde{F}_{10}\tilde{F}_{11}
+2\tilde{F}_{01}\tilde{F}_{10}^2 \ \ ,\\
\hs-5mm f_{13}& = &
\tilde{F}_{13}
-\tilde{F}_{01}\tilde{F}_{30}
-3\tilde{F}_{11}\tilde{F}_{20}
+6\tilde{F}_{01}\tilde{F}_{10}\tilde{F}_{20}
-3\tilde{F}_{10}\tilde{F}_{12}
+6\tilde{F}_{10}^2\tilde{F}_{11}
-6\tilde{F}_{01}\tilde{F}_{10}^3 \ \ ,\\
\hs -7mm f_{22}& = &
 \tilde{F}_{22}
-2\tilde{F}_{10}\tilde{F}_{21}
-\tilde{F}_{02}\tilde{F}_{20}
+2\tilde{F}_{01}^2\tilde{F}_{20}
-2\tilde{F}_{01}\tilde{F}_{12}
-2\tilde{F}_{11}^2
\nonumber\\& &\mbox{}%
+8\tilde{F}_{01}\tilde{F}_{10}\tilde{F}_{11}
+2\tilde{F}_{02}\tilde{F}_{10}^2
-6\tilde{F}_{01}^2\tilde{F}_{10}^2\ \ .
\end{eqnarray}

The quantities $\tilde{F}_{0i}$, $\tilde{F}_{i0}$,
$f_{0i}$ and  $f_{i0}$ are equal to the single-cell
factorial moments and factorial cumulants, respectively.
Expressions for  $\tilde{F}_{ji}$ ($f_{ji}$)  are obtained from
the corresponding expression for $\tilde{F}_{ij}$
($f_{ij}$)   by permutation of the subscripts. By definition,
$f_{01}=\tilde{F}_{01}$ and $f_{01}$ is equal to the average multiplicity
in  cell 2.

It may be noted that the bivariate relations reduce to the
univariate ones (\ref{dr:37a}) by simply amalgamating the indices. For
example, from
\begin{equation}
\tilde{F}_{12}  =
f_{12}
+f_{01} f_{20}
+2 f_{10} f_{11}
+f_{01} f_{10}^2, \label{eq:b8a}
\end{equation}
one recovers, by summing the indices
\begin{equation}
\tilde{F}_3=f_3 +3 f_1f_2 +f_1^3.
\label{dr:54}
\end{equation}

It is shown in~\cite{kendall} (Sect.~13.12) that  the above relations, while
seemingly complex, have in fact a surprisingly elegant structure, rooted in
simple algebraic properties of completely symmetric functions. Further
discussion on this point and  other useful properties  may be found
in~\cite{Cantrell:70}.

Extensions to more than two cells is straightforward, in principle, but
involves tedious algebra.

\section{Poisson-noise suppression}
To detect dynamical fluctuations in the density of particles produced in a
high-energy collision, a way has to be devised to eliminate, or to reduce as
much as possible, the statistical fluctuations---noise---due to the finiteness
of the number of particles in the counting cell(s). This requirement can to a
large extent be satisfied by studying
factorial moments and their multivariate
counterparts. It forms the basis of the factorial moment technique, known in
optics, but rediscovered in multi-hadron physics in~\cite{bialas1,bialas2}.
The method rests on the conjecture that the multi-cell
multiplicity distribution $P_M(\XVEC{n}{M})$ can be written as
\newcommand{\PMN}{%
\int \rd\rho_1\ldots \rd\rho_M P_{\rho}(\rho_1,\ldots,\rho_M)
\prod_{m=1}^M\frac{(\rho_m \delta)^{n_m}}{n_m!}\exp{(-\rho_m \d)}
}
\begin{equation}
P_M(n_1,\cdots,n_M)= \PMN . \label{s1:1}
\end{equation}
The Poisson factors represent uncorrelated fluctuations of $n_m$
around the average $\rho_m \d=\aver{n_m}$ in $m$-th interval; $\delta$ is here
the size of the interval.
This can also be written as:
\begin{equation}
P_M(n_1\cdots n_M)=
\aver{\prod_{m=1}^M \frac{\aver{n_m}^{n_m}}{n_m!}
\exp{(-\aver{n_m})}}_\rho\ , \label{s1:2}
\end{equation}
where the outer brackets mean that an average is taken over the probability
distribution of the
densities $\rho_m$, which are subject only to dynamical fluctuations.
If these are absent, $P_{\rho}(\rho_1,\ldots,\rho_M)$ is simply a
product of $\delta$-functions.

The formulae (\ref{s1:1}-\ref{s1:2}) are formally identical to
the expression for the multi-interval photo-electron counting
probability distribution in quantum optics and based
on the famous Mandel formula~\cite{Mandel:58:1,Mandel:59:1}. The latter
relates the probability distribution of {\em the number of detected
photo-electrons}  to the statistical distribution of the {e.m. field}.

In optics, $\rho_m$ has the meaning of a space- or time-integrated
field intensity.
The ensemble
average is calculated from the field density matrix which
describes its statistical properties.

Equations   (\ref{s1:1})-(\ref{s1:2}) express $P_M(n_1,\ldots,n_M)$ as
a linear transformation of \break
$P_{\rho}(\rho_1,\ldots,\rho_M)$ with a ``Poisson kernel''.
This transformation is known as the\break
``Poisson Transform'' of $P_{\rho}$~\cite{Wolf:64}.

The Poisson-transform of a single-variable function $f(x)$ is the function
$\tilde{f}(n)$ ($n$ integer)  defined by the linear transformation
\begin{equation}
\tilde{f}(n) =\int_0^\infty dx\,f(x) \frac{x^n}{n!} e^{-x}.\label{a:pt1}
\end{equation}
A trivial example is the function $\delta(x-\mu)$ whose transform
is the Poisson probability distribution.
The Bose-Einstein distribution
\begin{equation}
\tilde{f}(n)= \frac{\mu^n}{(1+\mu)^{n+1}} \ \ \ \ (n=0,1,\ldots),\label{a:pt2}
\end{equation}
is obtained as the Poisson-transform of the exponential function
$(1/\mu) \exp(-x/\mu)$.

For suitably-behaved functions, the inverse Poisson-transform
exists. It is closely related to the Laplace-transform of $f(x)$.
Several practical methods have been developed  to determine the
function $f(x)$ from its Poisson-transform. Besides
methods based on series expansions, the inversion
problem may be reduced to an inverse moment problem.
This follows from the equality between the factorial moments of
$\tilde{f}(n)$ and the ordinary moments of $f(x)$, as further discussed
 below.

A table of useful transforms for probability distributions
and further mathematical properties can be found in~\cite{saleh}.

{}From the basic Poisson transform equation (\ref{s1:2}) it is easily seen
that
the multi-fold factorial moment generating function has the simple form
\begin{equation}
G(\XVEC{z}{M}) = \aver{\prod_{j=1}^M
\exp{(z_j\rho_j \d)}}_\r,\label{s1:7}
\end{equation}
where the statistical average is  again taken over the ensemble of densities
$\rho_1,\cdots,\rho_M$, as indicated by the subscript.

On the other hand, the (ordinary) moment generating function of the
densities is given by:
\begin{eqnarray}
Q(\XVEC{z}{M})&=&\int P_\rho(\XVEC{\rho}{M})
\re^{\rho_1z_1+\cdots+\rho_Mz_M}\rd\rho_1\ldots \rd\rho_M\\
&=&\left\langle\prod_{j=1}^M \re^{\rho_jz_j}\right\rangle_{\textstyle
\rho}\ \ .
\label{dr:55}
\end{eqnarray}

Comparing (\ref{s1:7}) and (\ref{dr:55}), it follows that:
\begin{equation}
G(\XVEC{z}{M})=Q(\delta\rho_1z_1,\ldots,\rho_Mz_M\delta)\ \ .
\label{dr:56}
\end{equation}
This equation implies that the normalized multi-variate factorial moments
of the multiplicity distribution
\begin{equation}
F_{q_1\ldots q_M}=\frac{%
\tilde{F}_{q_1\ldots q_M}}{%
\aver{n_1}^{q_1}\ldots\aver{n_M}^{q_M}}\
\label{dr:56a}
\end{equation}
 are equal to the
normalized multivariate (ordinary) moments of the relative density
fluctuation $\rho_m/\aver{\rho_m}$. This is the ``noise-suppression'' theorem
{}~\cite{bialas1,bialas2}. It assumes that the noise is Poissonian
(cfr.~(\ref{s1:1})) and that the number of counts in all intervals (the total
multiplicity) is unrestricted\footnote{\ If the sum over all intervals of the
number of counts is fixed, a slightly more complicated relation can be
obtained
if the noise has a Bernoulli (multinomial) distribution~\cite{bialas1}}.

The property of Poisson-noise suppression has made measurement of factorial
moments a standard technique, e.g. in quantum optics, to study the statistical
properties of arbitrary electromagnetic fields from photon-counting
distributions. Their utility was first explicitly recognized, for the single
time-interval case, in~\cite{Bedard:67:1} and~\cite{Chang:69} and later
generalized to the multivariate case in~\cite{Cantrell:70}. The authors of
{}~\cite{Chang:69} further stress the advantages of factorial
cumulants compared to factorial moments, since the former measure genuine
correlation patterns, whereas the latter contain additional
large combinatorial terms  which may mask the underlying
dynamical correlations (however, see the discussion in Sect.~2.1.2).

Multivariate factorial cumulants are derived from the (natural) logarithm
of the factorial moment generating function.
Taking logarithms of both sides of (\ref{dr:56}), one finds that the
 multivariate normalized {factorial cumulants} of the counting
distribution are equal to the multivariate normalized {ordinary cumulants}
of the densities $[\rho \d]$. This
relation, therefore, extends the noise-suppression theorem to cumulants.
This property is exploited  in many fields from quantum
 optics~\cite{Cantrell:70} to
radar-physics and astrophysics (see e.g.~\cite{Fry:85}).

\section{Sum-rules}
In an interesting $\alpha$-model analysis of factorial
 correlators~\cite{pesch:seixas}, scaling relations are derived between
single-variate and 2-variate factorial moments which are independent of
 the dimension of the phase space. The result is stated as follows:
If a correlator $F_{11}(D,\delta)$ is effectively independent of
 $\delta$ in a range $\delta<D\leq\delta_0$, then
\begin{equation}
F_{11}(D)=2F_2(2D) -F_2(D).
\label{dr2:1}
\end{equation}
Here, $\delta$ is the interval size and $D$ the distance between the
intervals.

Similar types of relations---or sum-rules---are well-known
in optics since the early 1970's. They are exploited
in so-called Multi-Cathode  and Multiple-Aperture Single-Cathode
(MASC) photo-electron counting experiments (see
e.g.~\cite{cantrell:fields,bures} and refs. therein).

Consider again the multivariate multiplicity distribution
$P_M(\XVEC{n}{M})$ giving the joint probability for the
occurrence of $n_1$ particles in a cell $\Omega_1$, $\ldots,$ $n_M$
particles in cell $\Omega_M$, with $\Omega_i\cap\Omega_j=0$,
$\forall i,j$ and $i\ne j$.
Let $n$ be the number of particles counted in the union of the $M$ cells,
\begin{equation}
n=\sum_{m=1}^Mn_m\ \ .
\label{dr2:2}
\end{equation}
The probability distribution of $n$ is given by
\begin{equation}
P(n)=\sum_{n_1=0}^{n}\cdots\sum_{n_M=0}^{n} p_M(\XVEC{n}{M})
\delta_{n,n_1+\cdots+n_M}\, .
\label{dr2:3}
\end{equation}
Define  the single-variate factorial moment generating function
\begin{equation}
g(z)=\sum_{n=0}^\infty(1+z)^n\,P(n)\,.
\label{dr2:4}
\end{equation}
The function $g(z)$ can be expressed in
terms of  the multivariate generating function
(\ref{dr:51}) as:
\begin{equation}
g(z)=\left.G_M(\XVEC{z}{M})\right|_{z_1=z_2=\cdots=z_M=z}.
\label{dr2:5}
\end{equation}
Equation~(\ref{dr2:5}) allows to express factorial moments of $n$ in terms of
the multivariate factorial moments of $\{\XVEC{n}{M}\}$.
Application of the  Leibnitz rule
$$ \left(\frac{\rd}{\rd\!z}\right)^k f(z)=\sum_{\{a_j\}}\frac{k!}
{a_1\,!a_2\,!\ldots a_k!}
\left(\frac{\rd}{\rd\,z}\right)^{a_1}\,f_1(z)\cdots
\left(\frac{\rd}{\rd\,z}\right)^{a_k}\,f_M(z)
$$
to the function
$$f(z)=f_1(z)\cdots f_M(z)$$
leads immediately to the relation
\begin{equation}
\tilde{F}_q=\sum_{\{a_j\}} \tilde{F}^{(M)}_{a_1\ldots a_M} \frac{q!}%
{a_1!\,\ldots\,a_M!}.
\label{dr2:6}
\end{equation}
The summation is over all sets $\{a_j\}$ of non-negative integers such that
$$\sum_{j=1}^{M}a_j=q.$$
Formula (\ref{dr2:6}) may be looked upon as a generalization of
the usual multinomial theorem for factorial moments\footnote{See also
{}~\cite{eggers:phd}}.

Likewise, taking the natural logarithm of
both sides of (\ref{dr2:5}), one obtains an identical
relation as (\ref{dr2:6}) among
single-variate and multivariate factorial cumulants.

As an example, for two rapidity bins ($M=2$) of size $\delta$ separated
by a distance $D$, one finds:
\begin{eqnarray}
\tilde{F}_2 &=& \tilde{F}^{(2)}_{02}
+2\tilde{F}^{(2)}_{11}+
\tilde{F}^{(2)}_{20}\ ,\nonumber\\
\tilde{F}_3 &=& \tilde{F}^{(2)}_{03}
+3(\tilde{F}^{(2)}_{12}+\tilde{F}^{(2)}_{21}) +
\tilde{F}^{(2)}_{30}\ ,\label{fff} \\
\tilde{F}_4 &=& \tilde{F}^{(2)}_{04}
+4(\tilde{F}^{(2)}_{13}+\tilde{F}^{(2)}_{31})
+6\tilde{F}^{(2)}_{22}
+\tilde{F}^{(2)}_{40}\,.\nonumber
\end{eqnarray}

The factorial moments $\tilde{F}_{0i}$  are
determined from the single cell (marginal) counting distribution, whereas
the univariate factorial moments $\tilde{F}_q$ are obtained from the
sum of the  counts in the two cells.

The relations derived in \cite{pesch:seixas} follow immediately
from (\ref{fff}) by considering two adjacent cells and normalizing
properly.
Since the derivation of (\ref{dr2:6}) is completely general,
it obviously holds irrespective of the dimension of  phase space.

The relations (\ref{fff}) are trivially  extended
to  more than two cells. They allow  to measure high-order
correlators by varying the distances between the cells.
In optics and radar-physics, they are typically used  in determining
spatial coherence properties of arbitrary e.m. fields.

\section{Scaling laws}

A major part of this paper is devoted to  recent
experimental and theoretical research on possible  manifestations
of scale-invariance in high-energy  multiparticle production
processes. This work centers around two basic inter-related notions:
intermittency and fractality.
A review of the experimental data accumulated over the last years
will be given in Chapter 4. Theoretical work is discussed in Chapter 5.

In particle physics, intermittency is defined, in a strict sense,
as the scale-invariance of factorial moments
(\ref{dr:44})-(\ref{dr:46})
with respect to changes in the size of phase-space cells (or bins)
say  $\d y$, for small enough $\d y$:
\beq
F_{q}(\d y)\propto (\d y)^{-\f_{q}} \;\;\;\;\;\;  (\d y \rightarrow 0).
\label{fq113}
\eeq
The power  $\f_q>0$ is a constant at any given (positive integer)
$q$ and called ``intermittency index'' or ``intermittency slope''.
The form of (\ref{fq113}) strictly implies that
the inclusive densities $\rho_q$ and the connected correlation functions
$C_q$ become singular in the limit of infinitesimal
separation ($\delta y\rightarrow0$) in momentum space.

Inspired by the theory of multifractals, scaling behaviour of the
$G$-moments (\ref{gghwa}) has also been looked for in the form
\beq
G_q (\d y)\propto (\d y)^{\tau (q)} \;\;\;\;\;\; (\d y \rightarrow 0).
 \label{gq114}
\eeq
To describe the inter-relation of the two proposals, we
briefly discuss the formalism of fractals.

Power-law dependence is typical for fractals~\cite{Man82}, i.e. for
self-similar objects with a non-integer dimension. These range from purely
mathematical ones (the Cantor set, the Koch curve, the Serpinsky gasket
etc.) to real objects of nature (coast-lines, clouds, lungs, polymers etc).
For  reviews see~\cite{Zeld87,Pala87,PescTH5891}.

The fractal dimension $D_\rF$ is defined as the exponent
 which provides a finite
limit
\beq
0 < \lim_{\epsilon \rightarrow 0} N(\epsilon)\epsilon^{D_{\rF}} < \infty
 \label{n115}
\eeq
for the product of $\e^{D_\rF}$ and the minimal number of
hypercubes $N(\epsilon)$
of linear size $l=\epsilon$ (Kolmogorov definition) or $l \leq \epsilon$
 (Hausdorff
definition) covering the object when $\epsilon \rightarrow 0$.

To a physicist, the definition becomes more transparent  if one considers the
relation between the size $l$ of an object and its mass $M$ as a scaling law:
\beq
M \propto l^{D_{\rF}}.   \label{m116}
\eeq
For usual objects $D_{\rF}$ coincides with the topological
dimension (for a line
$D_{\rF}=1$, for a square $D_{\rF}=2$ and so on).
The condition $\epsilon \rightarrow 0$ means
in practice that such a law should hold  in
some interval of ``rather small'' $\epsilon$-values.

The probability $p_{i}(l)$ to be in a hypercube $N_i(l)$ is proportional
to $l^{D_{\rF}}$ at small $l$. Therefore, for a fractal the mean value of the
$q$-th order (ordinary) moment is given by
\beq
\lan p_{i}^{q}(l) \ran
\propto l^{qD_{\rF}} \;\;\;\; (D_{\rF}=\const)\ \ .  \label{p117}
\eeq
Multifractals generalize the notion of fractals, since for these holds
\beq
\sum_{i}p_{i}^{q}(l)=\lan p_{i}^{q-1}(l)\ran \propto l^{\tau (q)}  \ \ ,
 \label{p118}
\eeq
where
\beq
\tau (q)=(q-1)D_{q}.  \label{t119}   \eeq
The $D_{q}$ are called the R\'enyi dimensions~\cite{Renyi70,Feder88} and
depend
on $q$ (generally, for multifractals they are decreasing functions of $q$).

Sometimes it is more convenient to characterize multifractals by
spectral properties, rather than by their dimensions.

Let us group all the boxes with a singularity $\a$ ($p_{i}(l)\sim l^{\a},
l\rightarrow 0$) into a subset $S(\a )$, where $\a $ is called the local mass
dimension. The number of boxes $\rd N_{\a }(l)$  needed to cover $S(\a )$ is
\beq
\rd N_{\a }(l)=\rd\r (\a )l^{-f(\a )}\ \ ,  \label{dn120}
\eeq
where $f(\a )$ is the fractal dimension of the set $S(\a )$ related to
the R\'enyi dimension. For the sum of moments one obtains:
\beq
\sum_{i=1}^{N_{i}(l)} p_{i}^{q}(l)\propto \int \rd\r (\a )l^{\a q-f(\a )}\ \ .
  \label{dr:121}
\eeq
{}From (\ref{dr:121}), one gets by the saddle-point method:
\beq
D_{q}=\frac{1}{q-1} \mbox{min}_{\a}\,\left(\a q-f(\a )\right)=
\frac{1}{q-1} \left(\bar\a q-f (\bar \a )\right)  \label{dq122}
\eeq
with $\bar \a $ defined as
\beq
\left.\frac {\rd f}{\rd\a}\right|_{(\a =\bar \a )}=q(\bar \a ). \label{df123}
\eeq
The notion of R\'enyi dimensions $D_q$ generalizes the notion of fractal
dimension $D_{0}=D_{\rF}$, information dimension $D_{1}$ and correlation
dimension $D_{2}=\nu$. A R\'enyi dimension, therefore, is often called a
generalized dimension.

The difference between the usual topological dimension $D$ (i.e. the support
dimension) and the R\'enyi dimension is called the anomalous dimension (or
codimension)
\beq
d_{q}=D-D_{q}.  \label{dq124}
\eeq
The multifractal method is a widely used tool in many branches of physics
and science in general (cfr.~\cite{Zeld87,Pala87,DremSPU90}).

A direct relation may be established between the exponents of factorial
and generalized moments at
comparatively low values of $q$, much smaller than
effective multiplicities contributing to the sum:
\beq
\f _{q}+\tau (q)=(q-1)D.  \label{fq125}
\eeq
Then the exponents are related to R\'enyi dimension and to codimension as
\beq
\tau (q)=(q-1)D_q  \label{tq126}
\eeq
and
\beq
\f _q=(q-1)d_q. \label{fq127}
\eeq

According to the general theory
{}~\cite {sl,BouGeo}, there exists
``a class of multifractals exhibiting universal properties''.
They are called universal multifractals and
are classified by
a L\'evy index $0 \leq \mu \leq 2$ which allows
the codimension to be expressed as
\beq
d_q =\frac {C_1}{\mu -1}\cdot \frac {q^{\mu }-q}{q-1} \;\;\;\;\; ( C_1 =
 {\const} ).  \label{dq128}
\eeq
The L\'evy index $\mu $ is also known as
the degree of multifractality ($\mu =0$
for mono-fractals). Values  $\mu <1$ correspond to
so-called ``calm''
singularities, values $\mu>1$ correspond to ``wild'' singularities.

One can proceed further and try to analyse experimental data at
two different levels of
bin-splitting. For that purpose, it was recently
suggested~\cite{Ratti91,Ratti92} to study
Double Trace Moments (DTM). The procedure
is, first, to sum up $\nu $-th-order moments
of multiplicity distributions at some
bin-splitting level $\Theta $
within bins belonging to a single bin of one
of the previous steps (having  bins of size $\Delta$)
and then to calculate their $q$-th moments at that level
\beq\label{tr}
Tr_{q}^{\nu} \propto \sum_{\Delta}(\sum_{\Theta}n_{m}^{\nu})^{q}
\propto \Delta^{-K(q,\nu)+q-1}.\label{DTM}
\eeq
 It is claimed~\cite{Ratti91}
that ``the DTM-technique provides a robust estimate of $\mu $
and $C_1$'' for
universal multifractals.
According to
the theory of universal multifractals~\cite{sl,Ratti91},
one should observe the
following factorizable behaviour of ``double'' exponents $ K(q,\nu )$:
\beq
 K(q,\nu )=\nu ^{\mu}  K(q,1) \ \ ,\label{fq129}
\eeq
where $\mu $ is the same L\'evy index as in (\ref{dq128}).

Experimental results on multifractals and generalized
multifractals, as well as some theoretical implications are
discussed in Subsect.~4.7.7 and in Chapter~5.

\section{Bunching-parameter approach}

A simple mathematical tool alternative to the normalized factorial moments
(2.68-2.70) is the bunching-parameter approach, suggested for high energy
applications in \cite{chek94}. In order to reveal spiky structure of
the events, it is only necessary to study the behaviour of the probability
distribution near the multiplicity $n=q$ by means of the ``bunching
parameters''
\beq
\h_q(\d y) = \frac{q}{q-1}\frac{P_q(\d y)P_{q-2} (\d y)}{P^2_{q-1}(\d y)}\ \ ,
\ \ q>1\ .
\label{BB:128}
\eeq
As is the case for the normalized factorial moments, the bunching
parameters $\h_q$ are independent of $\d y$ if there are no dynamical
fluctuations. For example, $\h_q=1$ for all $q$ for the case of
a Poissonian probability distribution.

As the $F_q(\d y)$, the $P_q(\d y)$ can be averaged over a number $M$ of
bins. Assuming approximate proportionality of $\bar n_m$ and $\d y$ at
$\d y\to 0$ and $P_0(\d y)\to 1$ for $\d y\to 0$, one obtains
\beqa
\h_2(\d y) & \simeq & F_2(\d y) \nonumber \\
\h_q(\d y) & \simeq & \frac{F_q(\d y) F_{q-2}(\d y)}{[F_{q-1}(\d y)]^2}\ \ ,\
\  q>2
\eeqa
or
\beqa
\h_2(\d y) & \propto & (\d y)^{-\b_2} \nonumber \\
\h_q(\d y) & \propto & (\d y)^{-\b_q}
\eeqa
with
\beqa
\b_2 & = & d_2 \nonumber \\
\b_q & = & d_q(q-1) + d_{q-2}(q-3)-2d_{q-1}(q-2)\ \ ,\ \  q>2\ .
\eeqa
Expressing $d_q$ in terms of the L\'evy-law approximation (2.125),
\beq
\b_q=d_2 \frac{q^\m + (q-2)^\m - 2(q-1)^\m}{2^\m-2}\ \ \ .
\eeq
In case of monofractal behaviour $(\m=0)$, $\b_q=0$ for $q>2$. In the
limit of the log-normal approximation $(\m=2)$, on the other hand,
$\b_q=d_2$ and all bunching parameters follow the same power-law.

The L\'evy-law approximation allows a simple description of multifractal
properties of random cascade models using only one free parameter $\m$.
In the bunching-parameter approach, one can make an approximation of the
high-order bunching parameters to obtain a simple {\it linear} expression
for the anomalous fractal dimensions $d_q$, still maintaining the number
of free parameters at one.

Assuming that high-order bunching parameters can be expressed in terms of
the second-order one as
\beq
\h_q(\d y) = [\h_2(\d y)]^r\ \ ,\ \ q>2\ \ ,
\eeq
the linear expression becomes
\beq
d_q = d_2 (1-r) + d_2 r \frac{q}{2}\ \ .
\label{BB:134}
\eeq
The use of bunching parameters is interesting, because it gives a general
answer to the problem of finding a multiplicity distribution leading
to intermittency: according to (\ref{BB:128}), any multiplicity distribution
can be expressed as
\beq
P_q(\d y) = P_0 (\d y) \frac{ [P_1(\d y)/P_0(\d y)]^q}{q!} \prod^q_{\ell=2}
[\h_\ell(\d y)]^{q+1-\ell}, \ \ \ q>1\ \ .
\eeq
The possible forms of multiplicity distributions with multifractal
behaviour of $d_q$ (\ref{dq124}) are discussed in~\cite{chek95,CheKu}.

\section{The wavelet transform}

An increase of factorial and cumulant moments with decreasing
 bin sizes reflects
a widening of a multiplicity distribution, i.e. an increase of multiplicity
fluctuations in individual events. This phenomenon can be studied by other
methods, as well. In particular, the so-called wavelet transform seems to
be suited for that purpose.

The wavelet transform is of particular importance  in pattern
recognition.
This is a more general problem than the fluctuation study
itself, since it involves the analysis of individual event shapes, not
only the event ensemble, and may become of interest in the analysis of
very-high multiplicity events.

It is shown~\cite{ADREP} that, for  pattern recognition,  the wavelet
transform is about two orders of magnitude more efficient than
ordinary Fourier analysis.

An application of wavelets to multiparticle production processes
has been proposed in~\cite{CGL}.
The main principle of the wavelet transform is to study the dependence of
fluctuations on the phase-space bin size by the so-called difference method.
One considers the difference between the histogram of an individual
event at a definite resolution to the corresponding histogram at a (e.g.,
twice)
finer resolution. Proceeding step by step, one is able to restore the whole
pattern of fluctuations.

Let us explain how this procedure can be applied to an individual event. We
consider the one-dimensional projection of the event onto the rapidity
interval $\Delta Y$. Any $n$-particle event can be represented by the
histograms of particle densities $\rho = dn/dy$ at various resolutions. The
simplest information is obtained from the value of the average density
$\langle \rho \rangle = n/\Delta Y$. To consider the forward-backward
correlations, one splits the rapidity interval $\Delta Y$ into two equal parts
and gets the forward and backward average densities $\langle \rho _{f,b}
\rangle = 2n_{f,b}/\Delta Y$, where $n_{f,b}$ are the forward (backward)
multiplicities with $n_f + n_b =n$. Proceeding further to the $J$-th step,
we approximate the event in terms of the histogram with $2^J$ bins.

Let us construct now the difference of the two histograms described above.
Namely, we subtract the average density from the forward-backward histogram
and get another histogram with positive ordinate at one side and negative at
the other, demonstrating the forward-backward fluctuations in the event.

Splitting the forward and backward regions further into equal halves,
one gets the histogram at $J=2$. Its difference from the forward-backward
histogram at $J=1$ reveals the fluctuations at finer resolution.
Iterating
to higher values of $J$, one studies  how fluctuations evolve
at ever finer resolution. The set of difference histograms is called the
wavelet transform of the event. The above procedure
corresponds to the so-called Haar-wavelet transform. Those interested in
mathematical details are referred to~\cite{DMK}.

The wavelet transform provides direct information on the
evolution of fluctuations at different scales, i.e. on the dynamics of
individual high-multiplicity events revealing their clustering (and
subclustering) structure. A generalization to factorial (and cumulant)
wavelets is possible~\cite{CGL}. The simplest cascade models show such
remarkable properties of wavelet transforms~\cite{CGL} as
(quasi)diagonalization of their correlation density matrices, scaling
exponents etc. It is interesting to note that the
equations for the generating functions of wavelet transforms~\cite{CGL} look
very similar to the ``gain-loss'' equations (in particular, to QCD equations)
discussed at the  end of Chapter~5. All those features are yet
to be studied.

The very first  application to experimental data is presented in
{}~\cite{Suzu95}, where wavelet spectra of JACEE events are
studied.

\chapter{Experimental survey on correlations}

In this chapter, we review experimental results on ``classical''
correlations, a subject with a long history in particle physics.
It was instrumental in establishing fundamental concepts
of hadrodynamics, such as short-range order, which are  an
essential ingredient of all popular Monte-Carlo models of
hadronization. With the exception of Bose-Einstein interferometry,
the field lay dormant for several years, but was revived with the
introduction of generalized concepts. The data
cover a variety of multiparticle-production processes ranging
from $\re^+\re^-$ annihilation to nucleus-nucleus collisions.

In Chapter~4, we shall review material on factorial moments and related
quantities, obtained since 1986. At that time, a pioneering suggestion
was made to investigate the patterns of particle density fluctuations
in multihadronic events: the intermittency idea. Measurement of
factorial moments opened a way to establish possible scale-invariance
and fractal behaviour in hadrodynamics.

Interest in correlation functions received a vigorous boost
when their  intimate connection with factorial moments was
realized (see Chapter~2).
Both are now explored in parallel with novel techniques.
These offer  promising perspectives towards a long overdue
unified approach to correlation phenomena, including
Bose-Einstein interferometry.

Another obviously related subject, the phenomenology of
multiplicity distributions \cite{CarShih}, is not explicitly
covered here.
Multiplicity distributions inspired many early
ideas on scale-invariance and phase-transition analogies
in multiparticle production, such as Koba-Nielsen-Olesen scaling~\cite{KNO}
and the Feynman-Wilson liquid picture~\cite{FWfluid}.
However, the major part of the data relate either to full phase space
or to sizable portions of it.
It remains an interesting task for the future to
explain  the ``large-scale'' properties of multiplicity
distributions in terms of correlation function behaviour
at ``small distances'', the main subject of this paper.
Of course, the factorial moments discussed in
Chapter~4 are just another representation
of multiplicity distributions and their increase with decreasing bin
size reveals the evolution of the multiplicity distribution.

\section{Rapidity correlations}\\
The study of correlation effects in particle production processes provides
information on hadronic production dynamics beyond that obtained from
single-particle inclusive spectra. Correlations in rapidity $y$, as defined in
Sect.~2.1, have been studied in various experiments on $\E$, lepton-nucleon,
hadron-hadron, hadron-nucleus and nucleus-nucleus collisions. Strong
$y$-correlations have been observed in all experiments in one form or another,
depending on the specific form of the correlation function, type of
interaction, kind of particles, the kinematic region under consideration,
etc. The main conclusions were (for early reviews see~\cite{Foa75,Whit76}):
\begin{itemize}
\itemsep=-2mm
\item[1.] Two-particle correlations are strong at small interparticle
rapidity-distances $|y_1-y_2|$ (see Fig.~3.1).
\item[2.] They strongly depend on the two-particle charge
combination.
\end{itemize}

Rapidity correlations are now being studied with renewed attention.
One reason is that their structure at very small rapidity distances is
directly related to self-similar particle-density fluctuations
(intermittency), a topic to be covered in Chapter~4.

\subsection{Correlations in hadron-hadron collisions}\\
In Fig.~3.2 the pseudo-rapidity correlation function $C_2(\h_1,\h_2)$
as defined in (\ref{dr:ex1}) is given for
$\h_1=0$, as a function of $\h_2=\h$,
for the energy range between 63 and 900 GeV~\cite{Anso88}. Whereas
$C_2(0,\h)$ depends on energy, the short-range correlation  $C_\rS$
defined in (\ref{dr2.9}) does not strongly depend on energy and has a full
width of about 2 units in pseudo-rapidity.
The function $C_\rL$ is not a two-particle correlation, but derives from
the difference in the single particle distribution function for
different multiplicities. As can be seen in Fig.~3.2b, $C_\rL$ is considerably
wider than $C_\rS$ and
increases with energy (the 63 GeV data are from~\cite{Amen76}).

In Fig.~3.3, the semi-inclusive correlation $C^{(n)}_2(\h_1,\h_2)$ for
$\Pp\Pap$
collisions at 900 GeV~\cite{Fugl87} is compared to the UA5
Cluster Monte Carlo (MC) GENCL~\cite{Alner87}, as well as to the FRITIOF~2
{}~\cite{FRIT} and PYTHIA~\cite{PYTHIA} Monte Carlos, for charge
multiplicity $34\leq n\leq 38$. The Cluster MC is designed to fit just these
short-range correlations, but also FRITIOF~2 is doing surprisingly well
(see however Subsect.~4.4.4).

At lower energy, the NA23 Collaboration~\cite{Bail88} has studied the
short-range correlation of charged particles in pp collisions of $\sqrt s=26$
GeV in terms of $K_2(y_1,y_2)$ defined in (\ref{dr:2.7}). Only events with
charge multiplicity $n>6$ are used. The positive short-range correlations are
in agreement with those found earlier at $\sqrt s= 53$ GeV~\cite{Break82}.

The NA23 data are compared to single-string LUND~\cite{LUND} and to a
two-chain Dual-Parton Model (DPM)~\cite{DPM} in Fig.~3.4.
The one-string model (without gluon radiation) does not at all describe
the short-range rapidity correlation in the data. The two-chain model
does better, but remains unsatisfactory.
Somewhat better but still insufficient  agreement is obtained
by renormalizing the MC events
to the experimental multiplicity distribution (not shown). The effect of
Bose-Einstein correlations in the (++) and (--~--) data is found
to be insignificant, as may be expected for data integrated over
transverse momentum $p_\rT$ and azimuthal angle $\vf$. Obviously, more
chains, possibly with higher $p_\rT$, are needed to explain short-range
order with fragmentation models, even below $\sqrt s\approx30$ GeV.

NA22 results for $C_2(0,y_2)$ and $\tilde C_2(0,y_2)$
(Eqs.~\ref{dr:ex1}, \ref{dr:2.5b}) for $\p^+$p
and K$^+$p collisions at $\sqrt s$=22 GeV~\cite{Aiva91}
are compared with FRITIOF~2, a 2-string DPM
and QGSM~\cite{QGSM} predictions in Fig.~3.5a,b.  FRITIOF
and 2-string DPM largely underestimate the correlation.
QGSM reproduces $C^{--}_2(0,y_2)$ very well and even overestimates
$C^{++}_2(0,y_2)$ and $C^{+-}_2(0,y_2)$.
It has been verified that the differences between
QGSM and FRITIOF or DPM
are not due to the different treatment of tensor mesons
(only included in the latter two).

In Fig.~3.5c, FRITIOF and QGSM are compared to the NA22 data
in terms of the short-range contribution $\tilde C_{\rS}(0,y_2)$.
The $(+-)$ short-range correlation is
reproduced reasonably well by these  models.
For equal charges, however, the strong
anti-correlation predicted by FRITIOF is
not seen in the data. QGSM contains a small equal-charge correlation due to
a cluster component, but still underestimates its size.
Similar  discrepancies are also observed
in semi-inclusive (fixed multiplicity) data for each charge
combination (not shown here). They are even larger
than in the inclusive data, also in the QGSM model.

{}From this brief survey, we conclude that
in hadron-hadron collisions two-particle correlations are
badly reproduced and generally underestimated in currently used models.

\subsection{Correlations in $\protect\E$ and $\protect\m^+p$-collisions}\\
Fig.~3.6 shows $K^{+-}_2(y_1,y_2)$ and $K^{--}_2(y_1,y_2)$ for muon-nucleon
interactions at 280 GeV/$c$ \cite{Arne86}. A steep peak is seen at
$y_1=y_2=0$ for $K^{+-}_2$, with two shoulders along the diagonal $y_1=y_2$.
On the other hand, $K^{--}_2$ is below 0 for most of the distribution, but
we shall see that the most impressive correlation is in fact coming from
$y_1\approx y_2$, just for this case. As in hadron-hadron collisions,
correlations are strong and depend on the two-particle charge combination.

Fig.~3.7 shows $K_2(y_1,y_2)$ in $\m^+$p interactions at 280 GeV/$c$ with
$y_1$ $\e\ [-0.5,0.5]$, the hadronic invariant mass $W$ in the interval
$13<W<20$ GeV and for $n\geq 3$~\cite{Male90Fig88}, together with the NA22
non-single-diffractive M$^+$p sample, $n\geq 2$~\cite{Aiva91}. Correlations
in $\m^+$p seem smaller than in NA22, but one has to consider a possible
energy dependence. Indeed, extrapolating from the energy dependence of
$K_2(0,0)$ published in~\cite{Male90Fig88}, one finds quite
similar values for $\m^+$p at 22 GeV and M$^+$p in NA22.

In Fig.~3.8a,b we compare the function $\tilde K_2(0,y)$ for the NA22
non-single-diffrac\-tive M$^+$p sample (charge multiplicity
$n\geq 2$)~\cite{Aiva91} with that for $\E$-annihilation at the same
energy ($\sqrt s$=22 GeV)~\cite{Alth85Chwas88}. The values of
$\tilde K_2(0,y)$ are larger for $(++)$ pairs than for $(--)$ in
meson-proton (M$^+$p) reactions; for $(--)$ and $(+-)$ pairs they agree
with $\tilde K_2$ for $\E$ annihilation in the central region.

A comparison of the correlation functions for $\E$-annihilation and
non-single-diffract\-ive M$^+$p collisions throughout the full kinematic
region with $y_1$ $\e\ [-1,0]$ is shown in Fig.~3.8c for charged pairs.
The $\E$ data are given at $\sqrt s=14$ and 44 GeV~\cite{Alth85Chwas88}.
At $y_2=y_1$, the 22 GeV M$^+$p correlation lies between the $\E$ results.
The shape is, surprisingly, more symmetric than in $\E$.

For $\m^+\Pp$ ~\cite{Arne86,Male90Fig88} and $\E$ collisions
\cite{Alth85Chwas88,Podo91,Act92}, the LUND-type Monte Carlo is reported
to reproduce the majority of the experimental distributions. In~\cite{Bail88}
it is shown that this is mainly due to the inclusion of hard and soft gluon
effects. However, important underestimates of $K_2(y_1,y_2)$ are still
observable, in particular in the central and current fragmentation regions.
For $\E$~\cite{Podo91}, this is shown in Fig.~3.9, where $K_2(y_1,y_2)$ is
compared to the LUND model (JETSET 7.2 PS) as a function of $y_1-y_2$ (dotted
line), for the full sample (upper plots) and for a two-jet sample (lower
plots). In all cases, the LUND model underestimates the correlation at
$y_1-y_2=0$. In general, the disagreement becomes smaller
when  Bose-Einstein correlations are included (full lines).
The main feature to note is that correlations are much
weaker in the two-jet sample than in the full sample.
 Furthermore,  correlations are larger for $y\leq 0$ (left
plot), i.e. in the hemisphere opposite the most energetic jet,
than for $y>0$ (right plot). These two observations, again, point to
hard gluon radiation as the main source of two-particle correlation in
$\E$ collisions.

A systematic test of analytic QCD calculations and of  QCD
Monte-Carlo models for  two-particle correlations
has been performed by OPAL~\cite{Act92}. The authors study the function
\beq
R(\xi_1,\xi_2) = K_2(\xi_1,\xi_2)+1
\eeq
with $\xi=\ln(1/x_\rF)$, $x_\rF=2p/E_{\cm}$ being the Feynman variable,
i.e. the particle momentum $p$ in the cms normalized to half the
cms energy $E_{\cm}$. In Fig.~3.10, $R$ is plotted as a function of
$(\xi_1-\xi_2)$ for $(\xi_1+\xi_2)$ centered at the values 6, 7 and 8,
respectively. Fig.~3.10a proves that a next-to-leading order
calculation~\cite{Fo91} (full lines) is better than leading order (dashed),
but still overestimates the overall level of the correlation for
any reasonable value of $\La$. Since the next-to-leading correction
is large, still higher-order terms are  needed. It is therefore likely
that a satisfactory analytical treatment of correlations, even at the
parton level, will not be obtained in the very near future.

Higher-order effects are, in an average sense, included in the
existing Monte-Carlo models. In Fig.~3.10b, the same data are compared
to the coherent parton shower models JETSET~PS~\cite{LUND},
HERWIG~\cite{Mar88} and ARIADNE~\cite{Pet88}. The latter gives an
excellent fit to the data, JETSET lies slightly below (within
uncertainty of parameters), but HERWIG considerably above. The agreement
of JETSET could only slightly be improved by including Bose-Einstein
correlations. As far as incoherent parton shower models
are concerned, none of the various versions of COJETS~\cite{Odo84} gives
a particularly good representation of the correlation data.

All the models were tuned on the OPAL data in terms of event shapes and
generally describe single-particle distributions. It is clear that
correlations
allow better and more discriminative tests than more integrated quantities.

We have mentioned the difficulties string-hadronization models experience
in predicting like-sign correlations in hadron-hadron collisions.  It
is important to verify if the otherwise successful $\re^+\re^-$ models
are also able to reproduce correlations between charge-separated
systems such as $(+-)$ and $(\pm\pm)$ particle pairs.

\subsection{Charge dependence}
How $C_\rS$ and $\tilde C_\rS$ depend on the charge of the pairs
is shown in Fig.~3.11 for the combinations $(--),(++)$ and $(+-)$ in
NA22~\cite{Aiva91}. The short-range correlation is
significantly larger for $(+-)$ than for $(--)$
and $(++)$ combinations. This is  also seen in the EMC data~\cite{Arne86}.
Resonance production is a likely explanation of this
difference. For like charges, a small enhancement is seen near
$y_1\approx y_2\approx 0$ above a large negative background. This is
possibly due to  Bose-Einstein interference.

\subsection{Charge-multiplicity dependence}
The multiplicity dependence of $\tilde C_2^{(n)}(0,y)$ for the $(+-)$
 combination is shown in Fig.~3.12 \cite{Aiva91}. Near the maximum at
$y=0$ the correlation function is approximately Gaussian and narrows
with increasing $n$. In  Fig.~3.13a are presented  the values of
$\tilde C_2^{(n)}(0,0)$ as a function of $n$ for three charge combinations.
Within errors, $\tilde C_2^{(n)}(0,0)$ is independent of $n$, but
consistently higher for $(+-)$ and $(--)$ than for $(++)$. The reason for
the difference between $(--)$ and $(++)$ probably lies in the positive charge
of both beam and target.

On the other hand, an increase of $\tilde C_2^{(n)}(|\h_1-\h_2|)$ with
$1/(n-1)$ is found~\cite{EGGE75} when averaging over a region $|\h|<2$
(Fig.~3.13b). Since $\tilde C_2$ becomes smaller when moving away from the
center, and that may happen faster for higher than for lower $n$, this is not
necessarily in contradiction with the data in Fig.~3.13a.

\subsection{Transverse momentum dependence}
The search for density fluctuations, described in later sections, has
revealed the importance of correlations in  multi-dimensional phase space.
It is, therefore, of interest to gain insight into the transverse momentum
($p_\rT$) dependence of rapidity correlations. Early results  on this topic
can be found in~\cite{Biswas76}.
Recent data on $K_2(0,y_2)$~\cite{NA22} for all particles and for particles
with $p_\rT$ smaller or larger than 0.3 GeV/$c$, plotted in Fig.~3.14,
indeed reveal a strong sensitivity to transverse momentum.
The correlation function is largest,
and stronger peaked, near $y_2=0$ for $p_\rT<0.3$ GeV/$c$, in particular for
$(--)$-pairs. A similar effect was noted already in~\cite{Biswas76}.
The data of Fig.~3.14 were fitted with the functions
\begin{eqnarray}
f_1 &=& c \exp [-(y-y_0)^2/2\s^2]\ \ \ \ \hfil{\mb{(full\ line)}}\ \ , \\
f_2 &=& a \exp(-b|y|)\ \ \ \ {\mb{(dashed)}} \ \ ,
\end{eqnarray}
with $c$, $y_0$, $\s$, $a$ and $b$ as free parameters. Even though
for low $p_\rT$ the data point at $y_2=0$ lies systematically above the curve,
$K_2(0,y_2)$ is well fitted by the Gaussian $f_1$ but not by the exponential
$f_2$, in this one-dimensional projection on rapidity.

Changing to the variables $x_1=(y_2+y_1)/2$ and $x_2=(y_2-y_1)/2$, a
steepening is observed at small $x_2$ (not shown). For like-charge pairs,
this becomes particularly sharp when the bin size is
reduced to $\d x_2=0.1$. For the latter, $C_2(x_1,x_2=0)$ increases and
both a Gaussian and an exponential can fit the correlation function.

\subsection{Strange particles}
In string-fragmentation models, first-rank hadrons are formed from
neighbouring quark-antiquark pairs tunnelling out of the vacuum. The
hadronic final states, therefore, show short-range order due to local
flavour conservation. Using stable mesons only, this characteristic
property is difficult to study experimentally because of the large $\Pq\Paq$
combinatorial background. What is needed is a flag identifying the $\Pq\Paq$
pairs created together. A suitable choice is strangeness since the number
of $\rs\sbar$-pairs per event is small and the combinatorial background
strongly reduced.

Good strangeness identification is available for $\E$ annihilation
in the TPC detector at $\sqrt s=29$ GeV~\cite{Aiha84}. This collaboration
observes significant short-range $\PK^+\PK^-$ correlations in $y$,
well reproduced by the LUND model and by the Webber QCD model.

In hadron-hadron collisions, strange particle pairs have been studied by
the NA23 Collaboration~\cite{Asai87}. The distribution in the rapidity
difference $\D y$  for two $\PK^0$'s is given in Fig.~3.15a, for a $\PK^0$ and
a $\La^0$ in Fig.~3.15c.  The results are compared to the single-string LUND
model. As is the case for non-strange particles, the model slightly
underestimates the rapidity correlation.

\section{Azimuthal correlations}
In  interactions of unpolarised particles, no distinguished direction
exists in the plane transverse to the beam and the distribution in the
azimuthal angle $\vf$ is uniform. Still, a two-particle correlation exists
also in $\vf$ and is visible in the distribution $W(\D\vf)$ of
$\D\vf=|\vf_1-\vf_2|$, the azimuthal angle between two particles,
$\D\vf\in(0,\p)$. The  azimuthal correlation  may depend on the charge of
the  particles in the pair, on  the rapidity distance  $\D y = |y_1-y_2|$
between these particles and on their transverse momentum.

The first experiments to extensively study two-particle correlations as a
function of both rapidity and azimuthal angular separation~\cite{EGGE75,OH}
already showed that the correlation at small rapidity distance is strongest
when the two particles are produced in the same or opposite directions in
transverse momentum (see Fig.~3.16). The  correlation-length in rapidity
is larger towards $\D\vf=\p$ than towards $\D\vf=0$. Furthermore, significant
differences in the shape of the joint rapidity and azimuthal correlation
functions have been observed for pairs of like and unlike pions~\cite{OH}.

\vspace{2mm}
In Fig.~3.17, the distribution $W(\D\vf,\D y)$, normalized to unity, is
shown as a function of $\D\vf,$ for all charge combinations, in the intervals
$\D y<1, ~1<\D y<2$ and $2<\D y<3$~\cite{NA22}. A {horizontal line} at the
average value $1/\p$ corresponds to a flat distribution in $\D\vf$. The
distribution is influenced by conservation of transverse momentum, by the
decay of resonances (mainly for unlike-sign particles)
and by Bose-Einstein correlations (for like-sign particles).
In all cases, $W$ is larger than $1/\p$ for $\D\vf>\p/2$ and has a maximum
at $\D\vf = \p$.  Except for $(--)$ pairs at $\D y<1$, the $W$ function is
smaller than $1/\p$ for $\D\vf < \p /2$. Such a global
anti-correlation follows from transverse momentum conservation.

Model predictions are shown in Fig.~3.17 for FRITIOF~2 (dot-dashed),
two-string
DPM (full) and multi-string QGSM (dashed). The comparison with the data shows
that it is much easier to account for
azimuthal correlations at large than at small
$\D y$. At small $\D y$ the models differ from each other and from the
experimental data. The QGSM  shows somewhat better agreement with
experiment than the other models. This is a consequence of the
multi-string structure of QGSM, where strong azimuthal correlations in a
single  string are destroyed, with the result that the $\D\vf$-dependence
is weaker than in two-string models.

Differences between experiment and all models exist at small $\D\vf$ and
$\D y<1$, in particular for $(--)$ pairs. Bose-Einstein correlations, not
included in the models, may explain this disagreement. The influence of
Bose-Einstein correlation can also be observed in the $(++)$ combination, but
is smaller because of  the  influence of the (positive) beam particle.

Azimuthal distributions are shown in Fig.~3.18 for particles with
$\D y<1$, for
$p_\rT<0.30$ GeV/$c$ and for $p_\rT>0.30$ GeV/$c$, together with
model calculations. A comparison of these figures reveals that azimuthal
correlations have a strong $p_\rT$-dependence. Large {\it positive}
azimuthal correlations exist at small $\D\vf$ and $\D y<1$ for like-sign
particles with  small $p_\rT$. As the transverse momentum of particles
increases, the peak at $\D\vf = \p$ becomes more pronounced. This is
reproduced by the models and reflects momentum conservation.

For $\La\bar \La$ pairs, an azimuthal correlation has been observed in
MARK II at 29 GeV~\cite{Vais85}. Similar $\PK^+\PK^-$ correlations are seen
in the exclusive hh final state $\PK^-\Pp\to$\break
$\Pp\PK^+\PK^-\PK^-\p^+\p^-$ at 32 GeV/$c$~\cite{MaZP86}.

Azimuthal correlations between $(+ -)$ and $(++,--)$ charge combinations have
been studied in $\m$p collisions~\cite{Arne86} for $|\D y|<1$ and $|\D y|>1$.
The distribution $W(\D\vf)$ is described fairly well by the LUND model
including primordial $k_\rT$ and gluons, except that for $|\D y|<1$ it
slightly underestimates the anti-correlation for $(+ -)$ and overestimates
it for $(++,--)$.

In the azimuthal correlation of $\PK^0$ pairs (Fig.~3.15b) and of $\PK^0\La^0$
(Fig.~3.15d) studied by NA23~\cite{Asai87}, the data tend to show pairs of
small $\D\vf$ not pre\-sent in low-$p_\rT$ LUND (solid line).

By the same collaboration, the azimuthal correlation is studied~\cite{Bail88}
in terms of the asymmetry parameter
\beq
B=[N(\D\vf >\p/2) - N(\D\vf<\p/2)]/N_{\all}
\eeq
for hadron pairs with

\noindent
a) opposite charge ($\rh^+\rh^-$)\\
b) equal charge ($\rh^+\rh^++\rh^-\rh^-$)\\
c) possibly opposite strangeness ($\La^0\rh^+,x_\La<-0.2$)\\
d) no opposite strangeness ($\La^0\rh^-,x_\La<-0.2$),

\noindent
for $\D y<2$ and for $\D y>2$. No azimuthal correlation is seen for $\D y>2$
in all cases and for $\D y<2$ in case of no common $\Pq\Paq$ pairs
($\rh^+\rh^++\rh^-\rh^-,\La^0\rh^-$). For $\rh^+\rh^-$ and $\La^0\rh^+$, the
parameter $B$ is compared to low-$p_\rT$ LUND and DPM predictions in
Table~3.1.

\vspace{1mm}
\begin{center}
{\bf Table 3.1}  Asymmetry parameter $B$
\vs 4mm
\begin{tabular}{llll}
\hline
                   &Experiment   &LUND    &DPM \\
\hline
$\La^0\rh^+ (\D y<2)$ & 0.18~$\pm$0.03& 0.30~$\pm$0.01& 0.19~$\pm$0.01 \\
$\rh^+\rh^- (\D y<2)$& 0.066$\pm$0.003& 0.126$\pm$0.002& 0.106$\pm$0.002 \\
\hline
\end{tabular}
\end{center}

\vspace{2mm}
\noindent
The parameter $B$ is strongly overestimated in single-string low-$p_\rT$
LUND and still too large in the two-string DPM. Furthermore, $B$ increases
with the sum of the transverse momenta (Fig.~3.19) but less strongly than in
the models.

The azimuthal correlation has  also been studied for $\rc\bar \rc$ pairs in
$\rD\bar\rD$ production. An asymmetry is indeed observed in $\p^-\Pp$
collisions
at 360 GeV/$c$~\cite{Agui85}. Also there, the LUND model overestimates the
effect.

As shown on $\p^-$N interactions at $\sqrt s=26$ GeV \cite{Adamo95}, also
NLO perturbative QCD calculations overestimate the azimuthal asymmetry for
$\rD\bar \rD$ pairs (Fig.~3.20a). Agreement can be obtained with a
model \cite{Frixi94} where a (Gaussian shaped) transverse component is added
to the incoming parton momentum before performing the NLO perturbative
QCD calculation (Fig.~3.20b).

\section{Correlations on the parton level}

The OPAL collaboration \cite{Act93} has compared hadronic azimuthal
correlations to coherent and incoherent shower models (Fig.~3.21). The
coherent models JETSET~PS with angular ordering \cite{LUND}, HERWIG
\cite{Mar88} and ARIADNE \cite{Pet88} describe the azimuthal correlations
in hadronic $\PZz$ decays, but the incoherent models JETSET~PS without
angular ordering \cite{LUND} and COJETS \cite{Odo84} fail for
 $\vf\simgr\pi/2$.

The hadronization of a quark-antiquark pair at high virtuality is
currently thought to proceed via parton showering \cite{Dok92}.
QCD implies that this parton showering be coherent. The
coherence can be incorporated into Monte-Carlo programs as angular
ordering\cite{ao}, whereby for each successive branching the gluon is
emitted at a smaller angle.

Furthermore, the idea of local parton-hadron duality (LPHD)\cite{lphd}
suggests that features at the parton level survive the fragmentation process.
We can, therefore, expect that the coherence of the parton radiation
will be reflected in angular ordering of the observed particles.

As a method particularly sensitive to angular ordering,
particle-particle correlations PPC and their
asymmetry PPCA \cite{aleph_note,Syed94} are examined in a way analogous
to the study of energy-energy correlations \cite{basham},

\beqa
       \PPC (\chi)&=& \frac{1}{\Delta\chi}
       \lan 2 \sum^n_{i<j}
       \frac{1}{n^2}
       \delta_{\mathrm{bin}}(\chi-\chi_{ij})\ran
\\
       \PPCA (\chi) &=& \PPC (180^{\circ}-\chi)-\PPC (\chi)\ \ ,
\eeqa
where $\chi_{ij}$ is the full spatial angle between tracks $i$ and $j$,
$\lan~\ran$ is the average over all events in the sample, $n$ is the
number of charged tracks in an event, and $\Delta\chi$ is the bin
width. The function $\delta_{\mathrm{bin}}(\chi-\chi_{ij})$ is 1 if
$\chi_{ij}$ and $\chi$ are in the same bin and 0 otherwise.

At $\sqrt{s}=M_\rZ$, the fraction of two-jet events is very high. For
two-jet events, particles in different jets will in general be separated
by an angle $\chi$ greater than $90^\circ$. The PPC for $\chi > 90^\circ$
can, therefore, serve as an indication of what the PPC
{\em within}\/ a jet ($\chi < 90^\circ$)  would be
{\em in the absence}\/ of angular ordering. By forming the asymmetry,
these `uninteresting' correlations are effectively subtracted. The effects
of angular ordering should, therefore, be more  directly observable in the
PPCA than in the PPC. Note, however, that the sign convention following
\cite{basham} leads to a {\em negative} sign for a {\em positive} correlation.

Figures 3.22a and b show the PPCA distribution of L3 data (corrected for
detector effects \cite{Syed94}) compared to coherent and incoherent
Monte-Carlo models, respectively.

In Fig.~3.22b we see that for $\x\simkl 60^0$ JETSET~7.3 ~PS without angular
ordering (incoherent) disagrees strongly with the data, while being in fair
agreement at larger values of $\chi$. COJETS is seen not to reproduce the data
over the entire angular range. On the other hand, in Fig.~3.22a, the coherent
Monte Carlo models, JETSET with angular ordering, HERWIG, and ARIADNE
all reproduce the data reasonably well  over the full angular range. Note
that the disagreement of the incoherent models can not be due to the
Bose-Einstein effect. Turning this effect off in the non-angular ordered
JETSET model does not raise but lower its PPCA points.
So, the data from the L3 experiment strongly disfavour the incoherent models.

\section{Three-particle rapidity correlations}\\
Whether  dynamical  correlations exist beyond the two-particle correlations
discussed so far is of crucial importance for much of the present search for
scaling phenomena in multiparticle processes, a subject treated in Chapter~4.
With conventional techniques, this question is not easy to answer
and beyond the  sensitivity of many  experiments.

Nevertheless, three-particle
correlations in rapidity have been looked for in a number of experiments
{}~\cite{Whit76,Aiva91,Buma79,Azim80,BreakEP88}.
The third order normalized factorial cumulant is defined as
[cfr.~(\ref{a:4b})]:
\beq
K_3(y_1,y_2,y_3) = C_3(y_1,y_2,y_3)/\frac{1}{\s^3_{\inel}} \frac{\rd\s}
{\rd y_1} \frac{\rd\s}{\rd y_2} \frac{\rd\s}{\rd y_3}\ \ ,
\eeq
\beq
C_3(y_1,y_2,y_3)=\frac{1}{\s_{\inel}} \frac{\rd^3\s}{\rd y_1 \rd y_2 \rd y_3}
+ 2 \frac{1}{\s^3_{\inel}} \frac{\rd\s}{\rd y_1} \frac{\rd\s}{\rd y_2}
\frac{\rd\s}{\rd y_3} -
\eeq
\beqan
- \frac{1}{\s^2_{\inel}} \frac{\rd^2\s}{\rd y_1\rd y_2}
\frac{\rd\s}{\rd y_3} -
\frac{1}{\s^2_{\inel}} \frac{\rd^2\s}{\rd y_2\rd y_3}
\frac{\rd\s}{\rd y_1} - \frac{1}{\s^2_{\inel}} \frac{\rd^2\s}{\rd y_1\rd y_3}
\frac{\rd\s}{\rd y_2}
\eeqan
\beqan
{\mbox{with}} \ \ \ \ \ \s_{\inel} = \sum_{n\geq8}\s_n\ \ .
\eeqan

The $\tilde C_{\rS}(y_1,y_2,y_3)$ correlation function is determined as a sum
of topological correlation functions:
\beq
\tilde C_{\rS} (y_1,y_2,y_3) = \sum_{n\geq8} P_n \tilde C_3^{(n)}
(y_1,y_2,y_3)\ ,
\eeq
\beq
\tilde C_3^{(n)} (y_1,y_2,y_3) = \tilde\r_3^{(n)} (y_1,y_2,y_3) -
\tilde A_3^{(n)} (y_1,y_2,y_3)
\ , \eeq
\begin{eqnarray*}
\tilde A_3^{(n)} (y_1,y_2,y_3) = &\tilde\r_2^{(n)}(y_1,y_2) \tilde\r_1^{(n)}
(y_3)
+ \tilde\r_2^{(n)}(y_2,y_3) \tilde\r_1^{(n)}(y_1) + \tilde\r_2^{(n)}(y_1,y_3)
\tilde\r_1^{(n)}(y_2)\ - \\
& - 2 \tilde\r_1^{(n)}(y_1) \tilde\r_1^{(n)}(y_2)  \tilde\r_1^{(n)}(y_3)\ ,
\end{eqnarray*}
\beq
\tilde\r_3^{(n)} (y_1,y_2,y_3) = \frac{1}{n(1,2,3)} \frac{1}{\s_n}
\frac{\rd^3\s}{\rd y_1\rd y_2\rd y_3}\ \ ,\eeq
\ni
where $n(1,2,3)$ is the mean number of three-particle combinations in
events with charge multiplicity $n$.

The corresponding normalized function is defined as:
\beq
\tilde K_{\rS}(y_1,y_2,y_3) = \tilde C_{\rS}(y_1,y_2,y_3)/\sum_n
P_n\tilde\r_1^{(n)}(y_1) \tilde\r_1^{(n)}(y_2)\tilde\r_1^{(n)}(y_3)\ .\eeq

Because of small statistics, three-particle  correlations were not
observed in pp interactions at 200 GeV/$c$ at FNAL~\cite{Whit76}.
In $\PK^-\Pp$ interactions at 32 GeV/$c$~\cite{Buma79},
three-particle correlations were considered
using $\tilde C_{\rS}(y_1,y_2,y_3)$ and $\tilde K_{\rS}(y_1,y_2,y_3)$.
No positive short-range correlation effect was observed. Correlations
in the form of $K$ have been observed in the central region by the ISR
experiment for $n\geq8$~\cite{BreakEP88} .

Fig.~3.23 from NA22 shows $K_3(0,0,y)$ and $\tilde K_{\rS}(0,0,y)$ for
the combined M$^+$p sample at 250 GeV/$c$~\cite{Aiva91}. Also shown are the
values of $K_3(0,0,y)$ obtained in pp-interactions at $\sqrt s$=31-62 GeV
{}~\cite{BreakEP88} (lines). Inclusive three-particle correlations $K_3(0,0,y)$
are indeed seen  in the NA22 data.
They are strongest when a third particle partially compensates the charge of
a pair of identical particles. There are, however, no correlation effects
visible in
the function $\tilde K_{\rS}(0,0,y)$. In FRITIOF and QGSM, three-particle
rapidity correlations are absent
in both $K_3(0,0,y)$ and $\tilde K_{\rS}(0,0,y)$.

Recently, a factorization of the normalized
three-particle correlation function has been proposed
{}~\cite{CaSa89,CapFiaKrz89,WolfAPP90} under the form of
a ``linked-pair'' structure:
\beq
K_3(y_1,y_2,y_3)=K_2(y_1,y_2)K_2(y_2,y_3)+K_2(y_1,y_3)K_2(y_3,y_2).
\label{linked:pair}
\eeq

The comparison of the prediction of (\ref{linked:pair}) to the data is given
in Table 3.2, for $n\geq 2$, at a resolution of 0.5 rapidity units.
At this resolution, the linked-pair ansatz is in agreement with the
measured three-particle correlation within two standard deviations.
Note, that $y$-correlations are much stronger for low-$p_\rT$ particles
and that the linked-pair ansatz continues to hold.
\vs 2mm
With the accuracy presently attainable for three-particle correlations, it
is obvious that studies of still higher-order correlation functions require
better methods. The most successful ones will be discussed in Chapter~4.

\vspace{5mm}
{\bf Table 3.2}
The  3-particle correlation function compared to the
 prediction from the linked-pair ansatz,
 for non-single diffractive data ($n\geq 2$).

\vspace{5mm}
\begin{center}
\begin{tabular}{lllll}
\hline
       &\multicolumn{2}{c}{all $p_\rT$} & \multicolumn{2}{c}{$p_\rT<$0.15
GeV/$c$}\\
       &~~~~data          &~~~~LPA           &~~~data         &~~~LPA \\
\hline
$K_3^{---}(0,0,0)$ & 0.23$\pm$0.10 &0.30$\pm$0.03 &2.3$\pm$1.7
&2.0$\pm$0.4 \\
$K_3^{+++}(0,0,0)$ & 0.14$\pm$0.06 &0.21$\pm$0.02 &1.2$\pm$0.6
&1.0$\pm$0.2 \\
$K_3^{\rc\rc\rc}(0,0,0)$ & 0.39$\pm$0.04 &0.53$\pm$0.03 &1.9$\pm$0.5
&1.7$\pm$0.2 \\
\hline
\end{tabular}
\end{center}

\section{Summary and conclusions}
\begin{enumerate}
\i The main contributions to the correlation functions $C_2$ and $C_3$
come  from the mixing of events with  different multiplicity and different
single-particle density, but some effect remains in the so-called short-range
correlation part.
\i $C_2(0,y_2)$ increases much faster with increasing energy than its
short-range contribution.
\i The short-range correlation is significantly larger for $(+-)$ than
for the equal-charge combinations, and is positive over a wider rapidity
range in
$C_2(y_1, y_2=y_1)$.
\i The correlation functions
$\tilde C_2^{(n)}(0,y_2)$,
contrary to $C^{(n)}_2(0,y_2)$,
 are similar for different multiplicity $n$,
except that $\tilde C_2^{(n)+-}$ becomes narrower with increasing $n$.
\i In hadron-hadron collisions,
the correlation functions depend strongly on transverse momentum
and are largest for small-$p_\rT$ particles. Consequently, correlations are
stronger in multi-dimensional phase space than in a lower dimensional
 projection,
such as rapidity space. Further implications of this observation will be
discussed in Sect.~4.3.
\i In the central c.m.~region, and at comparable energy,
the correlation strength observed in M$^+$p collisions at
$\sqrt s$=22 GeV is of similar magnitude as in $\E$ collisions
 and as in $\m$p collisions, if the trend of the latter is extrapolated
to $W=22$ GeV.
Model predictions for $\re^+\re^-$ and $\mu$p interactions
slightly underestimate the correlation strength but give, nonetheless,
clear evidence that (hard) gluon effects are the main  source
of correlations in rapidity space.
\i Combinatorial background can be suppressed by studying the
correlation of strange particles. Data are scarce,
 but support the conclusions drawn from data on non-strange particles.
\i The UA5 cluster Monte Carlo and FRITIOF describe
$C_2^{(n)}(\h_1,\h_2)$ at CERN-Collider energies, at least in the
charge multiplicity range  $34\leq n\leq 38$.
At lower energies $(20\simkl \sqrt s \simkl 30$ GeV), the single-chain
LUND model shows a strong anti-correlation among like-charge particles.
The two-string FRITIOF model
and DPM predict  negative values for $C_2(y_1,y_2)$ or $K_2(y_1,y_2)$
in the central region for like charges.
They are positive but far below the data
for unlike-charge pairs.
QGSM reproduces $C_2(y_1,y_2)$ and $\tilde C_2(y_1,y_2)$
for all charge combinations, but cannot account for the short-range part
$\tilde C_\rS(y_1,y_2)$.
\i Positive correlations are observed at large values of the azimuthal
angle $\D\vf$, as expected from transverse momentum conservation.
The correlations among like-charge particle pairs at  small values of
$\D\vf$ and $\D y$, where Bose-Einstein effects should contribute, are
significantly larger than predicted by FRITIOF~2, DPM and QGSM.
The deviations are  stronger for particles with small transverse momentum.
\i In general, short-range correlations in $\E$ annihilation  are
reasonably well described by the LUND- and Webber-type models. To the
contrary, in models for hh collisions which contain  only one or two strings
without additional  $p_\rT$ effects or  gluons, correlations in rapidity
are  underestimated, those in azimuthal angle overestimated.

Models such as LUND and DPM are known to  underestimate the height
of the ``sea-gull'' wings (the particle average transverse momentum as
a   function of Feynman-$x$)~\cite{Ajin87}, a signal of semi-hard
interactions. The models neglect such processes.
This may partially explain why the models fail in both instances.

\i The distribution in the interparticle opening angle of $\E$ collisions
at LEP favours models with coherent parton showering.

\i Three-particle correlations are now observed in all charge combinations.
They are particularly large for low-$p_\rT$ particles. Within two standard
deviations, they satisfy the  linked-pair ansatz. No short-range contribution
$\tilde K_{\rS}$ is observed in three-particle correlations.
Other methods are needed to study higher-order correlations.
\end{enumerate}

\chapter{Multiplicity fluctuations and intermittency}

\section{Prelude}

The study of fluctuations in particle physics already has a
long history going back to early
cosmic-ray observations. To our knowledge, Ludlam and Slansky~\cite{Ludlam73}
were the first to advocate analysis of event-to-event
fluctuations in hadron-hadron collisions.
Comparing rapidity distributions of single events with the
sample averaged distribution, they put in evidence
strong clustering effects in longitudinal phase-space, indicating
``a remarkably structured phase-space density''~\cite{Slansky74}.
Fluctuations in individual events were also considered in the
context of Reggeon theory in the important paper establishing the
AGK-cutting rules~\cite{Abramov73}.

Early evidence for large concentrations of the particle number in small
rapidity regions for single events were reported in
cosmic ray experiments~\cite{Ale62,Ara78,Sla81} and
in pN collisions at 200 GeV beam momentum~\cite{Maru79}.
A further number of high density ``spikes" in
rapidity space have been reported during the last decade.
Fig.~4.1a shows the notorious JACEE event~\cite{Burn83} at a pseudo-rapidity
resolution (binning) of $\d\h=0.1$. It has
local fluctuations up to $dn/\d\h\approx300$  with a signal-to-background
ratio of about 1:1.
The NA22 event~\cite{AdamPL185-87} of Fig.~4.1b
contains a  ``spike"  at a rapidity resolution $\d y=0.1$
of $dn/dy=100$, corresponding to 60 times the average
density in this experiment. UA5~\cite{Carl87} has
reported ``spikes" in $dn/d\h$ up to 30 (10 times average) as early as
JACEE, but found these to be in agreement with a short-range cluster
Monte Carlo. Also  EMU-01~\cite{Adamo88} sees events with
$dn/d\h=140$ satisfactorily explained by FRITIOF.

{}From an experimental point of view, there is little doubt that events
with large local density fluctuations exist. The real
question is whether these are of dynamical or merely statistical origin,
whether the underlying probability density is continuous or intermittent.

Early attempts were made to answer the
question of non-statistical fluctuations employing
transform techniques~\cite{Takagi}, but these were not followed up.
The problem resurfaced
in the work of {Bia\l as} and Peschanski~\cite{bialas1,bialas2}, who
suggested that spikes could be a manifestation in hadron physics
of ``intermittency'', a phenomenon well-known in fluid-dynamics.
The authors argued that if intermittency occurs in particle production,
large density fluctuations are not only expected, but should also exhibit
self-similarity with respect to the size of the phase-space volume.

Ideas on self-similarity and fractals in jet physics had earlier been
formulated in~\cite{Fey79,Ven79}, rephrased in the
language of QCD branching processes in~\cite{Kon79} and in a
simplified form in~\cite{Gio79}.
For soft hadronic processes, fractals and self-similarity were
first considered in~\cite{Minh83} and their quantitative measures
in~\cite{DremJETP87,DremFest}.

In multiparticle experiments, the number of hadrons produced
in a single collision is small and subject to considerable
``noise''. To exploit the techniques employed in complex system
theory, a method must be devised to separate fluctuations of purely
statistical origin, due to finite particle numbers, from
the possibly self-similar fluctuations of the underlying particle
densities. The latter  are the quantities of  physical interest.
A solution, already used in optics and suggested for multiparticle
production in~\cite{bialas1,bialas2},
consists in measuring suitably normalized factorial moments
of the  multiplicity distribution in a given phase-space volume.

\section{Normalized factorial moments}
\subsection{The method}

The method proposed in~\cite{bialas1,bialas2}
consists in measuring the dependence of the normalized
factorial moments $F_q(\d y)$ defined
in (\ref{dr:44}-\ref{dr:46}) as a function of the resolution $\d y$.
For definiteness, $\delta y$ is supposed to be an interval in rapidity,
but the method generalizes to arbitrary phase-space dimensions.

In Sect.~2.2 we have pointed out that the scaled factorial moments
enjoy the property of ``noise-suppression".
It is easily verified that this crucial property does not apply
to ordinary moments $\lan n^q\ran/\lan n\ran^q$ (cfr.~Sect~4.7 below).
High-order moments further  act as a filter and
resolve the large $n_m$ tail of the multiplicity distribution.
They are thus  particularly sensitive to large density fluctuations
at the various scales $\d y$ used in the analysis.

As proven in~\cite{bialas1,bialas2}, a ``smooth" (rapidity) distribution,
which does not show any fluctuations except for the statistical ones,
has the property that $F_q(\d y)$ is independent of the resolution $\d y$
in the limit $\d y\to  0$. This follows directly from (\ref{dr:55}), if
$P_\r$ is a product of $\d$-functions in
$\r_m\ (m=1,\dots,M)$ centered around $\lan\rho_m\ran$.
On the other hand, if dynamical fluctuations exist and $P_\r$ is
``intermittent'' (i.e., regions of fluctuations exist at all scales of $y$),
the $F_q$ obey the power law (\ref{fq113}).
Equation (\ref{fq113}) is a scaling
law, since the ratio of the factorial moments at resolutions $L$ and $\ell$
\beq
R = F_q(\ell)/F_q(L) = (L/\ell)^{\f_q}
\eeq
only depends on  $L/\ell$.

As mentioned  in Sect.~2.4 and Subsect.~5.2.2, the ``intermittency indices''
$\f_q$ (slopes in a double-log plot) are related~\cite{Hen83,LiBu89,Hwa90}
to the anomalous dimensions $d_q=\f_q/(q-1)$, a measure for the
 deviation from an integer dimension.

We noted in Sect.~3.4 that the experimental  study of  correlations
is  difficult already for three particles. The close connection between
correlations  and factorial moments (Subsect.~2.1.4) offers a possibility to
measure higher-order correlations  with the factorial moment
method at smaller distances than previously feasible.
The method further relates possible scaling behaviour
of such correlations to the physics of fractal objects.
Despite the advantages, it should be remembered  that reliable data
can only be extracted if  factorial moments are averaged over a
large domain of phase space. This holds the
danger of obscuring  important dynamical effects.

The definition of ``intermittency" given in (\ref{fq113}), has its
origin in other disciplines\footnote{\ For
a masterly expos\'e of this subject see~\cite{Zeld90}}.
It rests on a loose parallel
between the high non-uniformity of the distribution
of energy dissipation, for example, in turbulent intermittency and the
occurrence of large ``spikes" in hadronic multiparticle final states
(Sect.~4.1). In the following  we  use
the term ``intermittency'' in a weaker sense, referring to
the rise of factorial moments with increasing resolution but not necessarily
according to a strict power law.

The suggestion  that normalized factorial moments of particle
distributions might show power-law behaviour has spurred a vigorous
experimental  search for (more or less) linear dependence of $\ln F_q$
on $-\ln\d y$. Within a surprisingly short time (one-dimensional) analyses
were performed for $\E$
{}~\cite{BuLiPe88Aba90,Brau89,Behr90,Abreu90,Akr91,Dec91,Mur93},
$\m$p~\cite{Dera90},
$\n A$~\cite{Verlu90},
hh~\cite{Ajin89-90,Alba90,Singh91,Are91,Bravi,Rimon91,Wang94},
h$A$~\cite{Holy89,Dera90Sing,Bott91,Ghosh92,Shiv93,Shiv94} and
$AA$~
\cite{Holy89,Dera90Sing,Adamo90,Seng90,AAke90,Ghosh91,Sark93,Abbo94,Albr94}
collisions.
With respect to the original objective, the early one-dimensional work
has remained inconclusive, but valuable information and experience was
accumulated. Much more promising insight has come from studies in two-
and three-dimensional phase space. This is discussed in Sect.~4.3.
Further extensions of this approach, concentrating on improved integration
methods and differential studies in Lorentz-invariant variables have lead
to further clarification of the issues involved in intermittency.
These very recent developments are presented in Sects.~4.8-4.10.

\subsection{Results on log-log plots (in one dimension)}

In this and the next few  sections we review experimental results
and model predictions obtained from  one-dimensional studies.
Due to the vast amount
of data available, we limit ourselves to an illustration of the
major characteristics of factorial moment behaviour in
various processes and at various energies.

In Fig.~4.2a, log$F_5$ is plotted~\cite{bialas1,bialas2}
as a function of -log $\d\h$ ($\h$ is the pseudorapidity)
for the JACEE event. It is compared with an independent
emission Monte-Carlo model tuned to reproduce the average $\h$ distribution
of Fig.~4.1a and the global multiplicity distribution, but has no short-range
correlations included. While the Monte-Carlo
model indeed predicts constant $F_5$, the JACEE event shows a first
indication for a linear increase, i.e. a possible sign of intermittency.

Further examples are given in Fig.~4.2b for KLM~\cite{Holy89}, again showing
an roughly linear increase for $\d\h<1\  (-\ln\d\h>0)$ instead of
the flat behaviour expected for  independent emission, and in Figs.~4.2c and d
for UA1 \cite{Alba90} in terms of $\d\h$ and $\d\f$, respectively.

Anomalous dimensions $d_q$ fitted over the range $0.1<\d y(\d\h)<1.0$
are compiled in Fig.~4.3~\cite{Bial90}. They typically range from
$0.01$ to $0.1$, which means that the fractal (R\'enyi) dimensions $D_q=1-d_q$
are close to one. The $d_q$ are larger and grow faster with increasing order
$q$ in $\m$p and $\E$ (Fig.~4.3a) than in hh collisions (Fig.~4.3b)
and are small and almost independent of $q$ in heavy-ion collisions
(Fig.~4.3c). For hh collisions, the $q$-dependence is considerably stronger
for NA22 ($\sqrt{s}=22$ GeV) than for UA1 ($\sqrt{s}=630$ GeV).

\subsection{Model predictions}

\subsubsection{Hadron-hadron collisions}

A comparison to NA22 data on slopes $\f_q$ (Fig.~4.4a) shows~\cite{Ajin89-90}
that intermittency is absent at $\sqrt{s}=22$~GeV in a two-chain DPM and
underestimated by FRITIOF. In Fig.~4.4b, PYTHIA is seen to stay below the
UA1 data~\cite{Alba90}, even after inclusion of Bose-Einstein interference
for identical particles. The UA5 cluster Monte Carlo GENCL, able to reproduce
conventional short-range correlations (at least in a certain range of
multiplicities cfr.~Fig.~3.3), follows the data down to a resolution
of $\d\h\approx0.3$, but completely fails for smaller $\d\h$.

Also, a multi-chain version of DPM including mini-jet production has
been compared to  NA22 and UA1 data. The slopes are found to be too small
by at least a factor of~2~\cite{Bopp90}.

With respect to intermittency analysis, the situation may improve
with the introduction of ECCO~\cite{Hwa92}, an eikonal cascade model
based on geometrical branching, which now can account for strong
fluctuations, in particular in higher dimensions (Sect.~4.3 below).
However, the present version of ECCO is still less refined
than the more conventional models with respect to other observables.

The above examples show that present models for multiparticle production
in hh collisions are unable to reproduce the magnitude and the growth
of factorial moments with increasing resolution. From the discussion in
Chapter~3, it is evident that model predictions for correlations in general
are quite unreliable. The two-particle correlation function, measured by
$F_2$, also determines to a large extent the higher-order factorial moments
(cfr. Eq.~\ref{dr:48}) because of the weakness of genuine high-order
correlations. It is, therefore, mandatory to improve the models before
evidence for ``new physics" at very small (rapidity) separation can be
claimed. We return to this important question in later sections.

\subsubsection{h$A$ and $AA$ collisions}

The intermittency indices are much smaller in h$A$ and $AA$ collisions
than in hh collisions, and the event samples are much smaller. Model
comparisons are, therefore, less conclusive than in hh collisions. FRITIOF
is found too low in NA22 \cite{Bott91} for $\p^+/\PK^+$ on Al and Au at
250 GeV/$c$, in E802 \cite{Abbo94} for central $^{16}$OAl and $^{16}$OCu at
14.6 $A$ GeV/$c$, in WA80 \cite{Albr94} for SS and Au at 200 $A$ GeV/$c$,
and in NA35 \cite{Dera90Sing} for pAu, OAu, SAu and SS at 200 $A$ GeV/$c$.
In WA80 it is shown that rough agreement can be obtained by renormalization
to the leftmost point of FRITIOF on the log-log plot (essentially the shape
of the overall multiplicity distribution) to the data. NA35 shows that
agreement can be obtained by adding Bose-Einstein interference for
like-charged particles (for a detailed analysis of the influence of BE
correlations see further below).

\subsubsection{Lepton-hadron collisions}

In Fig.~4.5a EMC data~\cite{Dera90} are compared to what is expected from
an extrapolation of conventional short- and long-range
correlations~\cite{CapFiaKrz89}. At small $\d y$, the data are consistently
above these expectations. As Fig.~4.5b shows, the slopes $\f_q$ in the same
data are considerably larger than predicted by the Webber and LUND models.
Similarly, Fig.~4.5c shows too low $\ln F_3$ from LUND, not only for
$\n$Ne but also for the ``simpler'' $\n \rD_2$ interactions~\cite{Verlu90}.

We tentatively conclude that also presently used {\it lepton-hadron} models
as such are unable to reproduce the intermittency observed in this process.

\subsubsection{$\E$ annihilation}

The annihilation of $\E$ into hadrons is by far the best understood of all
multihadron reactions. Creation of hadrons is traditionally pictured as a
multi-step process comprising a ``hard" parton evolution phase, described by
perturbative QCD---the parton shower---and a non-perturbative colour-confining
soft hadronization phase (Fig.~4.6). The former is a cascade process of nearly
self-similar type, and is expected to show characteristics typical of a
fractal object~\cite{Fey79,Ven79,Gio79}. In fact, already in 1979, in a
discussion of QCD jets, it was stated~\cite{Ven79} that ``the resulting
picture of a jet is formally similar to that of certain mathematical objects,
known as fractals, which look more and more irregular and complex as we
look at them with a better and better resolution". The expectation is,
therefore, that parton showers should exhibit intermittency at the parton
level. However, this is not sufficient to guarantee ``intermittency'' at
the hadron level. It is indeed difficult to imagine how the ``re-shuffling''
of the parton momenta during the hadronization phase with, e.g., the
formation of hadronic resonances and their subsequent decay would preserve
the (supposedly singular) nature of the correlations. A local parton-hadron
duality type of explanation is not satisfactory either, since ``it is merely
a name for a mechanism that is not at all understood''~\cite{Bial92}.

To describe the hadronization phase, all present Monte-Carlo codes rely
in last instance on a large amount of $\E$ data at different energies and
are carefully tuned to these. It came, therefore, as a surprise that a first
(indirect) analysis~\cite{BuLiPe88Aba90} of HRS results, shortly followed by
TASSO data~\cite{Brau89}, revealed deviations from model predictions quite
similar to those observed in lh and hh collisions (Figs.~4.7a,b).
More recently, CELLO~\cite{Behr90} and, in particular, the LEP
experiments~\cite{Abreu90,Akr91,Dec91}, claim ``reasonable''
agreement with the parton shower version of the LUND Monte Carlo
(Figs.~4.7c,d). Nevertheless, new DELPHI data now show, with ten times
larger  statistics, significant deviations even
with a ``re-tuned" version of the Monte Carlo (Fig.~4.7d).

The origin of intermittency in the models is not quite as clear as is often
stated. Indeed, comparison of the factorial moments on parton and hadron
level in Figs.~4.8a,b~\cite{BotBusch}, shows that in (standard) JETSET the
increase of $\ln F_q$ at small $\d y$ is not due to the parton shower, but to
hadronization! Only if the parton shower is allowed to continue down to very
low $Q^2_0$ values (Fig.~4.8c,d for $Q^2_0$=0.4 GeV$^2$), implying local
hadron-parton duality, is intermittency becoming visible also at the parton
level. It has been verified that the influence of $Q^2_0$ is, of course,
much less important at $1$~TeV.

On the other hand, intermittency seems to be fully developed on the parton
level already at 91 GeV in the Webber model, and is in fact smeared out by
hadronization~\cite{Jedr89}.

The sensitivity to the cut-off in the perturbative QCD cascade and the
role of hard and soft phases has also been  discussed
in terms of the dipole radiation model~\cite{Dahl89}.
Intermittency can be increased in the soft phase by an increase of the
$\p/\r$ ratio, also required from direct measurements by NA22~\cite{Agab90},
EMC~\cite{ArneC33-86} and in hA collisions~\cite{Walker91}. The direct pions
resolve the underlying parton structure better than the more massive
resonances. From a tunnelling production mechanism, these pions are expected
to have smaller $p_\rT$ than other particles, a property presently neglected
in the MC programs. A Goldstone-like mechanism causing additional soft
direct pion production at break-up points has recently been suggested by
the LUND group~\cite{AndGu94}.

For further progress, additional studies are needed.

$\bu$ One should identify the true causes of intermittency in present
Monte-Carlo models, preferably on more sensitive distributions, such as those
to be discussed below.
This should reveal the influence  of hard and soft gluon
emission at high energies where parton showering is fully developed and
dominates over the soft phase.

$\bu$ Intermittency is also particularly sensitive to the exact
treatment of the {\it soft} phase.
This phase can be studied with high statistics
at lower energies where parton showering is less important.

\subsection{A warning}

Before going into the necessary further detail,
we should mention the influence
of possible experimental biases. On purpose and by its very definition,
the higher factorial
moments are sensitive to a small number of events in the tail of the
 multiplicity
distribution in small phase-space bins.

Moments can be {\it reduced} by limited two-track resolution, by track losses
from limited acceptance or bad reconstruction, or simply due to truncation of
the multiplicity distribution in a finite event sample.

Moments can be {\it increased} due to double counting of tracks (track match
failures), Dalitz decays and nearby $\g$-conversions or K$^0/\La$ decays. A
dangerous increase comes from the commonly used ``horizontal'' averaging,
where a {\it constant} average (pseudo-) rapidity distribution is assumed
over the range $\D Y$. Contrary to first belief, this problem is {\it not}
completely solved by the  correction method  proposed in~\cite{Fial89} !

Further influence is to be expected from the choice of the sample, e.g.,
inelastic or non-diffractive, cuts on multiplicity, cuts on $p_\rT$,
all events or only those with $n \geq n_0$ in $\D Y$, etc., the size and
position of $\D Y$, the $\d y$ region chosen for the fit and the correlation
of errors.

Many of these effects have been studied in a number of experiments and we
refer to these and to~\cite{Wo90Ek90Sei90,Frie89} for more details.

\section{Higher dimensions}
\subsection{The projection effect}

So far, we have discussed factorial moments derived from one-dimensional
distributions in rapidity or pseudorapidity. The analysis can evidently be
extended to other 1D variables, such as the azimuthal angle $\vf$ in the plane
perpendicular to the beam or event axis, or the particle transverse momentum
($p_\rT$). Given sufficient statistics, distributions can be analysed in two-
and three-dimensional phase-space domains. Common choices are $(\D y,\D\vf)$,
$(\D y,\D\ln p_\rT)$, $(\D\vf,\D\ln p_\rT)$ and $(\D y,\D\vf,\D\ln p_\rT)$.

Fig.~4.9a gives an example of 1D-results from UA1~\cite{Alba90} showing that
intermittency is also present in $\vf$. The intermittency effect is larger
when two-dimensional cells $(\D y,\D\vf)$ are studied than in 1D
(Fig.~4.9b,c). This is particularly pronounced in $\E$ annihilations
(Figs.~4.10a,b), the measured slopes $\f_q$ being about six times larger
in 2D than in 1D. These observations
are now understood to imply that intermittency  ``lives in
3D"~\cite{Ochs,BiaSei90}. Projection onto lower-dimensional subspaces dilutes
the effect and leads to flattening of the factorial moments. This is most
pleasantly demonstrated by the fact that one can enjoy a continuous
(two-dimensional) shadow of a tree, in spite of the self-similar branching of
this tree in three dimensions.

The projection-effect is convincingly illustrated in Fig.~4.10a,b. The lines
in sub-fig. a) are fits by a 2D $\a$-model; the curves in sub-fig. b) are the
projections onto rapidity-space and show considerably less increase and even
a flattening for $\d y\to 0$ (note the difference in scale). Nevertheless,
Fig.~4.10a still shows saturation of $F_2$ at large $M$ even in 2D, an
indication that an analysis in three dimensions may be required.

\subsection{Transformed momentum space}

To study intermittency in three-dimensional phase space, one faces the
additional difficulty that the particle density is all but uniform in the
usual single-particle variables $y,\vf$ and $p_\rT$. The distribution in
$p_\rT$ is in fact falling exponentially. Uniformity of the density is,
however, an explicit assumption in the derivation of the power law
(\ref{fq113}). Violation of this condition renders an intermittency analysis
useless.

To circumvent this problem, the authors of~\cite{Ochs,BiGa90} have proposed
to use domains in a transformed momentum space with (almost) constant
density.
This is accomplished by a transformation of the original variables $y,\vf$
and $\ln p_\rT$ to ``cumulative" variables. Thus, for a single variable,
say $y$, one defines the new variable $X(y)$ as
\beq
X(y) = \frac{\int^y_{y_{\min}}\r_1(y')dy'}{\int^{y_{\max}}_{y_{\min}}
\r_1(y')dy'}\ \ .
\eeq
For higher dimensions, it is assumed in~\cite{Ochs} that the single particle
density factorizes as
\beq
\r_1(y,\vf,p_\rT) = \r_\ra(y)\r_\rb(\vf)\r_\rc(p_\rT)\ \ .\eeq
\ni
Under this rather strong hypothesis, one can transform each of the three
variables independently. The method proposed in~\cite{BiGa90} does not assume
factorization but is technically quite involved. In practice, the two
techniques give satisfactorily similar results~\cite{Bott92}.

Data on $F_2$ in various dimensions are shown in Fig.~4.11 for
$\E$~\cite{Abreu90} and hh collisions~\cite{Ajin89-90,Alba90}. In all cases,
the data behave more power-like in 2D than in 1D. From Fig.~4.11a, it is also
evident that JETSET PS remains in good agreement with $\E$ data in higher
dimensions.

At variance with power-law behaviour expected from intermittency, NA22 finds
that the 3D factorial moments show an upward bending (Fig.~4.11c). This
effect
persists after exclusion of Dalitz decays and $\g$-conversions. A rise faster
than power law is also observed in 3D for collisions of various projectiles
with $\mbox{Au}$ by NA35 (Fig.~4.12a)~\cite{Dera90Sing}. Following a
suggestion
in~\cite{FiMPI91}, NA35 finds that the normalized factorial cumulant
$K_2=F_2-1$ shows much better linearity in a log-log plot than $F_2$ itself
(Fig.~4.12b).

This observation, in fact, furthers  considerably our understanding of the
intermittency phenomenon. In~\cite{FiMPI91,Fial91} the author has compared
3D data on $F_2$ at $\sqrt s\simeq20$ GeV for $\m$p~\cite{Dera90},
$\p/\PK$p~\cite{Ajin89-90}, $\mbox{pAu}$, $\mbox{OAu}$ and
$\mbox{SAu}$~\cite{Dera90Sing} collisions using the parametrization
\beq
F_2 = 1 + c (M^3)^{\f_2} + c' \ \ ,
\label{eq:3.15}
\eeq
where $M^3$ is the number of 3D phase-space cells. The second term in
(\ref{eq:3.15}) is equal to  $K_2$. The constant $c'$ accounts for
long-range correlations, known to exist in hh collisions.

The comparison of (\ref{eq:3.15}) to the data is shown in Fig.~4.13; the
parameters are given in the figure caption. The parameter $c'$ is negligible
for $\m$p and heavy ion collisions, but non-zero for meson-proton and p$A$
collisions, in agreement with expectations. The most noteworthy result,
however, concerns $\f_2$, which is seen to have a value in the range 0.4-0.5
for all processes. This is remarkable in various respects.

Firstly, if confirmed in further studies, and in particular for $\E$
annihilation, it strongly suggests that the resolution-dependence of $F_2$
exhibits a high degree of ``universality", is independent of specific
details of the production process and thus reflects general features of
hadronization dynamics.

Secondly, such universality is at variance with the hitherto accepted idea
that the factorial moments and the anomalous dimensions become the smaller
the more complex the collision process, due to an increasing inter-mixing of
production sources~\cite{LiBu89}.

Thirdly, if ``universality" continues to hold in high energy $\E$
annihilation,
one must revise the commonly expressed opinion that the perturbative parton
evolution, and in particular hard-jet emission, is the primary cause of the
rise of factorial moments at high resolution. Needless to say, it would be
most interesting to verify systematically the universality conjecture in
other reactions and for three-particle correlations.

The experimental success of expression (\ref{eq:3.15}) becomes quite
intriguing when one realizes that the volume $\d\sim M^{-3}$ of a
phase-space cell (for sufficiently large $M$) is in fact related to
the invariant mass $M_{\inv}$ of the two-particle system or to $Q^2$,
the square of their four-momentum difference.  The form
(\ref{eq:3.15}) implies that the two-particle correlation function
behaves as a power law in $M_{\inv}$ or $Q^2$.
The data, therefore, seem to tell that an intermittency analysis
should be performed in (Lorentz-invariant) multiparticle variables,
rather than single-particle variables.  This will be further discussed
in Sect.~4.8-4.10.

As mentioned in Subsect.~4.2.3 above, ECCO~\cite{Hwa92} has some success
in describing the NA22 data on fluctuations in varying scales of
resolution, in particular when the analysis is done in three
dimensions (Fig.~4.14). The basis of this model is geometrical branching
for soft production at low $p_\rT$. The geometrical aspect of hadrons,
i.e.  the fact that they are extended objects, puts the impact
parameter $R$ in a pre-eminent role. The fluctuation in $R$ from event
to event leads to fluctuations in $p_\rT$ and explains the
non-vanishing intermittency in $\ln p_\rT$ reported by NA22. The
(stronger) intermittency in rapidity can be generated only with a
singular splitting function for branching in rapidity space. Since
there is no branching in $\vf$ in the model, intermittency is nearly
non-existent in this variable.  Still, the long-range correlation due
to $p_\rT$ conservation leads to a decrease of $F_q$ at large bin
size, a feature also observed by NA22.

\subsection{A generalized power law}

It has been pointed out~\cite{OchWo88-89} that
the one-dimensional moments follow the generalized power law
\beq
F_q \propto (g(\d y))^{\f_q}\ ,
\label{ochs:rel}
\eeq
in multiplicative cascade models.
In (\ref{ochs:rel}), $g(\d y)$ is a
general function of $\d y$. Expressing $g$ in terms of
$F_2$, one finds the linear relation
\beq
\ln F_q=c_q+(\f_q/\f_2)\ln  F_2 \ ,
\label{ochs:rel1}
\eeq
from which the ratio of anomalous dimensions is directly obtained.
This intriguing relation has successfully been confirmed by
experiment, not only in one dimension, but up to 3D~\cite{Ochs}.
Moreover, the ratios $\f_q/\f_2$ are found to be largely independent
of the dimension of phase space~(Fig.~4.15a) and of the type of collision
(Fig.~4.15b).

The ratio of the anomalous dimensions $d_q(=\f_q/(q-1))$
and $d_2$ are shown in Fig.~4.16b as a function of $q$.
The $q$-dependence is claimed to be
indicative of  the mechanism causing  intermittent behaviour.
For  a (multiplicative) cascade mechanism,
in the log-normal  approximation (long cascades), the moments satisfy the
relation~\cite{bialas1,bialas2}
\beq
\frac{d_q}{d_2}=\frac{\f_q}{\f_2} \frac{1}{q-1}=\frac{q}{2}.
\label{3:19}
\eeq
However, the use of the Central Limit Theorem for a multiplicative process,
such as in the $\a$-model, is a very crude approximation~\cite{AlbBi91}
particularly in the tails. As argued in~\cite{BrPe91}, a better description
might be obtained if the density probability distribution is assumed to be
a log-L\'evy-stable distribution, characterized by a L\'evy-index
$\mu$. In that case (\ref{3:19}) generalizes to
\beq
\frac{d_q}{d_2}=\frac{1}{2^\m-2}\frac{q^\m-q}{q-1}\ .
\label{3:19b}\
\eeq
For $\mu=2$, the Gaussian case, (\ref{3:19b}) reduces to (\ref{3:19}).

The multifractal behaviour characterized by (\ref{3:19}-\ref{3:19b})
reduces to a mono-fractal behaviour~\cite{Satz89,BialHwa91}
\beq
\frac{d_q}{d_2}=1
\eeq
for  $\m=0$, implying  an order-independent anomalous dimension. This
would happen if  intermittency were be due to
a second-order phase transition.
Consequently,  monofractal behaviour
might be a signal for a quark-gluon-plasma phase transition.

The  data are best fitted with the L\'evy-law solution with $\m=1.6$. This
value is  inconsistent with the Gaussian approximation,
and also  definitely higher than expected
for  a second-order phase transition.

The validity of the dimension-independent generalized power behaviour
has been questioned in a recent NA22 analysis~\cite{Ajin89-90} shown in
Fig.~4.16a. While a fit to the combined data on all variables and dimensions
(full circles), as well as a weighted average over all individual fits give
$\mu$ values in rough agreement with those of~\cite{Ochs}, the 3D-data have
$\mu>2$, not allowed in the sense of L\'evy-laws.

Even larger values of $\m$, ranging from 3.2 to 3.5, have been found for
$\mu$p deep-inelastic scattering in~\cite{BrPe91}. According
to~\cite{Ratti91,Ratti92}, this is evidence that the procedure to obtain
the L\'evy-index is used outside its domain of validity. An allegedly more
general method, based on Double Trace Moments (to be discussed in
Subsect.~4.7.7) indeed yields $\mu$-values within the mathematically allowed
boundaries. However, we shall see that the latter method may be criticized
on other grounds. A possible way out is self-affinity to be discussed in
Subsect.~4.3.5, below.

The linear $d_q/d_2$ behaviour in Fig.~4.16a) and b) gives some justification
for (\ref{BB:134}). Fig.~4.16c) and d) show~\cite{chek94} the slope $r$
of (\ref{BB:134}) for a number of experiments. All experiments, except
perhaps SAg/Br, show multifractal behaviour $(r>0)$.

Despite the confusion, it remains a noteworthy experimental fact that
the factorial moments of different orders obey simple
hierarchical relations of the type~(\ref{ochs:rel1}).
This  means  that correlation functions of different orders are
not completely independent but are somehow interconnected.
Such situations are commonly encountered in various branches of
many-body physics (see e.g.~\cite{CaSa89,WolfAPP90,Peeb80}), but a
satisfactory link with particle phenomenology, let alone QCD, remains
to be established. Nevertheless, on a simple example it was recently
shown~\cite{edw:sant} that a linear  relation between $\ln F_3$ and
$\ln F_2$ can be obtained  if the connected correlation functions are
assumed to be of a factorized Mueller-Regge power-law form in two-particle
invariant-masses squared $s_{ij}$,
i.e.~ $C_3(1,2,3)\propto (s_{12})^{1-\alpha_1}\,
(s_{23})^{1-\alpha_2} +\,\mbox{cycl. perm.}$. Note that this Regge-form
has the ``linking'' structure of (3.13).

\subsection{Thermal versus non-thermal phase transition}

\subsubsection{Second order phase transition?}

A simple model that can provide some
hint on the nature of a second-order phase transition is the Ising model
in 2D \cite{Ising}. Its intermittency behaviour has been studied both
analytically and numerically \cite{Satz89,Wosiek88}. The anomalous dimension
is found to be $d_q=1/8$, independent of $q$. Based on that finding, it has
been conjectured that intermittency may be monofractal if due to a QCD
second-order phase transition \cite{BialHwa91}. However, as mentioned in
Subsect.~4.3.3 above, all types of interactions, including heavy-ion
collisions, show multifractal behaviour.

Of course, the Ising model is very simple and the above conjecture has little
basis. In \cite{hwanaz}, intermittency is, therefore, studied in the
framework of the Ginzburg-Landau theory also used to describe the confinement
of magnetic fields into fluxoids in a type II superconductor. In the model,
the anomalous dimension is not constant, but follows
\beq
\frac{d_q}{d_2} = (q-1)^{\n-1}\ \ \ , \ \ \n=1.304\ \ ,
\label{4.10}
\eeq
with $\n$ being a universal quantity valid for all systems describable by the
GL theory, independent of the underlying dimension or the parameters of the
model. This is of particular importance for a QCD phase transition, since
neither the transition temperature nor the other important parameters are
known there.

In quantum optics, $\g$ production at the threshold of lasing is
discribable as a second-order phase transition. Indeed, a photo-count
experiment \cite{Young} has verified (\ref{4.10}) to high precision.
On the other hand, the current NA22 data on particle production in hadronic
collisions give $\n=1.45\pm 0.04$ \cite{Charl}, heavy-ion experiments
$\n=1.55\pm 0.12$ \cite{hwanaz} and $\n=1.459\pm 0.021$ \cite{Seng90}.

For a first-order phase transition, all $d_q$ are zero and no intermittency
would be observed \cite{BialHwa91}.
However, it has been shown in \cite{Babi95} that in a generalized GL model,
a first-order phase transition combined with the quantum optics analogy
of lasing at threshold can lead to intermittency behaviour in some regions
of the parameters, with approximately the same intermittency indices as
a second-order phase transition.

\subsubsection{Non-thermal phase transition?}

Of course, the phase transition does not need to be thermal, i.e.,
the new phase need not be characterized by a thermodynamical behaviour.
Such a transition could, e.g. take place during a parton-shower cascade
and has been formulated in \cite{VHov89} for a number of ``ultra-soft''
phenomena, including intermittency. It leads to the co-existence of different
phases, in analogy to different phases of the spin-glass systems. The
examples
of the JACEE event (Fig.~4.1a), which contains many ``spikes'' and
``holes'', and that of the NA22 event (Fig.~4.1b), which consists of just
one spike, indicate that such a possibility may be more than just a
speculation.

The condition for the existence of such different phases of a self-similar
cascade is that the function
\beq
\la_q = (\f_q+1)/q
\label{4.11}
\eeq
has a minimum at some value $q=q_\rc$ (not necessarily an integer)
\cite{PescTH5891,Pesc89,BrPe90,BiaZa}. The regions $q<q_\rc$ and $q>q_\rc$
are dominated by numerous small fluctuations and rare large fluctuations,
respectively. In the terminology of \cite{BiaZa}, the system resembles a
mixture of a ``liquid'' of many
small fluctuations and a ``dust'' of high density. We see either the
liquid or the dust phase, depending on whether we probe the system by a
moment of order $q<q_\rc$ or $q>q_\rc$, respectively.

In Fig.~4.17a, $\la_q$ is compiled \cite{BiaZa} from KLM, EMC and NA22
as a function of the order $q$. The low $p_\rT$ NA22 data \cite{Ajin89-90}
$(p_\rT<0.15$ GeV/$c$) indeed show a marked minimum with $q_\rc$ between 3
and 4, while the uncut data have not saturated at $q\leq 5$. Following
\cite{BiaZa}, the $\la_q$ behaviour has been studied by a number of
heavy-ion experiments \cite{Shiv93,Shiv94,Seng90,Sark93}. While a
saturation, but no clear minimum is seen by experiments stopping
their analysis at $q=5$ or 6, a minimum is now observed for $4<q_\rc<5$
in central C-Cu collisions at 4.5 $A$ GeV/$c$, where the analysis is carried
to $q=8$ \cite{Sark93} (Fig.~4.17b).

The observation of a minimum in the $\la_q$-distribution suggests a phase
transition \cite{PescTH5891,Pesc89,BrPe90}, but according to the
interpretation
\cite{BiaZa} it is merely the ``apparatus'' changing from a sensitivity
for the dominating small fluctuations at $q<q_\rc$ to an insensitivity for
those at $q>q_\rc$. The two phases could coexist without a transition being
necessary.

So, phase transition or not, two phases seem to coexist and it will be a
challenge to find
their physical interpretation in terms of the theory of strong interactions.

\subsection{Self-affinity}

Comparing log-log plots for one phase-space dimension, one notices that the
ln$F_q$ saturate, but at different $F_q$ values for different variables
$y,\vf$ or ln$p_\rT$. The saturation in one dimension can be
explained as projection effect of a three-dimensional phenomenon. However,
also in three dimensional analysis the power law (2.108) is not exact. In
Fig.~4.11c, the 3D data are seen to bend upward.

It has been shown in \cite{Wu93} (see also \cite{Wosi95}) that this can
be understood by taking the anisotropy of occupied phase-space (longitudinal
phase space \cite{Hove69}) into account. In view of this phase-space
anisotropy, also its partition should be anisotropic. In other words, the
density fluctuation in phase space should be {\it self-affine} rather than
{\it self-similar} \cite{Mand91}.

If the phase-space structure is indeed self-affine, it can be characterized
by a parameter called roughness or Hurst exponent \cite{Mand91},
defined as
\beq
H_{ij}=\ln \la_i/\ln \la_j \ \ \ (0\leq H_{ij}\leq 1)
\eeq
with $\la_i$ ($i=1,2,3;\ \ \la_1\leq\la_2\leq\la_3$) being the shrinkage
ratios in the self-affine transformations
\beq
\d x_i \to \d x_i/\la_i\ ,
\eeq
of the phase-space variables $x_i$.

The Hurst exponents can be obtained from the experimentally observed
saturation curves of the one-dimensional ln$F_q(\d x_i)$ distributions.
Using the NA22 curves for $y$ and ln$p_\rT$ (Fig.~4.11c), a Hurst exponent
of $H_{y,p_\rT}=0.516\pm 0.015$ is obtained for these two variables, in
agreement with self-affinity ($H<1$) rather than self-similarity $(H=1)$.

The upward bending for $F_q$ in the three-dimensional self-similar analysis
is then easy to understand:
Performing a self-similar analysis, phase space is not shrunk according
to the self-affine dynamical fluctuation. So, the real dynamic fluctuation
cannot be fully observed and the corresponding $F_q$ comes out smaller
at intermediate scales. At very small bins, however, this difference between
self-affine and self-similar space shrinkage disappears and the $F_q$ values
obtained approach each other. As a consequence, the slope on the log-log plot
has to increase at small bin sizes
and the self-similar analysis grants an upward bending if the
underlying structure is self-affine (i.e. corresponds to a power law).

On a self-affine Monte-Carlo branching model exactly reproducing the NA22
$d_q/d_2$ values of Fig.~4.16a, this upward-bending effect is shown to cause
the apparent violation of the L\'evy stability $\m\leq 2$ described in
Subsect.~4.3.3 \cite{Zhan95}.

\section{Dependences of the effect}
\subsection{Charge dependence}

A mechanism known to cause correlations  at small distances in phase space is
Bose-Einstein interference between identical
particles~\cite{CapFiaKrz89,CaFrie89,Gyul90}.
For reviews of the present status
of this field we refer to~\cite{Zaic,LPHEP}.
{}From the outset it must be realized, however, that the
conventional Gaussian- or exponential-type
parametrizations of the Bose-Einstein effect lead
to a saturation at $\d y\to 0$ and {\it not} to the power law (\ref{fq113})!

In~\cite{Gyul90} it is argued that the slopes should be  roughly a factor 2
larger for identical particles than for all charges combined. The experimental
situation  is less than clear, in particular for  1D analyses. Contrary to
the prediction, TASSO and DELPHI see less intermittency for identical
particles. EMC finds an enhanced effect for positive but not much for
negative particles in a one-dimensional analysis, and very similar slopes
in a 3D analysis. NA22 observes an enhancement for negatives, but not for
positives. UA1 sees no difference, whereas  NA35 sees an increase.

CELLO finds Bose-Einstein interference necessary to explain the residual
difference between data and JETSET 7.2, but needs an un-physically large
strength-parameter $\lambda$ to obtain agreement. In the DELPHI analysis,
Bose-Einstein interference is insufficient to explain the difference between
data and models,  even with an un-physically large value of
the coherence parameter $\lambda$.

Following a suggestion in~\cite{Biya90}, higher-order
Bose-Einstein correlations have been studied
by UA1~\cite{Neum91}, NA22~\cite{Agab} and DELPHI ~\cite{Abreu95}.
In this study, ``correlation functions'' of order $q$,
\beq
R_q(Q^2_{q\p})=N_q(Q^2_{q\p})/N^{BG}_q(Q^2_{q\p}),
\label{boseq}
\eeq
are defined as ratios of the distribution of like-charged $q$-tuplets
$(q=2,3,\dots,5)$ $N_q(Q^2_{q\p})$ and a distribution of reference
(background) $q$-tuplets $N^{BG}_q(Q^2_{q\p})$ obtained from random
event-mixing. The variable $Q^2_{q\p}$ is defined as a sum over all
permutations
\beq
Q^2_{q\p} = Q^2_{12} + Q^2_{13} + \dots Q^2_{(q-1)q}
\eeq
of the squared four-momentum difference $Q^2_{ij}=-(p_i-p_j)^2$ of particles
$i$ and $j$. Note that the functions (\ref{boseq}) are normalized
inclusive densities and not correlation functions in the proper sense
(cfr.~Sect.4.8).

The UA1 data are shown in Fig.~4.18. A good fit is obtained if in the
expansion
of $R_q(Q^2_{q\p})$ suggested in~\cite{Biya90}, Gaussians (dashed curve)
are  replaced by exponentials in $Q_{q\p}$ (solid curve). Since low
$Q^2_{ij}$ pairs are lost due to limited two-track resolution in the
detector, the  data at the smallest $Q^2_{ij}$ have to be regarded as a
lower limit. A power law as expected from intermittency cannot be excluded.

UA1 has further studied the distributions $R_2$ for all-charged-(cc),
$(\pm\pm)$- and $(+-)$-pairs as a function of $Q^2(\equiv Q^2_{2\p})$
(Fig.~4.19)~\cite{Alba90}. These results have important implications. The
charge-dependence of ``intermittency", controversial in single-particle
variable analyses (see before), is now quite clear in invariant-mass variables
$(Q^2=M^2_{\inv}-4m^2_\p)$. The data for $R^{\pm\pm}_2$ (dashed) has a much
stronger $Q^2$-dependence than $R^{+-}_2$ and effectively determines the
small-$Q^2$ behaviour of $R^{\rc\rc}_2$. This is unambiguous evidence that
intermittency at small $Q^2$ is predominantly due to like-sign particle
correlations. It does not necessarily imply, however, that Bose-Einstein
interference is the sole cause.

In~\cite{Arne86} it is shown on EMC data that, especially in 3D, $F^{--}_2$
deviates much more from LUND Model predictions than $F^{+-}_2$.
The LUND Model version used does not include  Bose-Einstein correlations.
The deviation from the data is indicative for the importance of this effect.

Bose-Einstein interference must thus play a significant role at least for
small $Q^2$. This seems in contradiction with claimed successes in
$\re^+\re^-$ annihilation of parton shower Monte Carlos which neglect
Bose-Einstein interference.

Finally, we reiterate our remark that a ``conventional'' Bose-Einstein effect
with exponential or Gaussian $Q$-dependence is incompatible with intermittent
power-law behaviour. We return to this point in Subsect.~4.8.4.

\subsection{Transverse-momentum dependence}

An interesting question is whether semi-hard effects~\cite{OchWo88-89},
observed to play a role in the transverse-momentum behaviour even at NA22
energies~\cite{Ajin87}, or low-$p_\rT$ effects ~\cite{VHov89,BialPL89} are
at the origin of intermittency. A first indication for the latter comes
from the most prominent NA22  ``spike" event~\cite{AdamPL185-87},
where 5 out of 10 tracks in the spike have $p_\rT<0.15$ GeV/$c$.

In Fig.~4.20a, NA22 data on $\ln F_q$ versus $-\ln\d y$ are given for
particles
with transverse momentum $p_\rT$ below and above 0.15 GeV/$c$, and with
$p_\rT$ below and above 0.3 GeV/$c$. For particles with $p_\rT$ below the
cut (left), the $F_q$ exhibit a stronger $\delta y$-dependence than for
particles with $p_\rT$ above the cut (right).

NA22 does not claim straight lines in Fig.~4.20a, but uses fits as an
indicative measure of the increase of $\ln F_q$ over the region
$1>\d y>0.1$. In the upper half of Fig.~4.20b, the fitted anomalous dimensions
$d_q$ are compared to those obtained in the full $p_\rT$-range. The
restriction to particles with $p_\rT < 0.15$ or $0.30$ GeV/$c$ indeed leads
to an $increase$ of $d_q$; a $decrease$ is observed for $p_\rT > 0.15$
or $0.30$ GeV/$c$. This observation is confirmed by IHSC~\cite{Are91}

FRITIOF predictions  are given in the lower part of Fig.~4.20b,
again for all tracks and for tracks  with restricted $p_\rT$.
It is known~\cite{Ajin89-90} that FRITIOF gives too small slopes for
factorial moments integrated over $p_\rT$. Here, one notices that it
also fails to reproduce their  $p_\rT$-dependence.

UA1 has a bias against tracks with $p_\rT<0.15$ GeV/$c$, but
gives the dependence of $\f_2$ on the average transverse momentum
$\bar p_\rT$ of the event (Fig.~4.20c) \cite{Wu94}. The data show a remarkable
decrease of $\f_2$ with increasing $\bar p_\rT$ and, after passing through
a minimum at $\bar p_\rT\approx 0.5$ GeV/$c$, a slight increase at higher
$\bar p_\rT$ values. Lower $\bar p_\rT$ events correspond to soft processes,
while higher $\bar p_\rT$ ones correspond to events with hard jet
subprocesses.
Both types of events have higher slopes $\f_2$ than their mixture at
intermediate $\bar p_\rT$ values. (See further in~\cite{WuLiu90} for a
possible connection to the multiplicity dependence to be described
in Subsect.~4.4.4 below.)

Fig.~4.20c also contains the results obtained from Monte-Carlo events
generated with PYTHIA~5.6~\cite{PYTHIA}. At low $\bar p_\rT$ values,
the PYTHIA $\f_2$ values are strongly suppressed as compared to those
of the data.

We conclude that the intermittency observed in NA22 and UA1 data is
enhanced at low transverse momentum and is not dominated by semi-hard
effects. Hard effects dominate in high energy $\E$ and lh collisions.
Data on the $p_\rT$-dependence of factorial moments in these processes
should help in clarifying the origin of intermittency. The effect of
$p_\rT$-cuts on $\E$ data has been studied by DELPHI~\cite{Abreu90}.
One-dimensional data are shown in Fig.~4.21 and provide
several important pieces of information:

i) The log-log plot for low-$p_\rT$ particles shows less saturation (i.e.
stronger intermittency) than for larger $p_\rT$ particles. So, intermittency
is strongest in the $p_\rT$-region where hard gluon effects are weakest!

ii) A discrepancy between data and models (only indicative in Fig.~4.7d above)
is observed  in the interval $0.255<p_\rT<0.532$ GeV/$c$. This looks
surprising at first, but we shall show in Subsect.~4.4.4 that the
intermittency
effect can be stronger for individual mechanisms than for a mixture.

iii) The factorial moments are larger for $p_\rT>0.532$ GeV/$c$ than for
$p_\rT<0.255$ GeV/$c$, opposite to the trend of the NA22 data (Fig.~4.20a).
Also this  seems contradictory, but it should be realized  that for NA22
transverse momentum refers to the beam axis, which is usually close to beam
and target jet-axes. In the $\E$ analysis, $p_\rT$ is calculated relative to
the global event axis which differs from the direction of individual jets.

\subsection{Dependence on jet topology}

In their recent analysis, DELPHI~\cite{Abreu90} selects 2-jet and 3-jet events
using the {JADE/E0} invariant-mass algorithm~\cite{Jade:clus}, with resolution
parameter values $y_{\cut}=0.04$ and 0.01, and with additional cuts to clean
the 2-jet and 3-jet sample. At large bin sizes, factorial moments rise
faster with decreasing bin size (and are, therefore, larger) in 3-jet
than in 2-jet events.
This is compatible with the (large bin-size) behaviour expected from hard
gluons. At small bin sizes the increase is similar for 2-jet and 3-jet events.

In 3-jet events, factorial moments were calculated for tracks belonging
to jet 1, jet 2 and jet 3 ordered in energy. The rapidity was defined with
respect to the individual jet axis. As seen in Fig.~4.22, intermittency is
weakest in jet~3 and strongest in jet~2. The deviation from JETSET is
also  strongest for jet~2.

\subsection{Multiplicity (density) dependence}
In general, a decrease of the intermittency indices $\f_q$  is found with
increasing energy, in particular for hh, h$A$ and $AA$ collisions.
As seen in Fig.~4.23a, a strong multiplicity dependence of the intermittency
strength is observed for hh collisions by UA1~\cite{Alba90}. The trend is
opposite to the predictions of the models used by this collaboration. This
decrease of the intermittency strength with increasing multiplicity is usually
explained as a consequence of mixing of independent sources of
particles~\cite{LiBu89}. The cross-over of data and FRITIOF at intermediate
multiplicity explains the apparent success of FRITIOF in Fig.~3.3,
for multiplicities close to 30 as being accidental.

Mixing of emission sources leads to a roughly linear decrease of the
slopes $\f_q$ with increasing particle density $\lan \r\ran$ in
rapidity~\cite{CapFiaKrz89,BiaFest89,Seib90}: $\f_q\propto \lan\r\ran^{-1}$.
This is indeed observed by UA1~\cite{Alba90}.
Multiple emission sources are  present in multi-chain Dual Parton models.
The calculated slopes indeed depend linearly on  multiplicity  but are
too small by a factor of two~\cite{Aur91}. Similarly, the model studied
in~\cite{Barsh90} with independent emission at fixed impact-parameter
finds decreasing $\f_q$ with increasing multiplicity.

Also here, a study of the multiplicity dependence in $\E$ data and JETSET
allows interesting comparisons. In fact, the LEP results~\cite{Abreu90}
suggest little or no $n$-dependence, except for  the lowest multiplicities,
where the slope is largest and also the difference with JETSET~PS is the
largest.

Fig.~4.23a helps in explaining  why intermittency  is so weak in heavy-ion
collisions (cfr.~Fig.~4.3): the density (and mixing of sources) is
particularly high there.
In Fig.~4.23b EMU01~\cite{Adamo90}, therefore, compares $\f_2$ for NA22 (hp
at 250 GeV) and heavy-ion collisions at similar beam momentum per nucleon,
as a function of the particle density. Whereas slopes averaged over
multiplicity are smaller  for $AA$ collisions than for NA22 in Fig.~4.3, at
fixed $\lan \r\ran$ they are actually higher than expected from an
extrapolation of hh collisions to high density and even grow with
increasing size of the nuclei. This may be evidence for re-scattering
(see~\cite{Verlu90}) or another (collective) effect, but, as shown by
HELIOS~\cite{AAke90} and recently confirmed by EMU-01~\cite{Adamo90}, one has
to be very sure about the exclusion of $\g$-conversions before drawing
definite conclusions.

We conclude this section with an additional warning. In Subsect.~4.3.2 we
mentioned the Fia\l kowski ``universality-conjecture'' and noted that it is
incompatible with the ``mixing'' hypothesis usually invoked to explain the
multiplicity dependence of factorial moments and slopes. A different
explanation of the multiplicity dependence may therefore be needed, especially
since intermittency and  Bose-Einstein effects are now known to be closely
related.

\section{Factorial cumulants}

Normalized factorial cumulant moments, first introduced in~\cite{Mue71} and
recently studied in~\cite{CaES91}, are defined in (\ref{dr:47}) as integrals
over the background subtracted correlation functions. They share with
factorial moments the property of ``noise suppression''. The normalized
factorial moments $F_q$ can be expanded in terms of normalized cumulant
moments $K_q$ as given in (\ref{dr:48}). This expansion has been found to
converge rapidly~\cite{CaES91}. The terms in the expansion correspond to
contributions from genuine $q,(q-1)\dots,2$-particle correlations.

An analysis of  factorial cumulant moments is presented in~\cite{CaES91}.
Roughly, it is estimated that $K_2\sim0.6, K_3\sim 0.7, K_4\simkl1.0,
K_5\simkl1.5$ for UA1 data at $\sqrt s=630$ GeV. (The inequalities for $K_4$
and $K_5$ are due to the approximation $\ol{AB}$ by $\bar A\cdot\bar B$ in
(\ref{dr:48}) since no direct measurements of these averages exist.)
Clearly, the two-particle contribution to factorial moments is large,
but higher orders are not negligible. At the energy of the NA22 experiment
$K_2$ is small ($\sim$0.2), but $K_3$ is significantly larger ($\sim$0.45).

{}From (\ref{dr:48}) it is seen  that the contribution $F^{(2)}_q$ to $F_q$
from
two-particle correlations alone  can be expressed as
\beq
F^{(2)}_3=1+3K_2\nonumber
\eeq
\beq
F^{(2)}_4=1+6K_2+3\overline{K^2_2}\ \ ;
\eeq
the  contribution $F^{(3)}_q$ from two- and three-particle correlations
to $F_4$ as
\beq
F^{(3)}_4=1+6K_2+3\overline{K^2_2}+4K_3\ \  .
\eeq
The difference $F_q-F_q^{(p)}$ is a measure for  the importance
of  higher-order correlations.

Fig.~4.24a shows a cumulant-decomposition of $F_3$ and $F_4$ in
UA1-data~\cite{Egg91}. The differences between the curves indeed indicate
large contributions from genuine higher-order correlations. Similar results
are observed for NA22~\cite{Ajin89-90} in Fig.~4.24b, for $p=$2 and 3 and
$q=$3 and 4, in one-, two- and three-dimensional phase space (transformed
$y, y-\vf$ and $y-\vf-\ln p_\rT)$. In general, the difference increases with
increasing ln$M$ (decreasing bin size). This means that the contribution of
higher-order correlations to the factorial moments increases at higher
resolution. An exception are factorial moments in the variable $\vf$,
for which only two-particle correlations are found to be non-zero (not
shown).\footnote{Absence of genuine higher correlations  has been reported
in~\cite{Wang94-2}, but at far too low statistics.}

The situation is completely different in heavy-ion collisions where, with
present accuracy,  $K_q\approx 0$ for $q>2$ (Fig.~4.24c). The factorial
moments are completely dominated by two-particle
correlations~\cite{Egg91,JaMuSi,Adam93},
implying that higher-order $F_q$ contain little or no further dynamical
information for this type of collisions (see~Eq.~\ref{dr:37a}).

Using the linked-pair ansatz~\cite{CaES91} (see further Sect.~5.1.1.),
higher-order cumulant functions can be expressed as products of $K_2$ (see
also~\cite{Dia90} for an interpretation in terms of
independent superposition of sources)
\beq
K_q = A_q K_2^{q-1},
\eeq
with free constants $A_q$.

For a negative-binomial (NB) multiplicity distribution, $K_2=1/k$ and the
linking parameters are fixed numbers given by
$A^{NB}_q=(q-1)!$~\cite{WolfAPP90}. A necessary condition is stationarity,
i.e. constancy of $1/k$. This works well for UA1. For NA22~\cite{Ajin89-90},
$A_q$ is observed to increase with decreasing bin size. Approximately constant
$A_q\approx (q-1)!$ are found if the data are averaged only over a narrow
rapidity region $(-0.75\leq y\leq0.75)$ and the most prominent spike event is
excluded. The linked-pair ansatz may thus be a valid approximation for
high-order correlations in small phase-space domains but not for the average
over phase space. This would be consistent with the well-documented
fact~\cite{GiovHove86} that the negative binomial is often a good
parametrization of multiplicity distributions in restricted $\delta
y$-intervals.

We shall come back to cumulants and genuine higher-order correlations in
Sect.~4.10, where they are studied by means of a largely improved
methodology.

\section{Factorial correlators}
\subsection{The method}

The moments defined
in (\ref{dr:44}-\ref{dr:46}) measure local density
fluctuations in
phase space. Additional information is contained in the correlation between
these fluctuations within an  event. This correlation can be studied
by means of the factorial correlators defined in (\ref{f4:2}).
Correlators are
typically calculated at a  given $\d y$ for each combination $mm'$ of bins
with size $\delta y$, and then
averaged over  all combinations separated by a given bin-distance $D$.
This is  illustrated  below.

\vspace*{1.3truecm}
\epsffile[20 60 80 100]{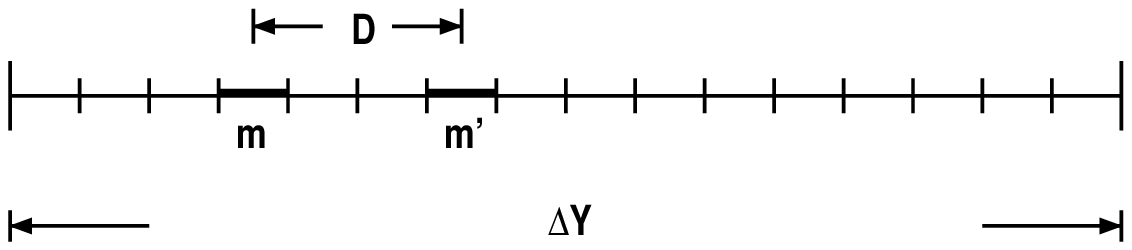}
\vs 2mm

In  the simple intermittency model ($\a$-model) described
in~\cite{bialas1,bialas2},  $F_{pq}$  depends on
$D$ but not on $\d y$ and follows the power law
\beq
F_{pq} \propto (\D Y/D)^{\f_{pq}}\ \ .
\label{3.27n}
\eeq
\ni
The powers $\f_{pq}$ (slopes in a log-log plot) obey
the relations~\cite{bialas1,bialas2}:
\beq
\f_{pq} = \f_{p+q} - \f_p - \f_q = pq\f_2\ \ ,
\label{3.28n}
\eeq
\ni
where the first equality sign is due to the $\a$-model proper,
the second to the
log-normal approximation. According to (\ref{3.27n}) $\f_{11}=\f_2$,
so that (\ref{3.28n})  can also be written in the form
\beq
\f_{pq} = pq \f_{11}\ .\eeq

\subsection{Results}

Preliminary results for pseudorapidity resolution $\d \h\geq 0.2$ have been
reported by the HELIOS Collaboration~\cite{HELIOS89}. There, however,
multiplicities $n_m$ had to be estimated from the transverse energy
$E_{\rT,m}$ in bin $m$ and the average transverse energy $\lan E_\rT\ran$
per particle: $n_m=E_{\rT,m}/\lan E_\rT\ran$ rounded to the nearest integer.
The first direct measurement is from NA22~\cite{Aiv91}. The $\ln F_{pq}$
are shown as a function of $-\ln D$ in Fig.~4.25a-d, for four values of
$\d y\geq 0.1$ (corresponding to $M=10$, 20, 30 and 40). Statistical errors
(estimated from the dispersion of the $F_{pq}$ distribution) are in general
smaller than the size of the symbols. $F_{pq}$ can be measured
up to third order in $p$ and $q$ for $\d y$=0.4 binning (Fig.~4.25a).
For $\d y$=0.1, the analysis is possible to first and second order only
(Fig.~4.25d). The smallest possible value for $D$ being equal to the bin size
$\d y$, Fig.~4.25a extends to $D=0.4$ and Fig.~4.25d to $D=0.1$. In all
cases,
an increase of $\ln F_{pq}$ is observed with increasing $-\ln D$. Very
similar results have recently been reported by EMC~\cite{Dera90},
EMU-01~\cite{Adamo90} and in \cite{Ghosh92}.

In Fig.~4.26a, the $\ln F_{pq}$ are compared at fixed $D=0.4$ for four
different
values of $\d y$. The dashed lines correspond to a horizontal line fit through
the data. In agreement with the  $\a$-model, the $F_{pq}$ indeed do not depend
on $\d y$. Also this result has been confirmed on EMC data~\cite{Dera90}
and in \cite{Ghosh92}.
The $\delta y$-independence of correlators holds exactly  in the $\a$-model.
Nevertheless, Fig.~4.26b shows that the $\d y$ independence is also valid in
FRITIOF. For the particular value of $D=0.4$, this even happens at very
similar values of $\ln F_{pq}$ as in the  data. In fact, this property is
far from unique to the $\a$-model, but holds approximately in any model with
short-range order~\cite{dewolf92}.

For $F_{11}$, the $\d y$ independence is easily derived from a parametrization
of the two-particle density, integrated over two regions of size $\d y$
separated by $D$. Using exponential short-range order~\cite{WolfAPP90}, this
gives
\beq
F_{11} -1~~ \propto~~\frac{1}{a^2}
\re^{-D/L}/ (\re^a-1)(1-\re^{-a}) \ \ ,
\label{3.29}
\eeq
\ni
where $L$ is a correlation length and $a=\d y/L$. According to (\ref{3.29}),
$F_{11}$ becomes independent of $\d y$ for $a\ll1$. Since $\re^{-D/L}\to 1$
as $D\to 0$, this form also leads to the deviations from (\ref{3.27n})
observed as a bending in Fig.~4.25.

Because of the bending, fitted slopes $\f_{pq}$ have no meaning, except
as an indication for the increase of $F_{pq}$ in a restricted range.
The slopes for two values of $\d y$ are compared to FRITIOF predictions
in Fig.~4.27a and 4.27b, respectively. As observed earlier for the case
of univariate  moments~\cite{Ajin89-90}, the FRITIOF slopes are too small
also for the correlators. This is not surprising since the model does not
succeed in reproducing even the lowest-order (i.e.~two-particle)
rapidity correlation function (Chapter~3).

\subsection{Interpretation}

Factorial correlators have been analysed in~\cite{eggers:correlators}
using a suitable parametrization of\break
$K_2(y_1,y_2)$ and the linked-pair
ansatz~\cite{CaSa89} for higher-order correlations.  The relations
(\ref{eq:b7}-\ref{eq:b10}) of Sect.~2.1.6 then allow to express all
correlators in terms of $K_2$ for arbitrary ($p,q$).  Note that the
expressions for $F_{pq}$ contain many lower-order ``combinatorial''
terms which effectively dominate and mask the contribution from
genuine ($p+q$)-order correlations.

A basically similar analysis is presented in~\cite{dewolf92}, inspired
by techniques used in quantum-optics.  The two analyses have no
difficulty to describe basic features of the NA22 data, including the
sum-rules discussed in Sect.~2.3 which were claimed to be a unique
test of random cascade models.  Fig.~4.28a compares NA22 data on
$F_2(\delta y)$ and $F_{11}(D)$ with the calculations
from~\cite{dewolf92}. $F_2(\delta y)$ is used to fix the parameters of
$K_2(\delta y)$ (assuming stationarity); $F_{11}(D)$ follows after
integration over the appropriate rapidity-domains.  With the
linking-ansatz of~\cite{WolfAPP90} all other correlators are
calculated without further assumptions. An illustrative example is
shown in Fig.~4.28b which compares $F_{12}(D)$ from NA22 to the
prediction.  The agreement is excellent in all cases. This observation
is confirmed in \cite{Ghosh92}.

According to (\ref{3.28n}), the ratio $\f_{pq}/\f_2$ is expected to grow with
increasing orders $p$ and $q$ like their product $pq$. In Figs.~4.27c and
4.27d
this is tested for $\d y$=0.4 and $\d y$=0.2, respectively. In both cases,
the experimental results lie far above the dashed line corresponding to the
expected $\f_{pq}/\f_2=pq$. Since the dependence of $\ln F_{pq}$ on $-\ln D$
is not strictly linear, this comparison depends on the range of $\d y$ and
$D$ used to determine $\f_2$ and $\f_{pq}$. In Fig.~4.27d one, therefore,
compares a number of fits. Slopes are smaller when the upper limit in $D$
is reduced, but do not reach the $\a$-model prediction (dashed line).

It can be verified that, at least for the higher orders, the
discrepancy with (\ref{3.28n}) is mainly due to the second equal sign,
derived from a log-normal approximation to the density
distribution. In a recent paper~\cite{AlbBi91}, this approximation has
been shown to be valid if the density fluctuations are weak or if the
density probability distribution is log-normal. The NA22 data
demonstrate that none of these conditions is fulfilled.

We conclude that the correlators $F_{pq}$ increase with decreasing
correlation length $D$, but do not really follow a power law for
$D\simkl 1$. For fixed $D$, the values of $F_{pq}$ do not depend on the
resolution $\d y$, a feature expected from the $\a$-model, but also
reproduced by FRITIOF and approximately true in any model with
short-range order.  When the increase of the correlators is roughly
approximated by a straight line in a restricted interval, the powers
$\f_{pq}$ increase linearly with the product $pq$ of the orders, but
are considerably larger than expected from FRITIOF and from the simple
$\a$-model.

The extension of single-variate factorial moments to the multivariate
case offers better insight into the complicated nature of the
correlations. However, the original expectation that correlators would
help in clarifying the issue of intermittency is not borne out by
present data.  Simple but reasonable models for higher-order
correlation functions which use the experimental 2-particle
correlations as input, have no difficulty in reproducing the behaviour
of factorial correlators measured e.g.~by NA22.

\section{Multifractal analysis}

Power-law dependence of normalized factorial moments on the
resolution $\delta$ (bin size) is a signature of self-similarity in the
fluctuation pattern of particle multiplicity. It suggests
that the probability distribution $P(\rho,\delta)$  of the
particle density $\rho$ has fractal properties.
For simple Widom-Wilson~\cite{Widom65} type scaling,
$P(\rho,\delta)$ is of the form
\begin{equation}
P(\rho,\delta)\sim\delta^{-\beta} P^{\star}(\rho/\delta^\nu)\ \ ,
\label{3:eq:1}
\end{equation}
where $\beta$ and $\nu$ are critical exponents.
All $q$-th order moments of $\rho$ ($q=1,2,\ldots$) obey
power laws in $\delta$ with inter-related exponents depending
on $q$ and on ($\beta$, $\nu$). This characterizes a simple or
mono-fractal. Another possibility is a multifractal behaviour, in which
$P(\rho,\delta)$ obeys a relation of the type~(cfr.~\cite{Kadanoff89})
\begin{equation}
\ln P(\rho,\delta)/\ln\delta =f(\alpha),
\quad \alpha=\ln\rho/\ln\delta.\label{3:eq:2}
\end{equation}
Multifractals, first introduced in~\cite{Mandelbrot74} represent
infinite sets of exponents---the multifractal spectrum---which
describe the power-law scaling of all moments of
$P(\rho,\delta)$. In principle, knowledge of the  multifractal spectrum
is completely equivalent to knowledge of the probability distribution.

Unlike geometrical or statistical systems, multiparticle production
processes pose special problems if a multifractal analysis is
to be considered. The most obvious one is the finiteness of particle
multiplicity in an event at finite energy.
Self-similarity, if existent, therefore cannot persist indefinitely to
finer and finer scales of resolution.

In multiparticle production $P(\rho,\delta)$ is not directly accessible.
At best one can construct, for a single event of multiplicity
$n$  and for given $\delta$, a frequency distribution
which approaches $P(\rho,\delta)$ only for $n\rightarrow\infty$.
For any finite (and usually small) $n$, the frequency distribution
and its moments will be subject to statistical fluctuations.

Since the data sample contains a large number of events, it is obviously
recommended to consider the event average. This averaging, however,
supposes ergodicity. The applicability of the multifractality concept
can, therefore, only be justified a posteriori.

\subsection{The method }

A multifractal analysis is based on the the properties
of $G$-moments whose definition is given in ~(\ref{gghwa}).
The moments $G_q$ (or more often $\ln G_q$) are obtained
for each individual event at a specified resolution $\delta y\sim1/M$ and
then averaged over the event sample\footnote{
Note that analyses based on $\lan G_q\ran$ or on   $\lan \ln G_q\ran$
in general differ and probe different aspects of the system
under study~\cite{Aharony89}.}.

In the theory of multifractals, the  $G$-moments share
with the scaled factorial moments the property that self-similar density
fluctuations lead, in principle,  to scaling behaviour
\beq
G_q \propto (\d y)^{\t_q} \ \ \ \ {\mbox{for}}\  \d y\to 0\ .
\eeq
In a fractal analysis (see also Sect.~2.4), one therefore determines
the slope
\beq
\t(q,M) = - \frac{\partial\lan \ln G_q(M)\ran}{\partial\ln M} \eeq
on a double-logarithmic plot, after averaging over all events in the
sample.

A multifractal spectral function is introduced via a Legendre
transform defined as
\beq
f(\a_q) = q\,\a_q - \t_q \ \ , \eeq
with
\beq
\a_q=\frac{\partial\t_q}{\partial q} \eeq
being the Lipschitz-H\"older exponents.
The spectral function $f(\a_q)$ is a smooth function, concave downward,
with its maximum at $q=0$. It gives a quantitative description of the
density fluctuations in the dense and in the sparse regions, corresponding to
its left and right wing, respectively. A wide spectrum reveals a non-smooth
density distribution.

The generalized (R\'enyi)-dimensions are given by
\beq
D_q = \frac{1}{q-1}\, \t_q=\frac{1}{q-1}\, \left[q \a_q-f(\a_q)\right]\ . \eeq

\subsection{Experimental results}

$G$-moments have been studied in a number of experiments~
\cite{Dera90,Ghosh92,Shiv93,Shiv94,Seng90,Ghosh91,Sark93,SugG,EMCG,IHCG}
\cite{NA22G,UA1G,CDFG,Wang95,Sark93-2,Shiv93-2,Jain90}.
As an illustrative example, we show  UA1 results~\cite{UA1G} in Fig.~4.29a,
where the event average $\lan \ln G_q\ran$ is plotted  as a function of the
resolution $(M=2^\m)$. Starting at a value of $0$ for $\m=0$
 according to definition (\ref{gghwa}), the moments grow
 for $q<1$ and fall for $q>1$ as $\m$  increases. The slopes decrease and
$\lan\ln G_q\ran$ tends to saturate for large $\mu$. The saturation is due
to
an increasing number of bins with content $n_m=0$ or 1, as $M$ becomes large,
\beq
G_q(M) \to n\left( \frac{1}{n}\right)^q = n^{1-q} {\rm \ \ \ \ for\ }
M\to\infty\ \ . \eeq

 Fig.~4.29b shows  $\lan\t_q\ran$ and $\lan\a_q\ran$ for $\m=1$ and 2
(small $M$). The corresponding spectral function $\lan f(\a_q)\ran$ is given
in Fig.~4.29c as a function of $\lan \a_q\ran$.
The fact that $\lan f(\a_q)\ran$ does not degenerate into a single point
implies multifractality in hadron production, at least for large bin
size $\d y=\frac{\D Y}{M}$ (small $\m$). However, for smaller bin sizes,
the function turns over (i.e., bends upward) and falls into the
non-physical region above the dashed line (not shown).

\subsection{Universality}

{}From Fig.~4.29c, it is clear that  $\lan f(\a)\ran$
depends on $\m$.
It also  depends on the cms energy $\sqrt s$ (or the multiplicity
$n=2^\n$).
In~\cite{Flor91} it is
conjectured  that $G$-moments (at fixed $q$) show universality in
$\xi=\m-\n$, however. The latter quantity is directly related
to the average particle multiplicity per bin,
$n/M=2^{\n-\m}=2^{-\xi}$.
Using a branching model, the authors of~\cite{Flor91} derive
the universality relation
\beq
\G_q(\xi) = \ln G_q(\m,\n)-\ln G_q(\n,\n).\label{gmunu}
\eeq
It expresses the scaling behaviour of a function of two variables
($\mu,\nu$) in terms of a function of one variable only.
The function  $\G_q(\xi)$ determines the $G$-moments as functions   of
$\mu$ for all values of $\nu$.
The validity of (\ref{gmunu}) is claimed to be a strong evidence for
self-similarity.

Fig.~4.30a shows $\G_q$ as a
function of $\xi$ for $q=\pm5$. All data points are close to universal
lines, thus indeed indicating universal behaviour.

A further prediction is that also  $\lan f(\a_q)\ran$ is universal
for fixed $\xi$. Fig.~4.30b demonstrates that this is not confirmed
by the UA1 data~\cite{UA1G}. In the EMC data~\cite{Dera90} the left
 branch shows universality, but not the right one.
Besides being more sensitive to universality breaking than $\G_q(\xi)$,
the function $\lan f(\a_q)\ran$  also reveals more clearly shortcomings in
the models. For PYTHIA and GENCL this is illustrated by Fig.~4.30b.

\subsection{Modified G-moments}

As stated earlier,  $G$-moments have the advantage  that not only spikes
are included in the analysis, but also non-empty valleys (for $q<0$).
 Disadvantages are that the moments saturate
at $\d y\to 0$ when the content of non-empty bins approaches unity and that
statistical fluctuations are not filtered out
(see also~\cite{ChiuFialHwa90,Seibert}).

In~\cite{HwaPan} a modified definition of the $G$-moments
is proposed in an attempt to circumvent the problem of statistical noise.
``Truncated'' $G$-moments are defined as
\beq
G_q = \sum^M_{m=1} p^q_m \O (n_m-q)\ ,
\label{ggmodif}
\eeq
where $\O$ is the usual step-function equal to 1 for $n_m\geq q$ and zero
otherwise. For very large multiplicity $n$ (as in a macroscopic statistical
system), $n/M\gg q$ and (\ref{ggmodif}) is in practice identical to
(\ref{gghwa}). In particle physics, $n$ is a relatively small number and the
$\O$ function exerts a crucial  influence on the $G$-moments. It imposes
non-analytical cut-offs at positive integer values of $q$. With the help of a
Monte Carlo (ECCO) based on the Geometrical Branching Model, the authors show
that $\ln\lan G_q\ran$ now exhibits a linear dependence on ln$M$ for $q>1$,
without saturation.

For $q>1$, the linearity of $\ln\lan G_q\ran$ with $\ln M$  has
been verified  on $\m$p, $\Pap\Pp$ and $\E$ data~\cite{EMCG,UA1G}.
The slopes $\t_q$ turn out to be very similar in all three reactions
and roughly equal to $\tau_q=-0.9(q-1)$.

\subsection{Bernoulli trials and $G$-moments}

Before we  conclude the  discussion of experimental characteristics
of $G$-moments, it is of interest to inquire in more detail about the
dynamical content revealed in multifractal analyses.
This is best done in a comparison to a model without dynamics.

Let $P(n)$ be the probability distribution for  observing $n$ particles
in an initial wide interval $\Delta Y$. Let this interval be
subdivided into $M$ smaller intervals of size $\delta y=\Delta Y/M$, each
of which contains $n_m$ particles ($n_m=0,1,\ldots,n$; $m=1,2,\ldots,M$) with
$\sum_m n_m=n$.
Assume that for every subdivision of $\Delta Y$ the $n$ particles
are independently distributed over the intervals with probability $1/M$.
For fixed $n$, the joint occupation probability in $M$ cells is
given by~(see also~\cite{ChiuFialHwa90})
\begin{equation}
P_n(n_1,\ldots,n_M)=P(n)\,\frac{n!}{n_1!\ldots n_M!}\,
\left(\frac{1}{M}\right)^n \,\delta(\sum_j n_j -n).\label{3:eq:3}
\end{equation}
With (\ref{3:eq:3}) the $G$-moments (\ref{gghwa}) at fixed $n$ are given by
\begin{equation}
G_q(n,M)= M\,n^{-q}\,\sum_{i=1}^n i^q \,B(n,1/M;i),\label{3:eq:4}
\end{equation}
where $B(n,1/M;n)$ is the binomial distribution.
For integer $q\geq1$   one obtains
\begin{equation}
G_q(n,M)= M\,n^{-q}\,\sum_{j=1}^q {\cal S}_q^{(j)}\,n^{[j]}. \label{3:eq:5}
\end{equation}
${\cal S}_q^{(j)}$ is a Stirling number of the second kind
(cfr.~(\ref{dr:43})) and $n^{[j]}=n(n-1)\ldots(n-j+1)$.
In the Poisson limit of the binomial ($n$ large and $1/M$ small with
$n/M$ fixed), (\ref{3:eq:5}) simplifies further to
\begin{equation}
G_q(n,M)= M\,n^{-q}\,\sum_{j=1}^q {\cal S}_q^{(j)}\,(n/M)^j. \label{3:eq:6}
\end{equation}
In inclusive analyses the average over $P(n)$ has to be taken in
(\ref{3:eq:5}-\ref{3:eq:6}). This introduces the (inverse) moments
$\lan n^{j-q}\ran$ and $\lan n^{[j]}/n^q\ran$ of the multiplicity
distribution in $\Delta Y$.

Numerical studies indicate that (\ref{3:eq:4}-\ref{3:eq:6}) reproduce and
explain many of the multifractal and universality properties
seen in the data. Here we can only give a few examples.

With respect to the structure of (\ref{3:eq:4}) it should be noted that
the binomial distribution is in fact a multifractal\footnote{This is
easily verified using Stirling's approximation to $n!$ which is notoriously
accurate even for quite small $n$.} in the sense of (\ref{3:eq:2})
\cite{Billingsley65,Kadanoff89}. Consequently, ``proper'', but quite trivial
binomial multifractal behaviour will be seen if the data are noise-dominated.
This is the case in practically all analyses referred to before. This point
was recognized in~\cite{ChiuFialHwa90}, but its full consequences were not
further studied.

{}From (\ref{3:eq:6}) follows immediately that the function $\Gamma_q(\mu,\nu)$
defined in (\ref{gmunu}) depends only on the ratio $n/M=2^{-\xi}$. Thus, the
parameter-free function $\Gamma_q(\mu,\nu)$ is indeed, but trivially,
universal in the Poisson limit. It describes accurately the data in
Fig~4.30a. Being a purely mathematical property of noise, it is not
surprising that the usual Monte-Carlo models also show this type of
universality. $\Gamma_q(\mu,\nu)$ ceases to be universal in the (more
general) binomial case, although the deviations remain small in cases of
practical interest. This probably explains the universality breaking observed
in $\re^+\re^-$ Monte-Carlo simulations at $1-10$ TeV in~\cite{Chiu91}.

The above considerations can be extended to the modified $G$-moments defined
in (\ref{ggmodif}). In particular $\Gamma_q(\mu,\nu)$ remains universal in
the Poisson limit. Further numerical properties of modified $G$-moments are
illustrated on Fig.~4.31(a-c). The results shown are based on (\ref{3:eq:6})
further averaged over $n$ with a negative binomial distribution truncated at
$0$. They depend only on $\lan n\ran$ in $\Delta Y$ and on the NBD parameter
$k$.

Fig.~4.31a demonstrates that the modified $G$-moments can be well
approximated
by power laws. The pseudo-linearity extends over a much larger interval
in $M$ than for usual $G$-moments obtained from (\ref{3:eq:6}). The improved
linearity is due to negative terms in the expressions of truncated moments
of the Poisson (or  binomial) distribution. The calculations displayed in
this figure coincide (up to an overall normalization factor) nearly exactly
with the EMC data for $M\geq8$ shown in~\cite{EMCG}. This proves that the
``clear asymptotic power-law behaviour of $\lan G_q\ran$ characteristic for
a self-similar system'' \cite{EMCG} is in fact due to Bernoulli noise.

Fig.~4.31b shows the ``Ochs-Wo\v siek'' plot for modified $G$-moments. Here
again, the quasi-perfect linear relation between $\ln G_q$ and $\ln G_{2}$
is seen to be a characteristic of Bernoulli trials. This linearity property
holds in fact for any combination $\ln G_q$ and $\ln G_{q'}$. The exponents
$\tau_q$ derived from power-law fits to the ``data points'' in Fig.4.31a are
shown in Fig.~4.31c. The line is a fit with the form $\tau_q=-C(q-1)$.
The slope $C$ is a slowly-changing function of the NBD-parameter $k$ with a
value around $0.9$, as experimentally observed in Subsect.~4.7.4 above!

\subsection{Evaluation of noise and connection between $F_q$ and $G_q$}

The self-similar property of multiparticle production at high energy can, in
principle, be investigated by $F$-moments and by $G$-moments. The power-law
behaviour of the scaled $F$-moments provides evidence for a self-similar
cascading process of dynamical origin. The $G$-moments, as an ingredient
of fractal theory, are designed to describe the multifractality aspect of
high multiplicities. In the real environment of high energy collisions,
however, the multiplicities are rather low and the $G$-moments
are dominated by statistical fluctuations.

The $F$-moments are defined for integer powers $q\geq 1$, the $G$-moments
for all real powers $q$. In order to establish a connection to
the $F$-moments, the powers $q$ are restricted to integer values of $q\geq 1$
here also for the $G$-moments.

The number $n_m$ of particles per subdivision $\d y=\D y/M$ has to be equal
to, or larger than $q\ (n_m=q+k,k=0,1,\dots)$ for the $F$- and $G$-moments.
Functions
\beq
B_{q,k}(M) = \lan\frac{Q_{q+k}(M,n)}{n^q}\ran
\eeq
are defined from the number of bins $Q_{n_m}(M,n)$ containing
$n_m=q+k$ particles in an event of multiplicity $n$ in the total phase-space
region $\D y$, normalized by $n^q$ and averaged over all events.
They express the basic fractal structure of the data, if they show
a power-law behaviour of the form
\beq
B_{q,k}(M) \propto M^{\la_{q,k}(M)}\ \ .
\eeq
In order to
suppress statistical fluctuations, the $G$-moments can be defined as the
event average over (4.33), or, equivalently, as
\beq
\lan G_q(M)\ran = \sum^\infty_{k=0} B_{q,k}(M)(q+k)^q\ \ .
\eeq
They are proportional to $M^{-\t_q}$ for large $M$.
The $F$-moments are defined as
\beq
\lan F_q(M)\ran = M^{-1} \sum^M_{m=1} \lan \frac{(n_m(n_m-1)\dots
(n_m-q+1)}{(n/M)^q}\ran
\eeq
or, equivalently, as
\beq
\lan F_q(M)\ran = M^{q-1} \sum^\infty_{k=0} B_{q,k}(M)\frac{(q+k)!}{k!}\ \ .
\eeq
They are proportional to $M^{\f_q}$ for large $M$
(note, however, that (4.41) is different from the form (2.68)
of the $F$-moments generally used).

When (4.33) and (4.40) to (4.42) are applied to the data they should show
a power-law behaviour for large $M$, if there are fractal structures present
in the data.

The dynamical contribution to the $G$-moments can be expressed by
\beq
\lan G_q\ran^{\dyn} = \frac{\lan G_q\ran}{\lan G_q\ran^{\st}} M^{(1-q)}\ \ ,
\eeq
where $\lan G_q\ran^{\st}$ can be determined by distributing the $n$
particles of an event randomly in $\D y$.
The randomization procedure destroys short-range particle correlations, but
does not alter the Bernoulli nature of the particle repartition in smaller
bins
discussed in Subsect.~4.7.5. As a result, this method does not eliminate the
binomial, noise-induced multifractal behaviour, but just gives its behaviour.

When $\lan G_q\ran^{\st}$ is equal to $\lan G_q\ran$, a trivial ``dynamical''
effect remains: a flat $dn/dy$ leads to a probability $1/M$ for a particle
to be in a given bin and $\lan G_q\ran^{\dyn}=M^{1-q}$.

The dynamical contributions to the slope $\t_q$ can be expressed by
\beq
\t^{\dyn}_q = \t_q-\t^{\st}_q + q - 1\ \ ,
\eeq
where $\t^{st}_q$ is the statistical part of the slope.
Subtracting the statistical contribution from $\t_q$ gives
\beq
\t_q-\t^{\st}_q = \t^{\dyn}_q - q + 1 \approx -\f_q
\eeq
which can be directly compared to the slopes $\f_q$ obtained from the
$F$-moments \cite{HwaPan}.

Fig.~4.32a gives a comparison \cite{HwaPan} of $\f_q$ (crosses) and
$(q-1-\t_q^{\dyn})$ (full circles) from ECCO simulation results and shows
that the deviation of $\t_q^{\dyn}$ from $q-1$ is indeed close to the
deviation
of $\f_q$ from zero. This observation gains support from the UA1 \cite{UA1G},
hA \cite{Ghosh92} (Fig.~4.32b) and AA \cite{Seng90} analysis.
The remaining difference can be attributed to the difference in the
definition of $\lan F_q\ran$ and $\lan G_q\ran$.

Fig.~4.32a, however, also shows that $\t_q^{\dyn}$ cannot be simply
replaced by $\t_q$ (open circles) in a quantitative analysis. This
is in agreement with the observation of Subsect.~4.7.5, but has not been
taken into proper consideration in
recent experimental application on lh \cite{EMCG},
hA \cite{Shiv93} and AA \cite{Sark93-2} collisions.

To summarize the present experimental findings, the data indicate that the
multifractal spectral function $f(\a)$ has, at least for large bin sizes,
the properties expected from the theory of multifractals. The function
$f(\a)$ is very sensitive to violations of universality and to details of
present Monte-Carlo models. However, with the methods used so far, the
multifractality analysis  breaks down at finer resolution. The finite
multiplicity effect--- statistical noise---overwhelms and it is difficult
to disentangle it from dynamical features. An advantage is that the
$\lan G_q\ran$ can probe holes in the distribution, not just spikes.
A recent extension of the definition of the $G$-moments filtering out
high multiplicities can claim some success in  extracting the dynamical
component. The advantage is that $\lan G_q\ran^{\dyn}$ lends itself more
readily to (multi)fractal interpretation and direct extraction of the
R\'enyi dimensions according to (2.114), while $\lan F_q\ran$ is more
closely related to the correlation function. Fig.~4.32 can serve as a rough
link between the two. Whereas a higher-dimensional analysis could be
useful also here, the low average multiplicity even in the highest
energy experiments presently precludes further progress in this direction.

\subsection{Universal multifractals}

At first sight interesting approaches, recently
applied~\cite{Ratti92} to obtain the degree of
multifractality (or L\'evy-index) $\m$ in multiparticle production, are
the methods of
Probability Distribution Multiple Scaling  (PDMS) and Double Trace Moments
(DTM)~\cite{Ratti91}.
In the first method,
the fundamental scaling law is written in
terms of the probability  for the number of particles $n_m$
in bin $m$ at resolution $M$ to be larger  than a certain threshold $
n_{\th}=M^\g$,
\beq
P(n_m(M)> M^\g)\propto M^{-c(\g)}\ .\label{large:dev}
\eeq
The statistical function $c(\g)$ is the codimension function describing the
sparseness of large intensities $n_m$. Like the factorial moments
(\ref{dr:44}-\ref{dr:46})
or the extended $G$-moments (\ref{ggmodif}), the DPMS method is
a straightforward filter for spikes of large $n_m$.

The PDMS method is closely related to Large Deviation Theory, a topic
in probability theory and of much theoretical interest in statistical
mechanics~\cite{Ellis:l}. Equation (\ref{large:dev}) expresses a
Level-2 Large Deviation property, describing deviations of the
``empirical measure'' $n_m$ from the infinite sample probability
density; $c(\gamma)$ is related to a generalized entropy.

Fig.~4.33a shows data~\cite{Ratti92} at different charge multiplicity
$(n=6$ and 14 are given as examples) presented in a double-logarithmic plot,
for various threshold values $n_{\th}=M^\g$. In spite of limitations on
statistics and on multiplicity $n_m$, a region of linearity
can be seen for all multiplicities and thresholds.
This is claimed to be evidence for PDMS.

When $c(\g)$ is smaller than the topological dimension $D$ of the embedding
space, it is possible to define a function $D(\g)=D-c(\g)$ corresponding to
the classical fractal dimension. If a sample of $N_\rs$ events is used in the
analysis (instead of one event), $c(\g)$ can become larger than the
topological dimension since different events can contribute to the same
bin $m$. In Fig.~4.33b the function $c(\g)$ is shown for the same multiplies.
The dotted line corresponds to $c(\g_\rs)=D+D_\rs$, where $D_\rs$ is the
sample dimension defined as $N_\rs=M^{D_\rs}$. For $n=20$, e.g., this limit
is crossed for a threshold of $n_{\th}$=3 with $N_\rs=15$. Singularities
of that type are called ``wild" singularities (not arising from
Poisson-like fluctuations).

A parametrization of $c(\g)$ in terms of two parameters is provided
by the theory of universal multifractals~\cite{Ratti91},
\beqa
c(\g) & = & C_1\left({\g\over \m'C_1}+{1\over\m}\right)^{\m'}
{\rm \ for\ } \m\not=1\nonumber \\
& = & C_1 \exp \left({\g\over C_1}-1\right)
{\rm \ for\ } \m=1\ ,
\eeqa
with ${1\over\m}+{1\over \m'}=1$ and $0\leq\m\leq2$;  $\m$ is the
L\'evy-index giving the degree of multifractality and $C_1$ is
the codimension of the average field. The two parameters
can be obtained from fits to the results given in Fig.~4.33a,
but turn out to be highly correlated.

The L\'evy-index $\m$ can, however, be determined independently of $C_1$
with the help of the Double Trace Moments~\cite{Ratti91}, a generalization
of the $G$-moments. They are defined in~(\ref{tr}). A DTM analysis of
$\sqrt{s}=16.7$ GeV data~\cite{Ratti92} yields $\m$-values ranging from
0.4 to about 0.9, increasing with multiplicity $n$. Such
values are far from monofractality $(\mu=0)$, but considerably below
$\m\sim1.6$ obtained in Subsect.~4.3.3 from factorial moments.

For multifractal theory, it is important to know whether the limit
$\m=1$ is crossed (signalling ``hard unbounded'' singularities) or
asymptotically approached from below (indicating
``soft bounded'' singularities).
This question cannot be answered at low energies and
needs high-energy, high-multiplicity  data.

Extension to  higher-dimensional space, though difficult in practice,
is necessary since singularities can easily be washed out if a particular
variable is not sensitive to them.

In spite of the interesting potential of the ``Universal Multifractal''
idea, one should keep in mind that the method suffers from
the same limitations as the multifractal method based on $G$-moments.
This is easily illustrated by considering again the Bernoulli-trials
model discussed in Subsect.~4.7.5, above. For the pure binomial noise
(4.34), one has
\begin{equation}
P_B(n_m(M)\geq n_{\th})=I_{1/M}(n_{\th},n-n_{\th}+1),\label{mf:1}
\end{equation}
where $I_x(a,b)$ is the incomplete beta function~\cite{abramowitz}.

\noindent The logarithm of (\ref{mf:1}) is approximately linear
in $\ln M$ for reasonably large $M$. Equation (\ref{mf:1}) not only has
all the features of the data plotted in Fig.~4.33a, but even agrees
numerically quite well. The co-dimension function  $c(\gamma)$ is, for
this simple model, approximately equal to $n_{\th}$, implying constant
differences between the slopes for successive values of $n_{\th}$. The
data in Fig.~4.33b show exactly this property.

Double Trace Moments are easily calculated in the Bernoulli model. We find
that the ``L\'evy-index'' $\mu$ is a smoothly increasing function of the
event multiplicity $n$ crossing the ``hard unbounded'' value $\mu=1$ near
$n=30$. We conclude that the data in Fig.~4.33 are merely reflecting
statistical noise.

Dynamically useful information could possibly be extracted if
Double Trace {\it factorial moments} were used instead of
the usual moments. This is easily verified on the simple
Bernoulli model and, of course, applies to  $G$-moments as well.

\section{Density and correlation strip-integrals}
\subsection{The method}

A fruitful  recent development in the study of density fluctuations is the
density and correlation strip-integral method~\cite{Hen83,Drem88,Lipa91-53}.
By means of integrals of the inclusive density over a strip domain, rather
than a sum of box domains, one not only avoids unwanted side-effects, such
as splitting up of density spikes, but also drastically increases the
integration volume (and therefore the accuracy) at a given resolution.

Consider first the  (vertical) factorial moments $F_q$ defined,
for an analysis in one dimension, as
\beq
F_q (\d y)\equiv\frac{1}{M} \sum^M_{m=1} \frac{\lan n^{[q]}_m\ran}
{\lan n_m\ran^q}
= \frac{1}{M} \sum^M_{m=1} \frac{\ds \int_{\W_m}\P_i dy_i\r_q(y_1\dots y_q)}
{\ds \int_{\W_m}\P_i dy_i \r_1(y_1)\dots \r_1(y_q)}\ \ . \eeq
\ni
The integration domain $\W_{\PB}=\sum^M_{m=1}\W_m$ thus consists of $M$
$q$-dimensional boxes $\W_m$ of edge length $\d y$. For the case
$q=2$, $\W_{\PB}$ is the domain in Fig.~4.34a.
A point in the $m$-th box corresponds to a pair $(y_1,y_2)$ of
distance $|y_1-y_2|<\d y$ and both particles in the same bin $m$.
Points with $|y_1-y_2|<\d y$ which happen {\it not} to lie in the same
but in adjacent bins (e.g. the asterix in Fig.~4.34a) are left out.
The statistics can be approximately doubled by a change
of the integration volume $\W_{\PB}$ to  the strip-domain of Fig.~4.34b.
For $q>2$, the increase of
integration volume (and reduction of squared statistical error)
is in fact roughly proportional to the order of the correlation.
The gain is even larger when working in two or three phase-space
variables.

In terms of the strips (or hyper-tubes for $q>2$), we define as (vertical)
{\it density} integrals
\beq
F^\rS_q (\d y) \equiv \frac{\ds \int_{\W_{\rs}}\P_i \rd y_i
\r_q(y_1,\dots,y_q)}
{\ds \int_{\W_{\rs}}\P_i \rd y_i \r_1(y_1)\dots \r_1(y_q)}\ \
\label{3.45}
\eeq
and, similarly, the {\it correlation} integrals $K^\rS_q(\d y)$ by replacing
the density $\r_q(y_1,\dots,y_q)$ by the correlation function
$C_q(y_1,\dots,y_q)$. (Note that in the literature the term ``correlation
integral'' is often also used for the $F^\rS_q(\d y)$.)

These integrals can be evaluated directly from the data, after selection
of a proper distance measure $(|y_i-y_j|,[(y_i-y_j)^2+(\f_i-\f_j)^2]^{1/2}$,
or better the four-momentum difference $Q^2_{ij} = -(p_i-p_j)^2$) and after
definition of a proper multiparticle topology, the snake integral
{}~\cite{CaSa89}, the GHP integral~\cite{Hen83}, or the star integral
\cite{Egger93} as shown in Figs.~4.34c-e, respectively.

\subsection{Results}

As an example, $F_4(\d y)$ is compared to $F^\rS_4(\d y)$ (and
$F^\rS_2,F^\rS_3)$)
for  the NA22 spike event~\cite{AdamPL185-87} in Fig.~4.35a (no error-bars are
shown, because it is one event). Depending on whether the prominent spike lies
entirely in one bin or is split across two, $F_4$ shows large fluctuations.
These are practically absent in $F^\rS_4$ (Fig.~4.35b). Large improvement in
one-dimensional $(\h)$ and two-dimensional $(\h-\vf)$ analysis is also
observed in \cite{Jie95}.

How much the statistical errors are reduced can be seen on Fig.~4.36a
where the NA22 data~\cite{Ajin89-90} are plotted as a function of
$-\ln Q^2$, with all two-particle combinations in an $n$-tuple
having $Q^2_{ij}<Q^2$ ~\cite{Hen83}. The following
observations can be made:

i) the errors and fluctuations are indeed largely reduced, as compared e.g. to
 Fig.~4.20a.

ii) with the (one-dimensional) distance measure $Q^2$, the moments show a
similarly steep rise as in the three-dimensional analysis (e.g. Fig.~4.11c).

iii) Contrary to the results in rapidity, positives and negatives behave very
similarly here (only negatives are shown in Fig.~4.36a), but are now much
steeper than all-charged.

iv) $F^\rS_2$ is flatter for $(+-)$ than for all-charged or like-charged
combinations.

The first two observations demonstrate the strength of the new method
and the advantage of using the proper variable. The second two observations
directly demonstrate the large influence of identical particle correlations
on the factorial moments. These results agree very well with results from
the UA1 collaboration ~\cite{Alba90} shown in Fig.~4.36b and with lh results
\cite{Arne86,Adams94}.

Monte-Carlo simulations with FRITIOF~2 show the following (see Fig.~4.37
for the case of $F^\rS_2$). The default ``plain'' version is unable to
describe the all-charged NA22 data, but a ``biased'' version (including
misidentified Dalitz decay + 0.25\% undetected $\g$-conversions) comes
closer to the data. However, not unexpectedly, both versions fail completely
in describing the like-sign data, where the model stays way too low. On the
other hand, $F^\rS_2$ for the $(+ -)$ combination is largely overestimated
when $\gamma$-conversions are included, but saturates without.

\subsection{Transverse-momentum and multiplicity dependence}

As in Subsect.~4.4.2 (Fig.~4.20c), UA1 \cite{Wu94} has studied the $\f_2$
dependence on the average transverse momentum $\bar p_\rT$ of the event,
but now in terms of density integrals in $Q^2$. In contrast to the strong
decrease (and subsequent slight increase of $\f_2$) with increasing
$\bar p_\rT$ observed for the one-dimensional analysis in Fig.~4.20c,
a strikingly flat behaviour (and slight increase above 0.6 GeV/$c$) is
observed for the data (full circles) in Fig.~4.38a.
The discrepancy of PYTHIA (open circles) is even stronger here than in
Fig.~4.20c. The slope $\f_2$ starts at even negative values for small
$\bar p_\rT$, but increases fast with increasing $\bar p_\rT$ to reach values
overestimating $\f_2$ at $\bar p_\rT\simgr 0.5$ GeV/$c$.

A similar disagreement is observed for the multiplicity dependence in
Fig.~4.38b. While the UA1 data (full circles) decrease with increasing $n$,
PYTHIA predicts a strong increase. In Figs.~4.38~c) and d), it is shown that
this violent discrepancy between PYTHIA and data is mainly due to like-sign
pairs, so to the way Bose-Einstein correlations are incorporated
into the model.

\subsection{Bose-Einstein correlations versus QCD effects}

Of particular interest is a comparison of hadron-hadron to $\E$ results in
terms of same and opposite charges. This is shown in Fig.~4.40 for
$q=2$ UA1 and DELPHI data in~\cite{Mandl92} (note that in this figure the
derivative of (\ref{3.45}) is presented in small $Q^2$ bins). An important
difference between
UA1 and DELPHI can be observed on both sub-figures: For ``large''
$Q^2(>0.03$ GeV$^2$), where Bose-Einstein effects do not play a role,
the $\E$ data increase much faster with increasing $_2\log(1/Q^2)$ than
the hadron-hadron results. For $\E$, the increase in this $Q^2$ region is
very similar for same and for opposite sign charges. At small $Q^2$,
however, the $\E$ results approach the hadron-hadron results. The authors
conclude that for $\E$ at least two processes are responsible for the
power-law behaviour: Bose-Einstein correlations at small $Q^2$ following
the evolution of jets at large $Q^2$. In  hadron-hadron collisions at
present collider energies only Bose-Einstein effects seem relevant.

Since string fragmentation causes an anti-correlation between same-charged
particles, it is of interest to compare $\E$ results to JETSET in terms of
strip integrals for the different charge combinations, separately.
This has been done in~\cite{Mandl92} and, indeed, the Monte-Carlo
results level off at small $Q^2$ and fall below the data for the same-charge
results, while they describe the opposite-charge data perfectly well
(not shown here).

The exact functional form of $F^\rS_2$ is derived from the
data of UA1~\cite{Alba90} and NA22~\cite{Ajin89-90},
again in its differential form\footnote{In fact in this differential
form $F^\rS_2(Q^2)$ is identical to $R(Q^2)$ usually used in Bose-Einstein
analysis. The only difference is that it is plotted on a double-logarithmic
plot, here.}, in Fig.4.40. Clearly, the data favour a power law in $Q$
over an exponential, double-exponential or Gaussian law.

If the observed effect is real, it supports a view recently developed
in~\cite{Bial92}. There, intermittency is explained from Bose-Einstein
correlations between (like-sign) pions. As such,  Bose-Einstein correlations
from a static source are not power behaved. A power law is obtained i) if
the size of the interaction region is allowed to fluctuate, and/or ii) if
the interaction region itself is assumed to be a self-similar object
extending over a large volume. Condition ii) would be realized if parton
avalanches were to arrange themselves into self-organized critical
states~\cite{Bak87}. Though quite speculative at this moment, it is an
interesting new idea with possibly far-reaching implications.
We should mention also that in such a scheme intermittency is viewed as
a final-state interaction effect and is, therefore, not troubled by
hadronization effects.

The effect on the factorial moments of adding  Bose-Einstein correlations in
FRITIOF  is convincingly demonstrated for heavy-ion collisions
in~\cite{Kadi92}. Because of the large number of collision processes, other
correlation effects are expected to play a much reduced role for this type
of interaction and Bose-Einstein correlations, as a collective effect, can
become the dominant source of non-statistical fluctuations. Also from these
results it is clear that more than one fixed interaction-volume
radius is needed to reproduce the experimental results.

In perturbative QCD, on the other hand, the intermittency indices $\f_q$,
are directly related to the anomalous multiplicity dimension
$\g_0=(6\a_\rs/\p)^{1/2}$ \cite{Gust91,Oc92,OW92a,dd92,BrMeuPe93} and,
therefore, to the running coupling constant $\a_\rs$. In the same theoretical
context, it has been argued \cite{Oc92,OW92a,dd92,BrMeuPe93} that the
opening angle $\x$ between particles is a suitable and sensitive variable
to analyse and well suited for these first analytical QCD calculations of
higher-order correlations. It is, of course, closely related to $Q^2$.

A first analytical QCD calculation \cite{Oc92,OW92a} is based on the
so-called double-log-approximation with angular ordening \cite{ao} and
on local parton-hadron-duality \cite{lphd}. A preliminary comparison with
DELPHI data \cite{ManBu94} gives encouraging results, even including an
estimate for the running of the strong coupling constant $\a_\rs$.
\section{Correlations in invariant mass}

\def\MINV{M_{\mbox{\small inv}}}
\def\KTPM{K_2^{+-}(\MINV)}
\def\KTMM{K_2^{--}(\MINV)}
The previous section has illustrated the advantages of the correlation
integral method with a ``distance'' measure directly related to
the invariant mass of the particle system. The results give additional
support to the {Fia\l kowski} conjecture, mentioned  in Subsect.~4.3.2,
from which could be anticipated that dynamical effects are most clearly
 revealed if the correlation functions and factorial moments are
directly analysed in terms of Lorentz-invariant variables.

Evidently, there are many arguments in favour of invariant mass as a
dynamical variable rather than the single-particle variables
often used in early intermittency studies. Resonances, the cause of
most of the correlations among hadrons, and threshold effects
appear at fixed values of mass; Bose-Einstein interference correlations
depend on four-momentum differences; multiperipheral-type ladder
diagrams are functions of two-particle invariant masses, and so on.

The idea  to study correlations as a function of invariant mass was,
to our knowledge, first proposed in~\cite{berger,thomas}.
The authors introduce a method which is technically a differential
version of the correlation integral method. It focusses directly on
the correlation functions (cumulants) rather than on the inclusive
density as in (\ref{3.45}). Starting from the definition (\ref{dr:13}),
one defines the correlation function
\begin{equation}
C_2(\MINV)=\rho_2(\MINV)- \rho_1\otimes\rho_1(\MINV),\label{CM}
\end{equation}
obtained after integration (in a suitable region of phase space)
of $C_2(p_1,p_2)$ over all variables except $\MINV$. Here,
$\rho_2(\MINV)$ is the  familiar normalized
2-particle invariant-mass spectrum. The ``background term''
$\rho_1\otimes\rho_1(\MINV)$ is the  integral
of $\rho_1(p_1)\rho_1(p_2)$ with  $\MINV$ fixed.
For the data  shown below, it is obtained from ``uncorrelated'' (``mixed'')
events, built by  random selection from a track pool. The same method is
used in evaluating the denominator in (\ref{3.45}).
Higher-order correlations are obtained in a completely analogous manner.
We further utilize the  function
$K_2(\MINV)=C_2(\MINV)/\rho_1\otimes\rho_1(\MINV)$, the normalized
factorial cumulant of order two.

The analysis in \cite{berger}, based on low statistics pp data at 205
GeV/$c$, demonstrates that $K_2^{+-}(\MINV)$ and  $K_2^{\pm\pm}(\MINV)$
follow an approximate power law, written by the authors as
\begin{equation}
K_2(\MINV)=(\MINV^2)^{\alpha_\rX(0)-1}\label{power-law}.
\end{equation}
The notation reminds of the interpretation of (\ref{power-law})
in terms of the  Mueller-Regge formalism (for details see~\cite{berger}).
The power $\alpha_\rX(0)$ is the appropriate Regge-intercept, $\rX=\rR$
for non-exotic pairs and $\rX=\rE$ for exotic ones.
The ratio $K_2^{--}/K_2^{+-}$ was further seen   to fall  as
$\MINV^{-2}$, consistent with $\alpha_\rR(0)-\alpha_\rE(0)=1$.
Not relying on  Mueller-Regge theory, the authors
argued that most of the correlations at small $\MINV$
are due to resonance decays into three or more pions and
to interference of amplitudes~\cite{thomas}.

The results already obtained in~\cite{berger} clarify several issues which
have troubled the interpretation of intermittency data. Among others, they
demonstrate that different charge-states should be treated separately since
the $\MINV$ dependence is very  different. This fact, obvious in $\MINV$
but much less so in rapidity, was not fully appreciated in early
intermittency analysis and  the crucial importance of like-sign particle
correlations remained hidden in ``all-charged'' analyses.

The method of~\cite{berger} has now been applied by NA22 \cite{XX1}
and DELPHI \cite{XX2}. Figure~4.41 shows data
on $K_2(\MINV)$ for a combined sample of non-diffractive $\pi^+/\PK^+p$
collisions at 250 GeV/$c$  in the central c.m.~rapidity region $-2<y<2$.
$\KTPM$ has a prominent $\rho^0$ peak, but is quite flat near threshold.
The peak in the first bin of sub-figure~(a) is attributed to
contamination from Dalitz-decays  and $\gamma$-conversions.
$\KTMM$ falls much faster. A fit of $K_2\sim (\MINV^2)^{-\beta}$ yields
$\beta^{--}=1.29\pm0.04$, $\beta^{++}=1.46\pm0.03$, $\beta^{+-}=0.17\pm0.02$,
in agreement with~\cite{berger} and consistent with the relation
$\alpha_R(0)-\alpha_E(0)=1$.

NA22 also finds that cuts on transverse momentum or relative azimuthal angle
$\delta\vf$ strongly affect the shape of $\KTPM$, but have little effect
on $\KTMM$ for $\MINV<0.5$ GeV/$c$${}^2$. This means that $\KTMM$ at small
$\MINV$ is essentially a function of $\MINV$ (or $Q^2$) only, illustrating
once more   the advantage of $\MINV$  compared to other variables.

The data in Fig.~4.41 confirm the conclusion of Sect.~4.8 that
the correlations in like-charge systems are at the origin of
the strong increase of factorial moments for small invariant masses.
Whether Bose-Einstein effects are solely responsible for the
differences between ($\pi^{\pm}\pi^{\pm}$) and $(\pi^+\pi^-)$ pairs
is not so evident. It suffices to consider~\cite{thomas}
the contributions from decays of various resonances to realize that
 the $\MINV$-dependence near threshold for ``exotic'' particle systems
must be stronger than for ``non-exotic'' ones. In a dual Regge picture,
such differences translate into very different values of the
respective Regge intercepts as in (\ref{power-law}).
It remains, therefore, to be verified if the $\MINV$-dependence of
the data can be explained as a superposition of a ``standard'' Regge-type
power law and a  conventional  Bose-Einstein enhancement.

As pointed out in \cite{XX3}, there is a feasible way to test this and
even to give access to the relative strength of BE interference and exotic
like-charge $\p\p$ interaction. The idea is that particle combinations exist
which are either a) exotic, but not identical (e.g., $\PK^+\p^+$ or
$\PK^-\p^-$ pairs) or b) identical, but not exotic ($I=0$ $\p^0\p^0$ pairs).
NA22 \cite{XX4} and ALEPH \cite{XX5} data indicate that very-short-range
correlations are indeed absent in the exotic $\PK\p$ channel. This supports
Bose-Einstein correlations rather than exotic Regge behaviour, but the point
deserves further investigation.

The possibility that the correlation functions depend mainly on
invariant mass has interesting further consequences. These were
analysed in~\cite{edw:sant}. Taking in (\ref{CM})
\begin{equation}
C_2(\MINV)\propto (\MINV^2)^{\alpha-1}\,\mbox{BE}(\MINV),\label{edw:c2}
\end{equation}
with $\mbox{BE}$ a conventional Bose-Einstein factor, exponential in $Q$,
good agreement is obtained with the NA22 second-order correlation integral
data of ($--$)-pairs. Integrating the correlation function over all
variables except $\delta y$ gives $F^{--}_2(\delta y)$ which also fits the
data. Although $C_2$ does not explicitly depend on the transverse momentum
of the particles, it turns out that $F_2(\delta y,p_{\rT 1},p_{\rT 2})$ is
larger and more steeply increasing than $F_2(\delta y)$ for small
$\delta y$ and small $p_\rT$'s. The opposite happens for large $p_\rT$'s.
This is the ``low-$p_\rT$ intermittency effect'' seen in the NA22 and UA1
data (cfr.~Fig~4.20a and Subsect.~4.4.2). The explanation is simple: under
the stated hypothesis, small $p_\rT$ for the two particles in a pair
means, on the average, smaller invariant mass than for unrestricted
transverse momentum and, therefore, larger and shorter-ranged correlations
in rapidity. Enhanced intermittency follows as a consequence of kinematical
cuts! The influence of the Bose-Einstein factor is easily checked in this
simple model. It is found to be necessary in order to reproduce the
correlation integral data  and $F_2$ for restricted $p_\rT$ but has, as
expected a priori,  very little influence on the $p_\rT$-integrated
$F_2(\delta y)$. This explains early controversy over the role of
Bose-Einstein effects in one-dimensional factorial moment analyses
(Subsect.~4.4.1).

A study of the  invariant-mass dependence of the two-particle correlation
function has for the first time given clear indications as to why the
hadron-hadron Monte-Carlo models fare so badly when confronted with
factorial moment data. For example, Fig~4.42 shows NA22 data for
$K_2^{--}(\MINV)$ and  $K_2^{+-}(\MINV)$ compared to FRITIOF. The
predicted shape of $K_2^{+-}(\MINV)$ is very different from the data,
especially in the $\rho^0$ region. It shows an enhancement at low mass
which causes the correlation function to drop much faster than seen in
the experiment. In the model, this structure is traced back to reflections
from $\eta$, $\eta'$ and $\omega$ resonances. The model also fails to
describe  $K_2^{--}(\MINV)$ since correlations are very weak or even
negative, except for  a threshold enhancement due to $\eta'$-decays.

These examples suffice to demonstrate that FRITIOF (or rather JETSET)
has serious shortcomings and is unable to reproduce two-particle correlations
in invariant mass. For correlations in rapidity and azimuthal angle
this was seen earlier (Sect.~3.1), but the  reasons remained obscure,
mainly because of the  insensitivity of these variables to dynamical
correlations at small mass.

A study of the correlation function in terms of invariant mass clarifies
the situation considerably. For NA22, the model was known to
overestimate significantly the production rates of
$\rho^0$ and $\eta$ mesons~\cite{na22:atayan} and presumably also
those of $\eta'$ and $\omega$ for which no direct measurements
exist (see also~\cite{Walker91} for h$A$ collisions).
This is now seen to distort heavily the $\MINV$-dependence of $K_2$.
Also Bose-Einstein low-mass enhancements, most likely responsible
for the fast drop of $\KTMM$ in the threshold region, are not
included in the FRITIOF model commonly used.  Finally, we note
that the values of $K_2$ in the considered mass interval  are much smaller
than  the data. This is related to the width of the charged particle
multiplicity distribution which is known to be too small in FRITIOF.
It affects the global magnitude of factorial moments and cumulants.

In Fig.~4.43a, the discrepancy is shown to be quite similar for $(+-)$
correlations in $\E$ collision \cite{XX2} and JETSET. As in hh collisions,
the correlation is underestimated in the mass region below the $\r^0$. This
discrepancy can be cured by decreasing the $\h'$ and $\r^0$ production and
increasing $\w$ production in JETSET.

Fig.~4.43b gives the like-sign correlation for the data and JETSET without
BE correlation. For $M_{\inv}<0.6$ GeV/$c^2$. The experimental data are
considerably higher than JETSET. This can be attributed to Bose-Einstein
interference. However, it is striking that JETSET also predicts a strong
rise towards threshold even without Bose-Einstein correlations. This is
the tail of the QCD effect also seen in Fig.~4.39 and mainly due to
multijet events. The difference between JETSET and data can indeed be
removed by including BE correlations in the model, but the $Q^2$ cut used
$(Q^2>0.04^2$) is too high to be able to distinguish a power law from
an exponential or Gaussian, as is done in Fig.~4.40.

To summarize,  the above proves  that the failures of models such as
FRITIOF and JETSET with respect to factorial moment and correlator data
(Subsects.~4.2.3 and 4.6.3), are not necessarily due to ``novel'' dynamics.
They are in first instance a consequence of a variety of
defects---such as incorrect resonance production rates and
absence of identical particle symmetrization---which belong
to  ``standard'' hadronization phenomenology.
These defects should  be eliminated before ``new physics'' can be claimed.

For $\re^+\re^-$ annihilation at LEP energies, we have found that
models such as JETSET-PS are much more successful than for all other
processes.  Besides the evident fact that this process is much better
understood theoretically, QCD effects dominate and the model
parameters are much better tuned to the data. Still, serious, recently
observed discrepancies of e.g.~JETSET-PS with LEP measurements on
particle and resonance production rates are a clear sign that
hadronization in $\re^+\re^-$ is in fact less well understood than
commonly stated and needs improvement.  It could be rewarding to
investigate carefully and differentially the invariant mass dependence
of the correlations using the sensitive methods now available.

Originating  mainly from the low invariant mass region (typically
$<1.5$ GeV/$c$${}^2$), it is not impossible that the observed correlations
are quite independent of the process initiating the primary colour separation
in the collision, being dominated by strong final-state interactions.
This would explain ``universality'' in the sense discussed earlier.

Many authors argue that  ``intermittency'' is somehow connected to (nearly
scale-invariant) perturbative QCD-cascading. Others strongly contest this
view on the argument that QCD cascades have a limited extent even at LEP
energies and are dominated by a very small number of ``hard'' emissions.
In the former case, one may expect significant differences in the
correlation functions at low mass for $\re^+\re^-$, on the one hand, and
for hh, h$A$ and $AA$ collisions, on the other. Preliminary
$\re^+\re^-$ data, mentioned in Sect.~4.8, seem to support the last opinion.
Whatever the final outcome, if differences are found, they should be used to
clarify the respective roles of perturbative and hadronization phases
in the different types of collision processes.

\section{Genuine higher-order correlations}

Multiparticle production in high-energy collisions is one of the rare
fields of physics where higher-order correlations are directly accessible
in their full multi-dimensional characteristics, under well controlled
experimental conditions.

Three-particle correlations have been observed in the form of short-range
rapidity correlations and higher-order Bose-Einstein correlations, but
evidence for {\it genuine} higher-order correlations (i.e., after
subtraction of all lower-order contributions) is very limited.
While it was found to be completely absent in heavy-ion collisions, first
evidence was given in Sect.~4.5 for their existence in hh collisions.

The correlation integral method turns out particularly useful for the
unambiguous establishment of genuine higher order correlations in terms
of the normalized cumulants $K_q(Q^2)$, when using the star integration
\cite{Egger93}
\beq
K^*_q(Q^2) = \frac{\ds \int \P_i dy_i \O_{12}\dots\O_{1q}C_q(y_1,\dots,y_2)}
{\ds \int \P_i dy_i \O_{12}\dots\O_{1q} \r_1 (y_1) \dots \r_1(y_q)}\ \ ,
\eeq
with $\O_{1j} = \O(Q^2-Q^2_{ij})$ restricting all $q-1$ distances
$Q^2_{1j}$ to lie within a distance $Q^2$ of the position of particle 1.
The star-integral method combines the advantage of optimal use of available
statistics and minimal use of computer time. Since higher acccuracy is
abtainable, dynamical structures in the correlations can be studied in
greater detail than with conventional methods.

Non-zero values of $K^*_q(Q^2)$ increasing according to a power law with
decreasing $Q^2$ are indeed observed for E665 for third order \cite{Adams94}
and for NA22 up fifth order \cite{genuine} (see Fig.~4.44 for the latter).

\section{Summary and conclusions}

\begin{enumerate}
\i Intermittency, defined as an increase of normalized factorial moments
with increasing resolution in phase space, is seen in all types of
collision.  Intermittency is a 3D phenomenon. The anomalous dimensions
are small ($d_q$=0.01-0.1) in a one-dimensional analysis, but the
factorial moments are considerably larger, and their
resolution-dependence more power-like, in two- or three-dimensional
phase space.  Self-similarity in the dynamics of multiparticle
production is an attractive but not fully proven explanation.
\i The factorial-moment method is very sensitive
to biases in the data. These  have to be
studied in detail before final conclusions can be drawn.
Because of its  sensitivity, the method has in fact  proven to be
very helpful in detecting and tracing such  biases.
\i The logarithms of factorial moments satisfy a possibly
dimension-independent linear relation which allows to determine
directly ratios of anomalous dimensions. The observed order-dependence
of anomalous dimensions excludes a second-order phase transition
(as treated in~\cite{BialHwa91}) as the origin of intermittency. Valid
in random cascade models, the Ochs-Wo\v siek relation shows that
correlation functions of different order are inter-related in a specific
hierarchical structure. An explanation in terms of dynamics
has not yet been found.
\i In hadron-hadron collisions, factorial moments and intermittency
indices depend strongly on transverse momentum and are largest for
low transverse momentum hadrons. This effect, also seen in rapidity-
and azimuthal correlations, needs further study in $\E$ collisions
and in  parton shower Monte Carlo's. There are serious indications that
``low-$p_\rT$ intermittency'' is a reflection of the very strong
dependence of correlation functions for identical particles on
invariant mass. This has to be examined in more detail.
\i The multiplicity dependence in hh collisions agrees with what is
expected from mixing of independent sources. However, the {Fia\l kowski}
observation on possible universality casts some doubt on this type of
interpretation. For a given density, heavy-ion collisions show more
intermittency than hadron-hadron collisions, possibly as a result of
Bose-Einstein interference  or other collective effects.
\i Factorial cumulants are direct measures of genuine higher-order
correlations. These are present in hadron-hadron collisions, in particular
for small phase-space domains, but seem to be absent in heavy-ion collisions.
\i Factorial correlators reveal  bin-bin correlations.
The correlators $F_{pq}$ increase with
decreasing correlation length $D$, but only approximately follow
a power law for $D\simkl 1$. For fixed $D$, the values of
$F_{pq}$ are independent of  resolution $\d y$, a property
predicted in  the $\a$-model, but also shared by
models with short-range order such as FRITIOF. The powers
$\f_{pq}$ increase linearly with the product $pq$ of the orders,
but are considerably larger than expected from FRITIOF and from
the simple $\a$-model.
\i A recent extension of the definition of $G$-moments filtering out
high multiplicities, claims success in extracting  the dynamical component.
However, under the conditions prevailing in present hadroproduction
experiments, multifractal and generalized multifractal methods seem unable
to overcome the overwhelming dominance of statistical fluctuations.
\i The correlation (or density) strip integral strongly reduces
statistical errors, as well as fluctuations due to splitting of spikes.
Using the squared four-momentum $Q^2_{ij}$ as a distance measure, an
increase similar to that found in three-dimensional analyses
is observed. This increase is caused by correlations among like-charged
particles. Bose-Einstein interference  must contribute significantly
to the intermittency effect but is not power-behaved in the
conventional approach. Power-law behaviour in Bose-Einstein interferometry
would imply a random superposition of ``emission centres'' with possibly
fractal properties. Parton avalanches in a self-organized critical
state are an intriguing possibility.

\i The analysis of cumulants in terms of invariant mass, or  related
variables, reintroduced recently after early work, has helped in
 clarifying several issues in intermittency. The reasons behind the
 dramatic failures of models for hadron-hadron collisions are clearly
 revealed. They are in first instance incorrectly predicted particle
 and resonance production rates and the near absence of correlations
 (or even presence of anti-correlations) in identical particle systems
 with small invariant masses.  These defects are not easy to cure in a
 consistent manner by simple parameter tuning and ``new'' physics may
 be needed to restore internal consistency in
 e.g.~string-fragmentation models.

\i The hadronization mechanism  in  hadron-hadron collisions is based
on identical physical principles and Monte-Carlo algorithms as those
applied with apparently great success to ``hard'' processes, in particular
to $\re^+\re^-$ annihilation.
There is now growing experimental evidence that the mentioned discrepancies,
first seen in hadron collisions, are also present in such processes. These
should first be removed before commonly expressed claims that ``No new
physics is involved in intermittency'' can be accepted.

\i The relative importance of perturbative and non-perturbative QCD
contributions to hadron correlations in $\re^+\re^-$ remains controversial,
also theoretically. This can be studied further with the new techniques
now available.
\end{enumerate}

\chapter{Theoretical description}

In parallel to the extensive experimental effort in quest for power-law
behaviour, intense activity has developed on the theoretical side,
to find acceptable explanations of the rapidly accumulating collection of
data on factorial moments. The meaning of ``intermittency'' in
multiparticle processes is still the subject of much debate and no definite
consensus has as yet emerged. Let us remember once again that we are
dealing here with the problem of evolution of particle number
distributions (or multiparticle correlations) in ever smaller bins.

{\it A priori, }
the most direct road of attack starts from quantum chromodynamics (QCD),
now firmly established as the theory of strong interactions. Unfortunately,
since the problem of confinement is unsolved, QCD can only be used as a
guideline to build phenomenological models for soft hadronic phenomena.
While successful for $\E$-annihilation, such models remain at present
unsatisfactory for most other processes, and in particular for hh-collisions.
The model's deficiencies are often invoked in support for claims of
``new physics'', but also this matter is far from being settled.

{}From the outset it is clear that phenomena such as ``intermittency''
are manifestations of dynamics in a, most probably, strongly non-linear
regime of QCD. It is, therefore, quite likely that the observed phenomena
are not very sensitive to the precise form of the Lagrangian, even though
the general properties of the interacting fields (e.g., the vector nature
of gluons) surely play a crucial role. Hence, a satisfactory description
might be possible on the basis of quite general properties of non-linear
systems as revealed by complex systems in many other branches of physics.
This idea lies  at the origin of various attempts to establish connections
with models for turbulence, multiplicative cascade processes,
``effective'' field theory, the statistical mechanics of disordered systems,
fractals, phase transitions of various kinds and others.

A perturbative QCD approach to intermittency
would provide a viable explanation if the
hadronization of quark and gluon systems possesses
the property of ``early confinement''
such that local parton-hadron duality would hold.
Various efforts in this direction have
indeed established that parton (quark-gluon)
avalanches exhibit (multi)fractal properties. These
follow from the (fortunate) fact
that the process possesses Markovian properties. Further
developments along this line will
evidently improve our understanding of perturbative
cascades.
A direct connection with experimental observations is,
however, not yet established.

Numerous other interpretations of ``intermittency''
have been advanced and will briefly be
described further below.

\section{Simplest approximations}

\subsection{The linked-pair approximation}

If the rapidity distribution does not change appreciably within a bin interval
(this is justified for small intervals, at least)
one can rewrite (\ref{dr:47}),
approximating it by
\beq
K_q(\d y) \approx \frac{1}{M(\d y)^q} \sum^M_{m=1} \int\limits_{\d y}
\prod_i \rd y_i K_q(y_1,\dots,y_q)\ .  \label{kq1}
\eeq
Since there are no statistically independent contributions to the cumulant
functions $C_q$ and $K_q$, their arguments should be somehow linked.
Studying the correlation of galaxies~\cite{Peeb80}, it was noted  that
$K_q$ can be decomposed into sums of products of two-particle correlation
functions, $K_2$ with overlapping arguments, in such a way that
all multiparticle correlations are
expressible as successive two-particle ones, so that the
whole chain of particles becomes
correlated. The last condition is necessary since we have learned that
$q$-particle correlations ($q>2$) are indeed present in the data.

In this scheme, higher-order scaled cumulants~\cite{CaSa89} are written as
\beq
K_3(y_1,y_2,y_3) = \frac{A_3}{3} \sum_{\perm} K_2(y_1,y_2)K_2(y_2,y_3)\ ,
\label{k32}
\eeq
\beq
K_4(y_1,y_2,y_3,y_4) = \frac{A_4}{12} \sum_{\perm} K_2(y_1,y_2)K_2(y_2,y_3)
K_2(y_3,y_4) \ ,  \label{k43}
\eeq
etc. Here, all the permutations of indices $1,\dots,q$ are summed over;
the number of terms is equal to
 the denominator of the factor in front of the sum.
The numerator is an a priori arbitrary parameter
for  each order of correlation.

Even now, the numerical integration in (\ref{kq1}) is hard to perform.
For that reason it
was suggested~\cite{CaSa89} to assume translation invariance of the cumulants
$K_q$, and to use the strip approximation (i.e. instead of integrating
over a set of hyper-cubes
with linear size $\d y$, one integrates over a strip along the main axis
$Y$ and over the differences $\z_i=(y_{i+1}-y_i)$). In this
way (\ref{kq1}) reduces to a simple but approximate formula~\cite{CaES91}:
\beq
K_q\approx A_q(K_2)^{q-1}\ .  \label{kq4}
\eeq
Substitution of $K_q$ in (\ref{dr:48}) allows any factorial moment
to be expressed in terms of the
second moment so that, for example,
\beq
F_3 = 1+3K_2+A_3K^2_2\ .  \label{f35}
\eeq

As  mentioned in Sect.~4.5, one can describe ~\cite{CaES91} the UA1 and UA5
data on factorial moments at energies from $\sqrt s$=200 GeV
to $\sqrt s$=900 GeV with constant values of $A_q$ for all intervals
$\d y$. Still, the intermittency indices derived from the linked-pair
approximation remain somewhat below  the experimental ones. At the lower
energy of the NA22  experiment, one gets a much larger value of $A_3$ if
the above estimate of $F_3$ is used boldly~\cite{Ajin89-90}. However,
the assumption of translation invariance is not really justified there
and one can hardly use~(\ref{f35}) on data which are averaged over a large
region of phase space.

Apart from this problem, there are also more basic  questions which remain
unanswered. For instance, even within the framework of the linked-pair
ansatz, one could add to (\ref{k32}) a term containing a product of three
$K_2$'s (loops or ring-graphs), as well as further terms with multiple
links among the pairs.

Although a dynamical justification for the linked-pair approximation
is lacking at present, it should be remembered that similar approximation
methods have proven their utility, e.g.~in the theory of liquids.
Within the Born-Bogoliubov-Green-Kirkwood-Yvon (BBGKY) hierarchy scheme,
they allow to ``close'' the otherwise infinite sequence of equations
relating  correlation functions of all orders. That the linked-pair
ansatz may have more than accidental relevance is further indicated by
the non-trivial fact that (\ref{kq4}) with $A_q=(q-1)!$
corresponds to a multiplicity distribution in $\delta y$ which is
of Negative Binomial type with $k$-parameter
$k=1/K_2(\delta y)$~\cite{WolfAPP90}. This two-parameter distribution
satisfactorily describes a large variety of (non-averaged) multiplicity
data and also occurs as an approximation to the soft parton
multiplicity distribution in QCD-jets~\cite{malaza:webber,nbd:gio,LAnd}.
Further extensions of the linked-pair approach beyond
those of~\cite{CaSa89,WolfAPP90} are treated in~\cite{lvanhove90,bozek91}.

The structure of many-particle correlations has also been analysed in
partially coherent radiative systems~\cite{CaFrie89}. This
 approach is closely related to the linked-pair ansatz~\cite{WolfAPP90}.

Conformal theories, treated in connection with intermittency
in~\cite{dnL92},
provide an alternative~\cite{Dremextra} to the linked-pair ansatz.
In such theories the $q$-th order irreducible Green function is
written as a product of two-particle ones to the power $1/(q-1)$.
Taking into account the $q(q-1)/2$ permutations of all particle
indices, one finds
\begin{equation}
K_q\approx B_q (K_2)^{q/2},\label{drem:extra0}
\end{equation}
instead of (\ref{kq4}), which also fits the experimental data reasonably
well~\cite{Dremextra}.

\subsection{The singularities of the correlation functions}

According to relations (\ref{dr:44})-(\ref{dr:48}),
the singular behaviour of the
factorial moments at small rapidity binning implies that the
correlation functions are singular for small separation of their arguments.
In particular,
the leading singularities of the correlation functions $\r_2$ and $C_2$ should
coincide with the singularity of the corresponding factorial moment $F_2$, if
the formal mathematical limit $\d y\to0$ is considered. For
\beq
F_2 \sim (\d y)^{-\f_2}\ \ \ (\d y \to 0)  \label{f26}
\eeq
one should get
\beq
C_2(y_1,y_2)\sim C^{(\rL)}_2 (y_1,y_2)|y_1-y_2|^{-\b}+C^{(\rN)}_2(y_1,y_2)
 \label{c27}
\eeq
with $\b=\f_2,\ \ C^{(\rL)}_2$ is a regular function of $(y_1-y_2)$,
while $C^{(\rN)}_2$ can contain non-leading singularities (less singular
than the first term).

A two-component model of this kind has been used in~\cite{AnCo90}, where
$C^{(\rN)}_2$ (and similarly for the higher-order correlations) is chosen
to be a regular function which results in a constant additive term to $F_2$.
The origin of the singular term has been ascribed in~\cite{AnCo90} to a
phase transition. For a singular term to be dominant, it must overwhelm
$F_2$ even numerically. However, the experimental data discussed above
indicate that this is not the case. Linear dependence in a double-log
plot is observed over quite a large background in the available region
of $\d y$. This implies that in this region the integral contribution
$F^{(\rL)}_2$ of the singular terms in (\ref{c27}) to  $F_2$ is rather
small so that
\beq
F^{(\rL)}_2 \ll F_2.  \label{fl8}
\eeq
Such a situation provides~\cite{AnCo90,Fial91} new possibilities with
$\b$ different from $\f_2$, since in that case one gets
\beq
\ln F_2\approx \ln (1+F^{(\rN)}_2) + F^{(\rL)}_2/(1+F^{(\rN)}_2)\ .
\label{ln9}
\eeq
It could even suggest a logarithmic dependence of $F^{(\rL)}_2$ on $\d y$:
$F^{(\rL)}_2\sim \log\d y$, i.e. a logarithmic singularity of the correlation
function for coinciding rapidities. However, such a behaviour is
indistinguishable from a power-like one for small exponents $\b$ and the
rather restricted range of rapidity intervals (e.g., $0.1<\d y<1$) in which
one usually looks for this
dependence. Actually, if $\b\log\d y\ll 1$ one gets~\cite{Fial91}
\beq
\ln F_2 \approx \ln (1+F^{(\rN)}_2) + \frac{\bar C^{(\rL)}_2}{\r^2_1}
(\d y)^{-\b}\approx \a_2-\f_2\ln\d y  \label{ln10}
\eeq
where
\beq
\a_2 = \ln(1+F^{(\rN)}_2) + \frac{\bar C^{(\rL)}_2}{\r^2_1} \ {\rm \ and\ }\
\f_2=\b \frac{\bar C^{(\rL)}_2}{\r^2_1}\ .  \label{a211}
\eeq
Herefrom, one would expect the intermittency exponent to be much smaller
than the corresponding strength of the singularity of the correlation
function, i.e. $\f_2 \ll \beta$.

The difference between logarithmic and power-like singularities may become
observable for higher moments at small $\d y$ as an upward curvature
appearing on log-log plots for power-law dependence. This would be more
noticeable for smaller $\d y$, for higher $q$ and for larger values of $\b$.
Still, one should not enter into the region of extremely small $\d y$, where
the empty-bin effect may dominate and turns down all the curves.

The existing experimental data on factorial moments are not in contradiction
with the above, although no clear signal of an upward curvature
of higher moments is seen because of large irregularities appearing at small
$\d y$ and large $q$. These irregularities are suppressed if
the method of correlation integrals \cite{Hen83,Drem88,CaSa89,Alba90} is
used, where the binning procedure is
not fixed but naturally follows the event structure. This has been discussed
in Sect.~3.8. In fact, the singularities are still better exposed
if factorial cumulants (\ref{dr:47}) are used instead of factorial moments
{}~\cite{FiMPI91}.

\subsection{Intermittency and Bose-Einstein correlations}

One of the possible sources of increase of factorial moments at small bin
sizes is the well-known attraction of identical Bose-particles (pions)
when their momenta are very close. Therefore, one is tempted~\cite{Gyul90}
to the extreme supposition that "intermittency" is governed by Bose-Einstein
correlations, i.e. by symmetry properties of fields but not by their
dynamics. As was shown in Chapter~4, the experimental data do indeed give
some indications on the relevant contribution of such an effect. In general,
the introduction of BE correlations tends to reduce the disagreement between
experimental data and Monte-Carlo models.  However, it was also shown there
that the dynamical part is non-negligible and, consequently,
is of main concern to us.

Referring the reader to the more specialized review
of the subject in~\cite{abds95}, we would like just to point out that the same
physics effects can become more (or less) pronounced depending on the
corresponding choice of the variable in which it is displayed. In particular,
it is shown in~\cite{abds95} that the rapidity variable is suitable to reveal
the
dynamical intermittency and to suppress the Bose-Einstein contribution to
factorial moments. At the same time, when analysed as functions of squared
4-momentum transferred between pions, the factorial moments get the power-like
increasing share  of BE-correlations which shadows the true (dynamical)
intermittency. Therefore, the power-like behaviour in that variable neither
proves nor disproves dynamical intermittency. One should keep this in mind
when looking at the corresponding experimental data. Surely, for quantitative
estimates Monte-Carlo calculations are necessary with full account of
the indefiniteness in the description of the BE-effect itself.

\section{Dynamical approaches}

\subsection{Various theoretical models}

{}From a theoretical point of view, experimental results are best approached
via quantum chromodynamics (QCD). Attempts in that direction are further
described in Subsect.~5.2.6. Unfortunately, the
application of QCD to soft processes involving small momentum transfers is
quite limited since strong non-perturbative effects are involved
(unless we use additional assumptions, as the local parton-hadron
duality hypothesis generalized to correlations of any order). Hence, we
are compelled either to construct general relations like as of
Chapter~2 and the previous section, or to develop phenomenological models
that fit experimental distributions by adjusting a number of free parameters.

By now, many phenomenological models have been proposed. Ideas inspired
by QCD have been used in parton shower models and in their phenomenological
counterparts: the dual topological model and quark-gluon string
models~\cite{LUND,DPM,QGSM,CaTr88,Ande87,Kaid82}; in coherent gluon jet
emission (Cherenkov gluons, in particular)~\cite{Drem80}; in the cold
quark-gluon plasma model~\cite{VHov89}. Models of a still more
phenomenological nature have been tried, such as cluster
models~\cite{Fugl87,Fugl88,DrDu75,LiMe83}, clan models~\cite{GiVH88} and
narrow hadron jet emission~\cite{OchWo88-89}. Whereas in all these models
definite dynamical mechanisms are proposed for the origin of the fluctuations
in multiparticle production, they still suffer from one important
deficiency: they do not reveal the nature of the scaling laws observed for
factorial moments at small bin sizes.

{}From that point of view, one would prefer the random cascade models
{}~\cite{bialas1,bialas2,BiPe88} and/or the general approach of phase
transitions~\cite{AnCo90,Pesc89,CaSa87,DrNa91}. While the cascade
models rely on analogies with turbulence theories and lead to
phase transitions, general considerations of the transition from parton
to hadron phases of the process are based on important properties of strong
coupling field theories reminding of  QCD lattice computations and the
conformal group symmetry. Both approaches lead in quite a natural way to
scaling behaviour of factorial moments and are preferred as heuristic tools.
However, to date they cannot compete with phenomenological Monte-Carlo
models in providing computational results comparable with experimental data
at the same level of precision. We shall discuss them separately at some
length later, together with important ideas of intermittency and fractality,
but first we shall describe a variety of phenomenological models applied
to the fluctuation problem.

First of all, we should mention the furthest developed Monte-Carlo versions
of QCD inspired models~\cite{LUND,DPM,QGSM,Mar88,Odo84,CaTr88,Ande87,Kaid82}
of parton showers or quark-gluon strings. Qualitatively, these models describe
the behaviour of the two-particle correlation function $C_2(y_1,y_2)$
observed experimentally, showing the signature of short-range correlations.
However, they cannot pretend to fit them quantitatively in hh collisions
for all topologies and cut-offs. All the models predict noticeably smaller
values of intermittency indices than the experimental data. Such models
also fail to describe the probability distribution of maximum particle
number per event at a given resolution $\d y$~\cite{PeTr87,Ande88}.

As discussed in Subsect.~4.2.3, for $\E$-annihilation the situation is still
controversial. Initially, the DELPHI-collaboration, working at the $Z^0$
peak, claimed agreement with the LUND parton shower model. Later, increasing
statistics tenfold, it finds that the agreement only holds at the level
of 10-20\% (see Fig.~4.7d)~\cite{Abreu90}. However, in general the situation
is better here than in hadronic reactions. In nucleus-nucleus reactions,
these models fail to describe the damping of spectator particles observed
in experiment~\cite{ESt95}. Further, more detailed studies are needed.

Thus, we see that fluctuations appear to be a stumbling block for
phenomenological models. They meet with difficulties when confronted
with measured factorial moments, in particular in hadron-hadron collisions.

In earlier days, no data existed on factorial moments for small bins
and information on strong fluctuations in the number of particles
inside such bins existed only for individual events. A distinctive feature
of the fluctuations was the azimuthal symmetry of particles belonging to
the spike. The whole event showed a noticeable ring of particles in the
plane perpendicular to the collision axis. Precisely for this reason, the
very first attempt to explain such fluctuations was based on an analogy with
Cherenkov photon radiation.

The hypothesis of coherent emission of gluon jets~\cite{Drem80} (involving, in
particular, the Vavilov-Cherenkov mechanism) predicted that these jets should
be emitted in a narrow pseudo-rapidity bin at rather large angles in the
center-of-mass system of the colliding hadrons. All subsequent models did not
predict any particular polar angle dependence for the dense groups of produced
particles. This specific feature was experimentally verified in
pp-interactions at 205 and 360 GeV energies~\cite{DremSJNP90}. It
turned out that the distribution of centers of dense particle groups on the
pseudorapidity axis contained several peaks superimposed on a fairly strong
background. Consequently, the proposed mechanism of coherent jet
emission in hadron interactions probably does exist but is not dominant.
It could provide the only distinction~\cite{DremSJNP90} between pp and
$\Pp\Pap$ data available from ISR, but no analysis of that kind has yet been
performed. The ring-like events were observed in earlier cosmic-ray
experiments~\cite{Ale62}~-~\cite{Maru79} and in recent studies of
nucleus-nucleus collisions~\cite{OCher95}. Nevertheless, other mechanisms
must be involved since intermittency is also observed in $\E$-annihilation,
where the conditions for coherence seem unlikely to be satisfied.

Large fluctuations may arise in an extended blob of a cold quark-gluon plasma
{}~\cite{VHov89}. The appealing feature of such a model is the relationship
between this phenomenon and the production of soft and low-$p_\rT$ hadrons,
lepton pairs and photons. However, this model faces problems in explaining
the large values of intermittency indices in electron-positron annihilation
compared to hadronic and nuclear processes, since, contrary to present
experimental results, it leads one to expect that the effect is largest
in nucleus-nucleus collisions.

Models based on clusters~\cite{DrDu75,LiMe83} or clans~\cite{GiVH88} are
more flexible in fitting factorial moments, since they involve several
free parameters. It has been demonstrated~\cite{VHov89,Seib89} that the
existence of clans leads to a power-law increase of factorial moments for
smaller bin sizes. No quantitative comparison with experiment has been
attempted, however. As to cluster models, in some cases they succeeded
in describing the multiplicity distribution in symmetric rapidity bins
of various sizes~\cite{Fugl87,Fugl88} and of the two-particle correlation
function. The data available at that time were limited to rather large
rapidity interval sizes ($\d y \geq 0.5$). In smaller regions, however,
simple cluster model fail completely (see Figs.~4.4b, 4.20c and 4.23a).

Multiparticle production is described somewhat differently
 in~\cite{FoWe78,CaFrie89}, where particle emission is explained
by two types of sources - chaotic and coherent.

Some of the above approaches attempt to describe multiplicity distributions
in varying rapidity bins in terms of the negative binomial distribution
(or modifications thereof). In the cluster model, this is accomplished by
varying the cluster parameters. In the clan model, the negative binomial
distribution is obtained by compounding a Poisson distribution for the
number of clans with a logarithmic distribution for their decay. In the
statistical model with two types of sources, the negative binomial
distribution naturally describes the chaotic sources, while the coherent
sources contribute a Poisson component.

Even though the negative binomial distribution can be phenomenologically used
to fit experimental data in the very first approximation, there are definite
distinctions from it in experiment. Besides, the asymptotic QCD predictions
disfavour it revealing new features~\cite{Dr94} of multiplicity distributions.
In particular, the NBD cumulants are always positive, while perturbative
QCD predicts negative values (and oscillations) of the higher order
cumulants. The pQCD prediction is supported by experiment for the
case of the total rapidity range.

\subsection{Intermittency and fractality}

In most cases, the models considered above need to have their parameters
adjusted to be able to fit  (if at all) the  data  on factorial moments.
Power-law behaviour of factorial moments is
most  naturally obtained for cascading
mechanisms and in phase transitions, as we shall see later.

The concept of intermittency has been borrowed from the theory of turbulence.
There, it represents the following property of a turbulent fluid: vortices of
different size alternate in such a manner as to form a self-similar structure.
They do not fill in the whole volume, but form an intermittent pattern
alternating with regions of laminar flow.
Mathematically, this property is described  by a
power-law dependence of the vortex distribution moments on the vortex size.
This is the reason why the exponents
$\f_q$ in the power-law dependence of factorial
moments $F_q(\d y)\propto (\d y)^{-\f_q}$ in (\ref{fq113})  are  called
``intermittency indices''.

As mentioned in Sect.~2.4, the self-similar nature of vortices directly
implies a connection between intermittency and fractality. Fractals are
self-similar objects of a non-integral dimension. The fractal dimension is a
generalization of ordinary topological dimensionality to non-integers.

More complicated self-similar objects exist, consisting of differently
weighted fractals with different non-integer dimensions. They are called
multifractals and are characterized by generalized (or R\'enyi) dimensions
$D_q$ which depend on the rank $q$ of the moment of the probability
distribution over such objects. The analysis of multifractals according to
the L\'evy indices goes beyond the simple definition of intermittency.

The formal definitions are given in Sect.~2.4. For more details,
 we refer to the review papers~\cite{Pala87,PescTH5891,DremSPU90} and
references therein. In connection with multiparticle  production, fractals
were first mentioned in~\cite{Ven79,Gio79,Minh83,DremJETP87}.

Somehow, the concept of fractality goes beyond a purely formal definition of
intermittency, by connecting the observed dimensionality with the geometrical
and thermodynamical properties of an object, as well as with the properties of
the distributions over this object~\cite{Pala87,Hen83}. The power-law
behaviour of factorial moments reveals the fractal structure of rapidity
distributions as decomposed in individual events. Its relation to geometrical
and thermodynamical properties is discussed below.
In string models~\cite{LAnd}, the self-similar behaviour of fluctuations is
ascribed to the fractal nature of the phase space available for subsequent
branchings that is formulated in~\cite{Bj?} as a "plumber view" of
multiparticle processes.

Sometimes, intermittency is ascribed~\cite{ABi?} to the fluctuations of the
geometrical sizes of emitting sources.

Some caution is necessary when applying these concepts to multiparticle
production. As discussed in Chapter~4, finite statistics, the rather small
number of particles produced in an event and, especially, in a given cell,
the method of bin-splitting, the rather restricted range of bin widths over
which the power-law behaviour is observed, all influence the final
conclusion. This is described in more detail in~\cite{PescTH5891,LeTs91}.

\subsection{Random cascade models}

In turbulence, intermittency was first demonstrated in cascade
models~\cite{Kolm41}. Modifications of these, as applied to multiparticle
processes, are rather popular nowadays~\cite{bialas1,bialas2}.

In such models, one considers a series of self-similar steps in partitioning
phase-space. Let us denote by $M$ the number of bins obtained by breaking
up the total phase space into $\la$ parts at each of the $\nu$ iterations of
the self-similar cascade. Thus $M=\la^{\nu}$ ($\equiv\frac{\D Y}{\d y}$
for a total rapidity range $\D Y$ divided into bins of width $\d y$).
Random cascade models involve a probability distribution $r(W)$
with corresponding moments
\beq
\lan W^q\ran = \int \rd W\ \ r(W)W^q\ ,\ \ \ \  \lan W\ran = 1\ . \label{wq12}
\eeq
The function $r(W)$ induces density fluctuations as the rapidity window
is broken up into ever smaller bins. The density  $P_m$ in the $m$-th bin
is given by the product
\beq
P_{m}=\frac{1}{M}\prod^\n_{n=1} W_n\equiv\frac{1}{M}
\frac{\r_{(m)}}{\lan \r_{(m)}\ran},  \label{pm13}
\eeq
where the sequence of indices $n$ defines a path leading to a given
bin $m$ with density $\r_{(m)}$. One assumes that there exists a range
of scales inside of which the weights $W$ are constant, i.e., they do
not depend on the scale at which they operate.

Herefrom, the  intermittent character  of the models follows as:
\beq
F_q = \lan (MP_m)^q\ran=\lan\prod^{\nu}_{n=1}W_{n}^{q}\ran =
(\D Y/\d y)^{\ln\lan W^q\ran/\ln\la}\ .  \label{fq14}
\eeq
The intermittency indices are equal to
\beq
\f_q = \ln\lan W^q\ran/\ln\la\ ,  \label{fq15}
\eeq
i.e., random cascade models possess a multifractal spectrum~\cite{Pala87}.

The simplest type of distribution $r(W)$ is the subclass of so-called
$\a$-models~\cite{Scher84} given by the two-level probability distribution:
\beq
r(W) = p\d(W-W_-) + (1-p)\d(W-W_+),  \label{rw16}
\eeq
where $0\leq W_-<1<W_+$ and $pW_-+(1-p)W_+=1$ because of the normalization
condition (\ref{wq12}). The density enhancement $W_+>1$ occurs with
probability $(1-p)$ at each step of the cascade, while a depletion $W_-<1$ is
present with a probability $p$. Combined, they create ``spikes'' and
``holes'' in the rapidity distribution. The intermittency indices are given by
\beq
\f_q = \ln[pW^q_-+(1-p)W^q_+]/\ln\la\ .  \label{fq17}
\eeq
The study of moments and multifractal analysis reveal new interesting
features.
(Let us mention that the $\a$-model is reduced to the $\b$-model for $W_-=0$
and describes a mono-fractal in that case). As the parameters $p$ and
$\la$ are changed, the model predicts various phase transitions
{}~\cite{PescTH5891,BiSZ90,BrPe90}. The moments of factorial moments are
useful to reveal these transitions due to the fact that the distributions of
factorial moments are extremely irregular by themselves~\cite{DremHolm90}.
Introducing the normalized moments of moments and ascribing to these
a power-law behaviour at small bins of the form
\beq
\lan Z^p_q\ran = \frac{\lan F^p_q(\d y)\ran}{\lan F_q(\d y)\ran^{~^p}}\propto
(\d y)^{p\f_q} \lan F^p_q(\d y)\ran \propto (\d y)^{\ve_{p,q}} \ ,
\label{zp18}
\eeq
one can analyse, in the framework of $\a$-models,  the dependence of the
indices $\varepsilon_{p,q}$ on the parameters of the model. As discovered
in~\cite{BiSZ90}, this dependence defines four regions in the parameter
space, which are reminiscent of four different phases. The indices
$\varepsilon_{p,q}$  act as  order-parameters.

The same conclusions have been obtained when studying~\cite{BrPe90} the
normalized factorial correlators $F_{pq}/F_pF_q$. Another important property
of these correlators is their independence of the bin width
{}~\cite{bialas1,bialas2}. This property has been confirmed by experiments (see
Sect.~4.6).
However, the $\a$-model does not predict the  correct  dependence
on the distance between bins. It predicts power-law behaviour
with an  exponent
\beq
 \f_{pq} = \f_{p+q}-\f_p-\f_q  \label{fp19}
\eeq
related to the usual intermittency indices.  The experimental values
do not follow a straight line on a double-log plot.  Moreover, there
are no finite intervals where they satisfy the above relation.  When
roughly approximated by a straight-line fit, the experimental values
of $\f_{pq}$ are larger than those of the $\a$-model.

However, one should keep in mind that this is a toy-model, which could
pretend to be valid for asymptotically long cascades, i.e. for
extremely high energies and multiplicities. Otherwise, one should
develop Monte-Carlo programs~\cite{LeTs91} losing the beauty of
analytical formulae. In fact, it has been clearly shown for such a
well-known mathematical model as the Cantor set~\cite{Levt90}, that it
is not an easy task to reveal its fractal dimension (known {\it a
priori}) using factorial moments, if the number of iterations is
finite.

Nevertheless, the heuristic value of $\a$-models in describing
 qualitative features of the process is rather high since, in
 particular, they suggest the possibility of phase transitions.  The
 nature of the transitions and their relation to the quark-hadron
 transformation are not clear yet. An interesting observation is the
 existence of ``non-thermal'' transitions similar to those in spin-glass
 systems. They differ from the usual ``order-disorder'' transitions by
producing ``different order'' in different regions of phase space, so that
one may call them ``clustered order-disorder'' transitions.  In that case,
the intermittency indices increase with increasing rank faster than
linear.  Analogies to statistical-mechanics systems~\cite{Pesc89,BrPe90}
provide further insight into the nature of the transitions.

\subsection{Field-theoretical approach and phase transitions}

Besides the scenario of a self-similar cascade, as another extreme, a
higher-order quark-hadron phase transition has been proposed to
explain strong fluctuations leading to intermittency
patterns~\cite{AnCo90,DrNa91}.  Evidently, the statistical mechanics
description is most useful here. Hadronization of a quark-gluon plasma
becomes the origin of exceptional events with large fluctuations
observed above a strong background of conventional events. Such a
point of view is supported by studies of a two-dimensional lattice of
Ising spins~\cite{Wosi89,Satz89}, which shows that intermittency
appears at the critical temperature.  The intermittency indices are
directly related to critical exponents of the system.
Similar features have been found for the $q$-state Potts model on the
Bethe lattice ~\cite{Haj92} at the phase transition but without
long-distance correlation.  Clear fractal structure is also observed
for $SU(2)$ gluodynamics near the phase-transition point in lattice
calculations ~\cite{Polik} .

The scenarios of cascading and of phase transitions need not
contradict each other, if one accepts the point of view that the role
of a quark-hadron transition is to fix the fractal pattern formed by
cascading. The fluctuations are ``frozen'' at the transition point and
can be computed by just considering this point.

A well-defined field-theoretical procedure exists to treat
fluctuations and phase transitions in common media~\cite{PatPok}. It
requires, first of all, the definition of an order-parameter and its
treatment as an effective fluctuation field.  In case of multiparticle
production, one could choose the rapidity density distribution of
particles in individual events $\r_{(\re)}(y)$ (or a function of it) as
an order-parameter which fluctuates about its inclusive average
$\r(y)$ at each rapidity value $y$.  Its function as an
order-parameter is clarified by the local parton-hadron duality
hypothesis, which has been successful in describing the experimental
data for electron-positron annihilation processes. This hypothesis
states that the average values of $\r_{(\re)}(y)$ at partonic $\r_{(\Pp)}$
and hadronic $\r_{(\rh)}$ levels differ by a (for all rapidities) common
numerical factor. In particular, if one defines~\cite{DrNa91} the
fluctuation field as
\beq
 \varepsilon(y) = \frac{\r_{(\re)}(y)}{\r_{(\rh)}(y)} -1 \ , \label{e20}
\eeq
then its average in the hadronic phase is equal to zero, while it
differs from zero at the partonic level:
$\lan\varepsilon(y)\ran_{(\Pp)}=\r_{(\Pp)}(y)/{\r_{(\rh)}(y)}-1=\,
\mbox{const}$.
Other possible choices for the fluctuation field exist (for example,
$\r^{1/2}(y)$ ~\cite{AnCo90,CaSa87,ScSu73} or
$\r-\r_{(\rh)}$~\cite{dnL92}).  They have advantages and disadvantages
which we shall not discuss here.

The probability of a fluctuation is given by
\beq
W(\varepsilon)=Z^{-1}\exp[-F(\varepsilon)]\ ,  \label{we21}
\eeq
\beq
Z=\int \rD\varepsilon \ \exp[-F(\varepsilon)]  \label{z22}
\eeq
where $\rD\varepsilon$ refers to the functional differential,
$F(\varepsilon)$ is the free energy and $Z$ is the partition function.
Adding to $F(\varepsilon)$ a term $J(y)\varepsilon(y)$ with an external
current $J(y)$, one obtains the irreducible Green functions:
\beq
\lan \ve_1 \dots\ve_q\ran=\frac{\d^q\ln Z}{\d J(y_1)....\d J(y_q)}
\vert_{J=0} \ ,    \label{ve23}
\eeq
which are related to correlation functions, so that, for example (we omit
the index h),
\beq
\lan\varepsilon_1\varepsilon_2\ran=\frac{\r_2(y_1,y_2)}{\r(y_1)\r(y_2)}-1
\equiv K_2(y_1,y_2)\ .  \label{ve24}
\eeq
The factorial moments are easily obtained as
\beq
F_q(\d y)=(\d y)^{1-q}\int^{\d y}_0 \rd\z_1...\int^{\d y}_0 \rd\z_{q-1}
r_q(\z_1,\dots,\z_{q-1})   \label{fq25}
\eeq
where
\beq
\z_i=y_{i+1}-y_i \ \ and \ \ r_q=\r_q(y_1,...,y_q)/\r(y_1)...\r(y_q)\ .
 \label{zi26}
\eeq

At first sight, the fluctuation field theory is not directly related
to the underlying QCD.
Yet, these theories are connected through the fluctuation pattern of
individual events $\r_{(\re)}(y)$, which should be described by both of
them if they pretend to be valid. Thus, our guesses on the fluctuation
field $\ve(y)$ reflect special features of cascading and confinement in QCD.

For small fluctuations, one can represent $F(\ve)$ by a Taylor series
\beq
F(\ve) = F_0 + \int \rd y \left[ \frac{b}{2} \left[ \frac{\rd\ve}{\rd y}
\right]^2
+ \frac{a}{2} \ve^2 + c\ve^3 + \rd\ve^4 + \dots \right] \ ,  \label{fe27}
\eeq
which corresponds to the Ginzburg-Landau Hamiltonian when $c=0$,
$d\not=0$ and all higher terms are equal to zero.  It has been
found~\cite{hwanaz} that some scaling indices (but not the critical
exponents) have universality properties in this approach.

For free fields, i.e. $c=d=\dots=0$, one gets
\beq
\lan\ve_1\ve_2\ran_f = \g \exp \left[-|y_1-y_2|/\xi\right] \ ,  \label{e28}
\eeq
\beq
\g = \p\xi/b; \ \ \ \xi=(b/a)^{1/2} \ .  \label{g29}
\eeq
This exponential form fits the two-particle correlation function
qualitatively (and is often used, in particular, for
nucleus-nucleus collisions~\cite{Egg91}), but it does not provide
intermittency at small $\d y$.  One should remember that it is related
to the free-field Lagrangian but not to the Ginzburg-Landau potential
and, therefore, describes usual short-range correlations without any
phase transitions. One should also note that the approach is
formulated in momentum space and not in configuration space.

One is tempted to conclude that fluctuations are strong at small
rapidity intervals and that the perturbative approach fails. The
phenomenon of intermittency should be described by a strong coupling
field theory, where perturbative methods do not work. In particular,
the renormalization group approach and conformal theories have been
tried~\cite{DrNa91,dnL92} and have provided power-law behaviour of
Green functions and factorial moments at small bin widths. So, it
seems rather reasonable to fit the correlation function by an expression
\beq
C_2(y_1,y_2) \propto \frac{1}{|y_1-y_2|^\k} \exp [-|y_1-y_2|/\xi],\;\;\;\;
(\k<1)
\label{c229}
\eeq
for all rapidity separations. For rapidity separations smaller than
the correlation length $\xi$ one gets pure power-law dependence of the
correlation function due to a phase transition phenomenon.  The
associated singularity should soften energy and transverse momentum
spectra. In the simplest approximation, intermittency indices increase
linearly with their rank.

Let us stress an important difference of the above consideration with
the previous treatment of phase transitions. The correlation length
$\xi$ does not tend to infinity and the exponential law is not
replaced by a power law at large $\d y$, as one is accustomed
to. Instead, the power-law appears at small $\d y$ ($\d y \ll \xi $),
and does not influence the dependence at large rapidity.  This
happens, because rapidities play now the role of coordinates in the
usual treatment, so that one has to deal with the ultraviolet (not
infrared) stable point of the Gell-Mann-Low function.  One can
speculate that particles lying far apart on the rapidity scale reveal
the dynamics of the process with a finite correlation length related
to a particular form of the Lagrangian, while those at nearer points
remind of the self-organising critical processes with a scaling law
not tightly connected to a particular form of the Lagrangian (sandpile
phenomenon).  Thus, correlations appear as ``{\it frozen} (due to
hadronization) {\it sounds}'' of cascading.

 Similar problems have been discussed in the framework of
Feynman-Wilson fluid models~\cite{AnCo90,CaSa87}. One should introduce
the notion of temperature, additional assumptions on thermal
equilibrium, on Kadanoff scaling at the critical temperature, on the
relative role of conventional and stochastic (or quark-gluon plasma)
components, and so on.  Imposing special boundary
conditions~\cite{acpv} on the grand-canonical partition function, one
can relate Kadanoff scaling in the fluid to KNO-scaling in
multiparticle processes and describe the fractal properties of the
fluid in a wide range of scales.  Formula (\ref{c229}) appears to be
valid for correlation functions of the conventional hadronic system,
but for a system at the critical point, pure power-like behaviour with
a different exponent is restored.  One is therefore lead to consider
the whole process in the framework of a two-component (conventional
$+$ critical) model.

The same approach has been extended~\cite{amd} to a multidimensional
analysis of intermittency using, however, the assumption that the
correlation functions factorize in rapidity and transverse momentum.
The predictions for factorial moments differ from QCD predictions.
The factorization hypothesis allows to proceed analytically and to
relate intermittency to fractal properties of the system in original
space-time (this problem has been addressed also
in~\cite{DremJETP87,dl92,Bial92}) but looks rather artificial for any
field theory (QCD included).  For instance, if one assumes that
conformal symmetry is responsible for
intermittency~\cite{DrNa91,dnL92}, one obtains non-factorizable Green
functions and the predictions differ from the above-mentioned ones due
to the mixing of longitudinal and transverse momentum components,
inherent in field theories.  We shall see, however, that conformal
theory and QCD also differ.

Numerical values of intermittency indices can be calculated in the
conformal scheme and agree qualitatively with experimental findings.
Again, the phase transition plays a crucial role.

Studies of the role of phase transitions in multiparticle production
are still in their infancy and have, until now, provided qualitative
results only.  Also the relation between hadronization and phase
transitions in the simple cascade models treated in the previous
sub-section is not yet clear.

The relative role of parton cascading and hadronization is another
matter of debate.  The problem would be solved if extreme proposals
were valid.  Indeed, parton cascading with an ``infinite'' number of
steps would provide intermittency indices quadratically increasing
with their rank (this corresponds to ``wild'' or ``hard''
singularities).  Phase transitions, on the other hand, yield
monofractal behaviour with a linear increase of the indices.
In reality, the two extremes must be modified
so that finite cascading
would lead to a slower increase, while the next operator-product
expansion terms for  a phase transition
would induce a faster than linear rise of the intermittency exponent.
Such problems have as yet not been treated.

\subsection{The statistical-mechanics formalism}\\
Statistical analogy is  a powerful way to  analyse  properties of
chaotic dynamical systems~\cite{Sinai72}, in general, and of
multifractals and  cascade models,
in particular~\cite{Tel88,Pesc89,PescTH5891}.
It rests on the  possibility to define a
partition function $Z(q)$ of the system in the following way:
\beq
Z(q) = \sum^M_{m=1} p_m^q \ ,  \label{zq30}
\eeq
where
\beq
p_m = \frac{\r_{(m)}}{M\lan\r_{(m)}\ran}  \label{pm31}
\eeq
is a normalized probability weight on the ensemble of bins $(m)$ and
$\r_{(m)}$ is a random (rapidity) density registered in each bin $m$.

The relation to multifractal (or intermittent) properties of a system is
established if one considers systems for which the probability inside
a box $m$ is proportional to a power $\a_m$ of the box size and the number of
degenerate boxes (with the same value of $\a_m$) follows a power law as well.
Then, assuming a continuous limit, one finds  as in (\ref{dr:121})
\beq
Z(q)\simeq \int\limits^{a_+}_{a_-} M^{f(\a)-q\a} \rd\a \ ,  \label{zq32}
\eeq
(integration running between maximum and minimum zeros of $f(\a)$),
wherefrom one easily obtains the multifractal spectrum of the system $f(\a)$
(see, for example,~\cite{Pala87,DremSPU90,DremFest}).

Remembering the definition of the intermittency indices $\f_q$, one relates
them to the spectrum $f(\a)$ via the relation
\beq
f(\bar\a) = -q^2 \frac{\rd}{\rd q} \left[ \frac{\f_q+1}{q}\right] \ ,
\label{fa}
\eeq
where $\bar\a$ is defined as
\beq
\bar\a(q) = 1 - \frac{\rd\f_q}{\rd q}\ .  \label{aq}
\eeq

The interpretation of $f(\a)$ is transparent since it weights the number of
degenerate boxes. It, therefore, corresponds to the entropy in statistical
mechanics. In a similar way, the rank $q$ may be interpreted as an inverse
``temperature'' $\b=1/T$ and the relation (\ref{fa}) corresponds to the
usual thermodynamical formula
\beq
S = - \frac{dF}{dT},  \label{s35}
\eeq
where $S$ is the entropy and $F$ is the free energy, whose
``temperature'' dependence is provided now by
\beq
\la(q) = \frac{\f_q+1}{q}=1-F\ .  \label{laq}
\eeq

Two features of this analogy are particularly useful. On the one hand,
application of the thermodynamical formalism allows for coverage of
multifractals by boxes of different sizes, i.e. for a more precise
description of individual events.  This has been used in proposals of
correlation measures with non-uniform
coverage~\cite{Hen83,Drem88,Alba90}.  On the other hand, the
multifractal treatment admits an extension of the thermodynamical
formalism to non-equilibrium systems. It has been used to classify the
phase transitions in $\a$-models and to demonstrate that, in
multiparticle processes, phases could exist similar to spin-glass
states ~\cite{Pesc89,PescTH5891}.

It is important to note that the minimum of the $\lambda(q)$
(\ref{laq}) corresponds, according to (\ref{fa}), to zeros of the
fractal spectrum.  This is a signal for a phase transition in
thermodynamical systems.
One should stress, however, that the similarity of the distributions
is in itself not sufficient to justify use of statistical physics
terminology in its original meaning for the quantum field systems we
are interested in here.  Besides, the analogy breaks down completely
for values of $q$ exceeding the multiplicities effectively
contributing to the moments. Nevertheless, this analogy is used
in~\cite{Band,HeKr} to derive the dependence of the pressure in
a Feynman-Wilson liquid on its chemical potential and some peculiar
features are found in hadron-hadron reactions (see also the review
paper~\cite{DrLe95}).

It is evident that further explorations of statistical mechanics
approaches to multiparticle production are needed.  In particular,
analytical properties of a partition function, often useful in
connection with phase transitions, have not been much analysed.

The location of the (complex) roots (zeros) $z\nu$  of the multiplicity
generating function (\ref{dr:15}) has recently been
studied in~\cite{zeros:edw,ID94} after earlier suggestions
by Biebl and Wolf~\cite{Brown}. This work is based  on the
analogy with the famous Lee-Yang  zeros\cite{Yan52},
whose location fully characterizes
the thermodynamic properties of the physical system.

For  multiplicity distributions, the strength of the fluctuations of the
multiplicity in an event is  directy related to
the location of the zeros in the complex $z$ plane:
the magnitude  of the factorial cumulants, and thus the strength of the
correlations, is determined by the roots closest to the origin.

In the discrete version of QCD, developed in LUND, is was
demonstrated that the zeros of (\ref{dr:15}) belong to a fractal
Julia-set~\cite{zeros:bo} with intruiging properties.
Detailed studies  of this set, and various connections with standard
phenomenology, such as  KNO-scaling, remain to be worked out.

\subsection{Intermittency, evolution equations and QCD }

In the previous section we considered two rather extreme
possibilities, simple cascade models and phase transitions, as
possible mechanisms leading to scale-invariance in particle production
processes.

Further interesting results have been derived from studies of
simplified kinetic branching evolution equations for ``birth-death''
(or ``gain-loss'') processes.  Many of these are treated in textbooks
{}~\cite{Bartl55,Sevas71} and were applied to multiparticle production
{}~\cite{Giovan79,LamWal84,Shih86,Batu88,Chliap90}.  In general, the
time-evolution of the number of ``clusters'' (partons, resonances, etc.)
 is described by forward or backward (retrospective) Kolmogorov
equations, which relate the time derivative of the generating function
to some combinations of that function. A particular example is the
Smoluchowski equation treated in ~\cite{Meun92}.  The terms
``forward'' and ``backward'' imply, that each tree graph may be viewed
either as a splitting to ever ``thinner'' branches, or as a
convolution to ever ``thicker'' branches.

For linear evolution equations, the solutions depend directly on the
initial conditions~\cite{Biya90}.  Non-linear equations often have
solutions asymptotically independent of
pre-history~\cite{Batu88,Meun92}.  Stationary regimes may appear if
the annihilation of clusters is stronger than the ``birth-rate''.  In
that case, the dispersion is proportional to the average multiplicity
and intermittent behaviour can be obtained, if the proportionality
factor is larger than one and the mean multiplicity decreases
correspondingly for smaller bins.  This can easily be proven by means
of the definition of the second-order factorial moment.
Let us note that systems obeying non-linear evolution could exhibit
quite general properties, independent of the detailed form of the
equations, such as period-doubling. Some ideas along this line of
thought, using properties of stochastic systems and Feigenbaum
attractors, have been formulated in~\cite{Dias87,Batu92}.

More detailed analysis of intermittency in the framework of the
Smoluchowski equation~\cite{Meun92} reveals various regimes of
time-evolution and cascading, depending on the parameters of the
model.  The Smoluchowski equation is of the backward type.  It
contains terms linear and quadratic in the generating function $G$
with opposite signs.  Formally it looks like
\beq
\frac{\rd G}{\rd t}=G*G - G*G_{1}  \ ,  \label{dg37}
\eeq
where  $G(u,t)=\sum_{n\geq 1}  N(n,t)u^{n},\ \ \ G_{1}=G(1,t)$  with
$N(n,t)$ representing the number of clusters  of  (integer)  mass
$n$ at time $t$ and  the  convolution  $*$  is defined  through  the
aggregation coefficients of clusters. The fractal  properties  of
aggregates  and  the  occurrence of phase  transitions  have  been
analysed in~\cite{Meun92}.

Obviously, it would be desirable if an explanation of scaling phenomena
in multiparticle production could be derived from, so to say,
first principles, i.e. in the framework of QCD. Multiparticle production
in QCD is the result of quark-gluon branching and the subsequent transition
to hadrons. As such, the self-similar multiplicative branching (or cascade)
process could give rise to a scaling regime. The perturbative QCD parton
shower picture is justified for interactions with large transferred momenta
(or virtualities), but in hadronic reactions one mostly has to deal with
soft processes. Perturbative QCD is valid in the initial
stages of high-energy cascades in electron-positron annihilation
and could, therefore, be used as an explanation for intermittency.

It is well known that the perturbative QCD cascade gives rise~\cite{Alta77}
to a mean multiplicity of partons increasing rapidly with energy.
Equations for higher moments of parton multiplicity distributions are rather
complicated~\cite{Andre81,Dok92}, but reveal in any case that
the parton number distributions are much wider than a Poissonian. The infrared
limit becomes very important and one should consider infrared-safe properties.
Assuming that the singularity is avoided in a way similar
as in an electromagnetic cascade in a medium,
one can estimate the fractal dimension of internal
motion of partons in a jet and it turns out to be quite
low for a single jet in $\E$-annihilation~\cite{DremJETP87}.

The simplest theoretical models, such as the tree diagrams of
the $\f^3$ model, simplified QCD~\cite{Hwa89,ChiuHwa89,Hwa90-91} also based
on tree diagrams, the Schwinger tunnelling transition~\cite{Bial89} or the
effective Lagrangian approach~\cite{ScSu73},  indicate that
the totality of all produced partons exhibits intermittency.

 It is not yet clear what modifications of QCD equations going beyond
the tree graphs (so-called Double Logarithmic Approximation - DLA)
 would fit this region best. An approach to that problem has been
proposed~\cite{GribLev} in deep inelastic processes where the transition
from the Bjorken limit to the Regge domain proceeds through some
intermediate region in which quadratic terms (in the fields) appear and
cause some recombination  of partons at high densities. Here, the
evolution equation for the number of gluons $xG(x,q^{2})$
in   a hadron  with a transverse size $q^{-1}$ and for  small values of
the Bjorken $x$-variable is taken to be
\beq
\frac{\partial^2 xG(x,q^2)}{\partial \ln (1/x) \partial \ln q^2}
=\alpha_sxG(x,q^2)-\frac{C\alpha_s^2}{q^2}
[xG(x,q^2)]^2 \ ,  \label{xg38}
\eeq
where $\alpha_{s}$ is the QCD coupling strength and $C$ is a
constant. It is inspired by QCD ideas and Regge phenomenology, but
has not been derived rigourously. No analysis of the intermittency
property has been attempted so far. One should note, however, that
the general effect of such a quadratic damping is to narrow  the
multiplicity distribution (see, for example~\cite{Batu89}), which
leads to decreasing factorial moments.

One should, however,  not rely on the similarity of
 equations (\ref{dg37}) and (\ref{xg38}), since
this could be misleading. As is well known~\cite{Andre81,BrPe92,Dok92}, the
equations for the generating functionals for gluon and quark jets in QCD are
non-linear, while the corresponding equations for the structure
functions in DLA are just linear GLAP equations. The generating
functional for a parton p is given by
\beq
G_{\Pp}(u,v,x,Y)=\sum_{n_{\rq},n_{\rg}}\frac {1}{n_{\rq}!n_{\rg}!}\prod
\int \rd x_{\rq,i} \rd x_{\rg,j}u(x_i)v(x_j)W^{\Pp}_{n_{\rq}n_{\rg}}(x,x_{i},
x_{j},Y)  \label{gp39}
\eeq
where $W$ is a differential probability to create $n_{\rq}$ quarks and
$n_{\rg}$
gluons with an evolution parameter $Y\sim \ln\ln Q^{2}$ in a p jet.
The equations for the generating functionals are
\beq
\frac{\partial G_{\rq}(x,Y)}{\partial Y}=\int_{0}^{1}\rd x' P_{\rq\rq}(x',x)
[G_{\rq}(x',Y)G_{\rg}(1-x',Y)-G_{\rq}(x,Y)],  \label{gq40}
\eeq
\beqa
\frac{\partial G_{\rg}(x,Y)}{\partial Y}=& \ds\int\limits_{0}^{1}\rd x'
[P_{\rg\rg}(x',x) G_{\rg}(x',Y)G_{\rg}(1-x',Y)-G_{\rg}(x,Y)) \nonumber\\
&+n_{f}P_{\rq\rg}(x',x)(G_{\rq}(x',Y) G_{\rq}(1-x',Y)-G_{\rg}(x,Y))]  \ ,
 \label{gg}
\eeqa
where the $P$'s are the corresponding GLAP kernels, $n_{\rf}$ is the number of
flavours and the initial conditions are such that at $Y=0$, $G_{\rq}=u$ and
$G_{\rg}=v$ for a single jet.

The status of the equations for generating functionals is not
completely clear up to now. They are able to reproduce the higher
order graphs of the perturbation theory far beyond the tree
level~\cite{Dok92}. Their success in predicting the tiny features of
multiplicity distributions in total phase space (for a review
see~\cite{Dr94}) encourages speculations about their quite general
status, with some confinement properties taken into account already at
that stage.

A simplified version of (\ref{gg}) with the quark term omitted
(gluodynamics) has been studied in~\cite{BrPe92}.  Intermittency for
the factorial moments of the gluon cascade is claimed to be in
qualitative agreement with experimental data, as well as evidence for
a structural phase transition. However, the formula for the
intermittency indices contradicts other QCD results. This is not
surprising, since the generating function technique should here
be applied to the subjet hitting the bin under investigation,
not to the whole jet as done in~\cite{BrPe92}.

The treatment of QCD cascades has been taken further in~\cite{Gust91}
within the framework of the dipole formalism including coherence effects.
The multifractal dimension of the parton cascade for high order q is
found to be equal to the QCD anomalous dimension
$\gamma_0=(6\alpha_{s}/\pi)^{1/2}$ and a first pre-asymptotic
correction has been calculated. Moreover, a geometrical interpretation
of the anomalous dimension of QCD is proposed.

A direct solution of QCD evolution equations has
 been attempted for the second correlator in~\cite{Oc92}.
The behaviour of factorial moments of any rank (as well as of double
trace moments - see below) in small phase-space windows
 for $\E$-collisions is treated both in DLA and in the next Modified Leading
 Logarithmic Approximation (MLLA) of QCD in~\cite{dd92}.
 Similar results for factorial moments in DLA are obtained
 in~\cite{OW92a,BrMeuPe93}. They  are closely related to the previously
 derived formulae of~\cite{dmo}.

In the approach described above, one considers three stages of the process:\\
1. the initial quark emits a hard gluon,\\
2. the gluon evolves into a jet consisting of several subjets,\\
3. one of the subjets hits the phase space window chosen.\\
Integrating over all the stages one gets the final multiplicity distribution
(for details see~\cite{dd92}). For comparatively large windows one can use
the fixed coupling constant, while for smaller bins its running should be
taken into account.

 Fixed coupling QCD factorial moments reveal~\cite{Oc92,dd92} the
intermittency phenomenon  with intermittency indices equal to
\beq
\phi_{\QCD}(q)=(q-1)(D-\frac {q+1}{q}\gamma_{0}),  \label{ddphi}
\eeq
in  DLA. This formula is valid if, for the $D$-dimensional analysis with
$M$ bins along each axis, one defines $F_q \propto M^{\phi (q)}$.
For large $q$ the indices increase linearly. The term with negative
sign in the second bracket is proportional to the QCD multiplicity
anomalous dimension $\gamma_0$. From (\ref{ddphi}), one would conclude
that QCD prescribes fractal behaviour with codimension
\beq
\left.d_{q}\right|_{\QCD}=D-\frac {q+1}{q} \gamma_{0}.  \label{ddcod}
\eeq
The phase space term $D$ is obviously non-fractal.  The
$\gamma_{0}$-term in (\ref{ddcod}) is due to the energy dependence of
multiplicity and gives monofractal behaviour.  The gluon energy
spectrum contribution, represented by $\gamma_{0} /q$, gives
multifractal behaviour.  The next-to-leading corrections to
$\gamma_{0}$ also provide $q$-dependent terms.

The calculated values should be compared to the slopes in the region
 $\d y>1$ (which are rather large) since (\ref{ddcod}) is derived in
 fixed-coupling QCD (the intermittent behaviour in that region was
 discussed first in~\cite{SaSa}).  In smaller bins, the QCD running
 coupling becomes important and modifies the above relations
 \cite{dd92}.  The factorial moments now behave in a semi-power-like
 manner so that there is no strict intermittency even though an
 approximate one can still be claimed.
The second term becomes very close to 1 for the low-rank moments and
the low-rank intermittency indices turn out to be very small for
one-dimensional analyses, in accordance with experiment.  Corrections
to $\gamma_{0}$ of the order of $\gamma_{0}^{2}$ can be taken into
account~\cite{dd92}. The influence of confinement seems negligible.
Thus small as well as large $\d y$-intervals may be described in a
unified treatment, at least at a qualitative level.
The transition point from "large" to "small" bins depends on the rank
of the moment in QCD, in full accordance with experimental findings.

We should further mention that the ratio $d_{q}/d_{2}$ depends explicitly
on $D$, contrary to experimental claims (see Subsect.~4.3.3).

The ``free energy'' $F(q)$ is related to $\lambda (q)$ given by (\ref{laq}) as
\beq
F(q)=1-\lambda(q)\approx\gamma_0-\frac{\gamma_0}{q^2}\ \ .\label{lamb}
\eeq
In DLA, it is a steadily increasing function of $q$. However, with
corrections to $\gamma_{0}$ taken into account~\cite{dd92} $F(q)$ becomes
non-monotonic. This happens for rather large values of q, with the result,
that the distinction between factorial and usual moments becomes crucial and
statistical analogies inapplicable. These findings show the limitations of
perturbative QCD and provide further insight into the properties of
multiplicity distributions, such as KNO-scaling,  in full phase
space~\cite{doka,dokb}.

In particular, QCD gives rise to the prediction~\cite{dokb,Dr94} of a
negative value  of the cumulant of rank 5 confirmed by experiment and
to the general conclusion of a non-infinitely-divisible nature of total
multiplicity distributions in QCD (that prohibits, e.g., the one-ladder
multiperipheral cluster models).

The increasing branch of the multifractal spectrum $f(\a )$
may be easily calculated using (\ref{fa}), (\ref{aq}) giving
\beq
f(\a )=2\gamma_{0}^{1/2} (\a -\gamma_{0})^{1/2}.   \label{ddfa}
\eeq

The double trace moments (DTM), redefined in analogy with factorial moments as
\beq
F_{q,\nu} \equiv \frac {1}{\Delta }\left(\sum_{\Theta_{m}\in \Delta}\frac
{n_{m}(n_{m}-1)...(n_{m}-\nu+1)}{n^{\nu}}\right)^{q}\ \ ,
\eeq
behave \cite{dd92} in QCD as
\beq
F_{\nu ,q} (\Delta ) \propto \Delta^{\frac {\gamma_{0} (q^{2}-1)}{q\nu }}
 \propto \Delta ^{q\phi_{\nu }
-\phi_{q\nu } +q-1} \equiv \Delta^{-\phi (q,\nu )+q-1} ,  \label{DTMnuq2}
\eeq
wherefrom one finds
\beq
\phi (q,\nu )=\phi (q \nu )-q \phi (\nu ) = (q-1)\left(1-\frac {q+1}{q}\cdot
\frac {\gamma_{0}}{\nu }\right) .  \label{pdif}
\eeq
The second factor in (\ref{pdif}) may be called ``double codimension''.
It is not symmetric in $q$ and $\nu$ and shows that increasing $\nu $
one decreases effectively the anomalous dimension. For $\nu =1$,
as required, the double codimension becomes equal to the usual codimension.
The scaling exponent (\ref{pdif}) is not factorizable in $\nu $ and $q$.
The above redefinition of DTM is aimed at reducing the Poissonian noise
and the role of phase-space factors, otherwise very important.

In fact, it is surprising that the above expression describes qualitatively
the general trends and even the absolute normalization of the
functions $K(q,\nu )$ shown in~\cite{Ratti91,Ratti92}.
For large $\nu$, the ratio $K(q,\nu )/(q-1)$
is completely determined by the phase space factor  and
should tend to 1. This is seen in the experimental data.
In the region of small $q\sim 1$ and $\nu \sim 1$ the strong
compensation in (\ref{pdif}) prevents its use, but even there it
gives quite reasonable values of $K(q,\nu )$. This probably
indicates that DTM defined as powers of multiplicities are not
sensitive enough to dynamics, a suspicion raised  earlier (Sect.~4.7).

On the other hand, the parton cascading picture employed
in~\cite{dd92} may be applied, strictly speaking, only to hard
processes at extremely high energies and one should not rely much on
the asymptotic estimates of QCD when considering experimental data.
Also, the difference between usual and factorial moments becomes
extremely important at large values of $q $ or at sufficiently small
bins with low multiplicities.  One may suspect that relation
(\ref{pdif}) and, especially its first part, has a much wider range of
applicability than just for $e^{+}e^{-}$-collisions and some universal
relation could be valid for other reactions.  If so, it will be
important to understand whether these common features are due to
common dynamics or to insufficient sensitivity of the proposed
measures.

The  large fluctuations, {\it e.g.} those observed by the NA22 collaboration,
have raised the suspicion that at small bins one is dealing with unusually
wide distributions, which could have infinite moments. This has led one to
consider L\'evy-stable probability distributions. The L\'evy indices
derived from QCD factorial moment indices or DTM exponents show
\cite{dd92} no sign of ``wild'' singularities.

The DTM technique has been first applied to experimental data from
hadron-hadron reactions (described in Subsect.~4.7.7) where direct QCD
arguments are invalid.  For $\E$-collisions, however, it may be
worthwhile to analyse  the data with this method.

Let us further note that there is a difference between predictions
of QCD and those of variants of conformal theories considered
in~\cite{dnL92}, or of multiplicative models~\cite{amd}. For example,
the intermittency indices derived in a rapidity or azimuthal angle
analysis should be equal in QCD.
This is a consequence of the symmetrical form (in pseudorapidity $\eta
$ and azimuthal angle $\varphi $) of the gluon propagator, which can be
written as
\beq
k^{2}=(p_{\rT,1}+p_{\rT,2})^{2}=4p_{\rT,1}p_{\rT,2}(\sinh ^{2}
\frac{\eta_{12}}{2}+\sin^{2}\frac{\varphi_{12}}{2}) ,  \label{k2}
\eeq
where $p_{\rT,i}$ is the transverse momentum of $i$-th parton.
 In a conformal theory, the intermittency indices for the second factorial
moment are given by
\beq
\phi_2 (\d y)=2\eta,  \;\;\;\;\; \phi_2 (\d \varphi )=0\ \ ,  \label{p2dn}
\eeq
respectively. Here, $\eta $ is the conformal anomalous dimension estimated
to be $\eta \approx 0.07-0.1$. For multiplicative models, one finds
\beq
\phi_2 (\d y)=1-D_{y}, \;\;\;\; \phi_2 (\d \varphi )=D_{\varphi }-1
 \label{anp2}
\eeq
with $0 < D_{y} < 1, 1 < D_{\varphi } < 2$.

The emergence of intermittent behaviour in solutions of non-linear
equations and in perturbative QCD encourages further studies along
these directions.  At the least, they hold a promise of further
theoretical insight.  One should keep in mind, however, that a direct
comparison of QCD-based asymptotic results with present-day
experiments is not justified.  Some effects revealed by the data seem
to be of a different, as yet unsatisfactorily explained, origin.

\chapter{Conclusions}

Developments in physics --- and in science in general ---
over the last decade, have brought exciting new discoveries and
deeper insight into the dynamics of complex systems.
Studies of classical and quantum chaos, non-equilibrium
dissipative processes, random media, growth phenomena and many
more have all contributed to reveal  the pervasive importance
of self-similarity, of power-laws and of fractals in nature.
Research in these fields is  still in full evolution and continues
to uncover intriguingly simple and often surprisingly
universal behaviour in complex, non-linear systems.

The suggestion of Bia\l as and Peschanski to look for
scaling in particle fluctuations was one of the first attempts
to apply modern ideas and techniques from complex-system
research to multihadron-production processes.
In preceding pages, we have presented a critical
 overview of the impressive amount of  experimental and theoretical
work this proposal has generated since its formulation in 1986.
The continuing interest in the field testifies of the
growing conviction that new avenues need to be explored
for progress in strong-interaction physics.

Impressive as it may be, this work has not yet led to
final answers concerning the fundamental issues.
Approximate power-law scaling of  particle density and
correlation functions is now indeed observed,
especially in two- or three-dimensional phase-space.
However, so far it can be explained from an interplay between jet
formation and more or less ``conventional'' correlations
among identical particles due to quantum interference.

Nevertheless, as often happens, the detailed scrutiny of
data (and detectors) on the full variety of collision processes in the search
for power behaviour has led to many new observations
of interest in their own right.
It has helped to recognize the importance
of detailed studies of correlation phenomena at large and small
distances in momentum space and new sensitive and general techniques
have been developed for their analysis.

Standard hadronization models, all too often accepted as satisfactory,
have been exposed to severe and sometimes even painful tests.
Intermittency analysis has revealed deficiencies in our understanding
of the hadronization process. These defects are not easy to cure in a
consistent manner
by simple parameter-tuning and ``new'' physics may well be needed
to restore internal consistency in, e.g.,~fragmentation models of the LUND
type. Present work on this subject starts to provide hints
that purely probabilistic treatment of
the break-up of colour fields has to be supplemented
with  effects deeply connected with the structure of the non-perturbative
QCD vacuum. Progress in this direction would in itself be ample
compensation for the efforts spent on attempting to establish
fractality in multihadron production.

Data obtained in the last years have shown the overwhelming importance of
correlations among identical particles in the ``intermittent'' regions
of phase space. This quantum mechanical phenomenon,
discovered in particle physics in 1959, still awaits
satisfactory incorporation into present hadronization phenomenology, if it
is to be used as a reliable interferometric tool, e.g.,~in
studies of quark-gluon plasma formation.

Theoretical work has developed along a large variety of directions.
Fractal properties have been discovered in string-fragmentation models.
Within the realm of perturbative QCD, parton correlations and
emergence of power behaviour are now studied with
increasing sophistication. The relevance for present-day
phenomenology remains doubtful, however.

The powerful methods of statistical mechanics have been intensively
exploited in studies of random cascades as well as in equilibrium and
non-equilibrium critical phenomena.
Results of real intrinsic value have been obtained, with potentially
interesting applications in other fields.

In search for an explanation of ``unusually large'' density
fluctuations, ``intermittency'', in analogy with fluid turbulence, has
progressively led to appreciate the importance and often spectacular
manifestation of non-linear strong-coupling dynamics.

Experiments in deep-inelastic scattering are now starting to probe
hadron structure in a regime where the perturbative parton-cascade
picture becomes blurred.  Non-linear perturbative evolution and
confinement play an increasingly important role in the very low
Bjorken-$x$ region now accessible in HERA.  Present attempts to
understand this region invoke QCD Reggeon-theory.

It is intriguing to speculate that a power-law dependence of the
low-$x$ parton correlation functions could manifest itself as ``gluonic''
intermittency in virtual parton cascades, with the occasional creation of
regions with very large, or very small gluon density.

\newpage

\subsubsection{Acknowledgements}\\
It is a great pleasure to thank the many physicists who have
contributed enthusiastically to this new field with experimental data,
theoretical or methodical ideas or constructive criticism. Among those
from which we have gained most considerably are
B.~Andersson,
I.V.~Andreev,
A.~de Angelis,
A.~Bia\l as,
J.~Bjorken, F.~Botterweck, B.~Buschbeck,
P.~Carruthers, A.~Capella, M.~Charlet,
S.~Chekanov,
H.~Dibon,
I.~Derado,
H.~Eggers,
E.L.~Feinberg,
K.~Fia\l kowski,
A.~Giovannini,
E.~Grinbaum-Sarkisyan,
G.~Gustafson,
R.~Hwa,
G.~Jancso,
V.~Kuvshinov,
A.V.~Leonidov,
P.~Lipa,
F.~Mandl,
H.~Markytan,
W.~Metzger,
J.-L.~Meunier,
W.~Ochs,
R.~Peschanski,
O.~Podobrin,
S.~Ratti,
I.~Sarcevic,
H.~Satz,
N.~Schmitz,
J.~Seixas,
E.~Stenlund,
A.~Syed,
F.~Verbeure,
R.~Weiner,
B.~Wo\v siek,
J.~Wo\v siek,
Y.F.~Wu,
K.~Zalewski, and many others to be found in the references.
It is a  great favour to be able to work with these
scientists and to share their pioneering spirit in difficult territory.

\chapter{Figure Captions}
\vs 1cm

Fig.~3.1 Contours of the two-particle correlation function,
$R^{\rc\rc}(y_1,y_2)$, from 205 GeV/$c$ pp interactions~\cite{Wink75}.

Fig.~3.2 a) The charge correlation function $C_2(\h_1,\h_2)$ plotted
for $\Pp\Pap$ collisions at fixed $\h_1=0$ versus $\h_2$ at 63, 200,
546 and 900 GeV, b) the "long-range" contribution $C_\rL$ and c) the
short-range contribution $C_\rS$~\cite{Anso88}.

\vs 2mm
Fig.~3.3 The semi-inclusive correlation function
$C^{(n)}_2(\h_1,\h_2)$ for $34\leq n\leq38$ $\Pp\Pap$ collisions at
900 GeV, compared to the UA5 Cluster MC, PYTHIA and FRITIOF
2.0~\cite{Fugl87}.

\vs 2mm
Fig.~3.4 Normalized correlation function $K_2(y_1,y_2=0)$ for (CC),
(--~--), (+~+) and (+~--) combinations in $n>6$ pp collisions at 360
GeV/$c$, compared to predictions from single chain LUND and a
two-chain DPM~\cite{Bail88}.

\vs 2mm
Fig.~3.5 a,b) Correlation functions $C_2(0,y)$ and $\tilde C_2(0,y)$
for M$^+$p reactions as compared with calculations in FRITIOF
$(-\cdot-\cdot-)$, DPM (-------) and QGSM $(---)$
(non-single-diffractive sample). c) Correlation functions $\tilde
C_{S}(0,y_2)$ for M$^+$p reactions as compared with FRITIOF
$(-\cdot-\cdot-)$ and QGSM $(---)$~\cite{Aiva91}.

\vs 2mm
Fig.~3.6 Normalized two-particle correlation function $K_2(y_1,y_2)$
for pairs of a) oppositely and b) negatively charged hadrons produced
in muon-nucleon scattering at 280 GeV/$c$ \cite{Arne86}.

\vs 2mm
Fig.~3.7 Normalized correlation functions $K_2(0,y)$ for the M$^+$p
non-single-diffraction sample and $\m^+$p-interactions at 280 GeV/$c$
$(13<W<20$ GeV)~\cite{Male90Fig88}.

\vs2mm
Fig.~3.8 a,b) Normalized correlation functions $\tilde K_2(0,y)$ for
the M$^+$p non-single-diffraction sample and $\E$-annihilation at
$\sqrt s$=22 GeV.  c) Normalized correlation function $\tilde
K_2^{cc}(y_1,y_2)$ at $y_1=-1\div0$ in the non-single-diffraction
M$^+$p sample at 22 GeV and $\E$-annihilation at 14 and 44
GeV~\cite{Aiva91}.

\vs 2mm
Fig.~3.9 Normalized correlation functions $K_2(y_1,y_2)$ for the
entire data sample (upper plots) and for selected two-jet events
(lower plots) in $\E$ annihilation at 35 GeV. The left plots show the
average over the region $-1\leq y_1+y_2\leq0$ and the right plots
correspond to the average over $0\leq y_1+y_2\leq 1$. The data (open
circles) are compared to the JETSET 7.2 PS model with (solid lines)
and without (dotted lines) Bose-Einstein correlations~\cite{Podo91}.

\vs 2mm
Fig.~3.10 OPAL data on $R(\xi_1,\xi_2)$ as a function of
$(\xi_1-\xi_2)$, for $(\xi_1+\xi_2)$ between 5.9 and 6.1 (IV), 6.9 and
7.1 (V), 7.9 and 8.1 (VI), respectively, compared to a) analytic QCD
calculations (the dashed curves indicate leading order QCD
calculations for $\La$=225 MeV, the three solid curves represent
next-to-leading QCD calculations with, from top to bottom, $\La=1000$,
255 and 50 MeV, respectively), and b) coherent parton-shower
Monte-Carlo models ARIADNE (solid), JETSET (dashed) and HERWIG
(dotted);\break c) incoherent parton-shower Monte-Carlo models JETSET
(solid) and two versions of COJETS (dashed and dotted)~\cite{Act92}.

\vs 2mm
Fig.~3.11 Correlation functions $C_{\rS}(0,y)$ (a) and $\tilde
C_{\rS}(0,y)$ (b) in M$^+$p interactions at 250 GeV/$c$ (open circles
correspond to non-single-diffractive events and are shifted to the
right to avoid overlap) ~\cite{Aiva91}.

\vs 2mm
Fig.~3.12 Topological correlation functions $\tilde C_2^{(n)}(0,y)$ in
M$^+$p reactions at 250 GeV/$c$ for $(+-)$ pairs~\cite{Aiva91}.

\vs 2mm
Fig.~3.13 a) $\tilde C^{(n)}_2(0,0)$ dependence on $n$ and $z=n/\lan
n\ran$ for M$^+$p interactions at 250 GeV/$c$. The last $(--)$ point
corresponds to $n\geq16$, the last $(+-)$ and $(++)$ points to
$n\geq18$ ~\cite{Aiva91}. b) $\tilde C^{(n)}_2(|\h_1-\h_2|)$ as a
function of $1/(n-1)$ for pp collisions at ISR energies~\cite{EGGE75}.

\vs 2mm
Fig.~3.14 Normalized correlation functions $K_2(0,y)$ for particles
with all $p_\rT$, $p_\rT<0.30$ GeV/$c$, $p_\rT>0.30$ GeV/$c$ as
compared to functions (3.2) and (3.3), solid and dot-dashed lines,
respectively~\cite{Aiva91}.

\vs 2mm
Fig.~3.15 Distribution in a) the rapidity gap $\D y$ and b) the
azimuthal distance $\D\vf$ for $\PK^0_\rS\PK^0_\rS$ pairs in pp
collisions at 360 GeV/$c$.  c) and d) same for $\PK^0_\rS\Lambda$
pairs~\cite{Asai87}. The lines correspond to LUND.

\vs 2mm
Fig.~3.16 The multiplicity-averaged angular correlation function\\
$\lan (n-1)\tilde C^{(n)}_2(\h_1,\vf_1;\h_2,\vf_2)\ran$ for CC
combinations in units of $10^{-3}$~\cite{EGGE75}.

\vs 2mm
Fig.~3.17 $W(\D\vf, \D y)$ for inclusive non-single-diffractive
$\p^+$p interactions at 250 GeV/$c$ as compared to FRITIOF2.0
(dot-dashed), DPM (full) and QGSM (dashed)~\cite{NA22}.

\vs 2mm
Fig.~3.18 $W(\D\vf,\D y,p_\rT)$ as compared with calculations in
FRITIOF2.0 (dot-dashed), DPM (full) and QGSM (dashed) for $\D y<1$ and
$p_\rT$ cuts as indicated~\cite{NA22}.

\vs 2mm
Fig.~3.19 The $p_\rT$ dependence of the azimuthal correlation
parameter $B$ for h$^+$h$^-$ pairs in pp collisions at 360 GeV/$c$
compared to LUND and DPM~\cite{Bail88}.

\vs 2mm
Fig.~3.20 Charmed pair azimuthal correlation: WA92 data are compared
with predictions from (a) an NLO perturbative QCD calculation and (b)
a model where a parton transverse momentum, $p_\rT$, is added to the
NLO perturbative QCD predictions (dotted line: $\lan p^2_\rT\ran=1.0$
(GeV/$c)^2$, solid line: $\lan p^2_\rT\ran=0.3$
(GeV/$c)^2$~\cite{Adamo95}.

\vs 2mm
Fig.~3.21 Two-particle azimuthal correlations with respect to the
sphericity axis in OPAL~\cite{Act93} compared to coherent and
incoherent MC models.

\vs 2mm
Fig.~3.22 The PPCA distribution for corrected L3 data compared to a)
coherent and b) incoherent Monte-Carlo models \cite{Syed94}.

\vs 2mm
Fig.~3.23 Three-particle rapidity correlations a) $K_3(0,0,y)$ [the
lines correspond to the ISR results at 31 GeV (full) and 62 GeV
(dashed)] and b) $\tilde K_3(0,0,y)$ for M$^+$p interactions at 250
GeV/$c$.  [the FRITIOF (dot-dashed) prediction is indicated for the
charge combination $(---)$, QGSM (dashed) for $(---)$ and
$(--+)$]~\cite{Aiva91}.

\vs 2mm
Fig.~4.1 a) The JACEE event~\cite{Burn83}; b) The NA22
event~\cite{AdamPL185-87}.

\vs 2mm
Fig.~4.2 a) log$F_5$ as a function of $-\log\d\h$ for the JACEE
event~\cite{bialas1,bialas2} (full circles) compared to independent
emission (small crosses); b) $\ln F_2$ and $\ln F_4$ as functions of
$-\ln\d\h$ for O$^{16}$ Em at 200 $A$ GeV (KLM)~\cite{Holy89}, c) log
$F_q$ on $_2\log\d\h$ and d) $_2\log\d\vf$ at 630 GeV
(UA1)~\cite{Alba90}.

\vs 2mm
Fig.~4.3 Anomalous dimension $d_q$ as a function of the order $q$, for
a) $\m$p and $\E$ collisions, b) NA22 and UA1, c) KLM~\cite{Bial90}.

\vs 2mm
Fig.~4.4 a) Slope $\f_q$ as a function of order $q$ for
NA22~\cite{Ajin89-90}, two versions of FRITIOF and a two-chain dual
parton model, b) $F_3$ vs. $\d\h$ from UA1~\cite{Alba90} compared to
GENCL, PYTHIA and PYTHIA + Bose-Einstein Monte Carlos.

\vs 2mm
Fig.~4.5 a) EMC results~\cite{Dera90} compared to expectations from
{}~\cite{CapFiaKrz89}, b) slopes $\f_q$ for EMC data as well as Webber
and LUND models, c) $\n A$ data~\cite{Verlu90} in comparison to LUND
model expectations.

\vs 2mm
Fig.~4.6 Jet evolution: the self-similarity in the parton cascade
 derives from the similarity of each step in the
 evolution~\cite{Ven79}.

\vs 2mm
Fig.~4.7 $\ln F_3$ (and $\ln F_2$) as a function of $-\ln \d y$ (or
$\ln M$, $M=\D Y/\d y$) for a) HRS~\cite{BuLiPe88Aba90}, b)
TASSO~\cite{Brau89}, c) CELLO~\cite{Behr90} and d)
DELPHI~\cite{Abreu90} data compared to LUND (and Webber) parton shower
models .

\vs 2mm
Fig.~4.8 $\ln F^\rH_q$ as functions of $-\ln \d y$ for JETSET 6.3
parton shower at $\sqrt s$=91 GeV at the a) parton, b) hadron level,
both with cut-off $Q^2_0$=1 GeV$^2$, c) d) with cut-off $Q^2_0$=0.4
GeV$^2$~\cite{BotBusch}.

\vs 2mm
Fig.~4.9 a) $\log F_2$ as a function of $-\log\d\vf$ for UA1
data~\cite{Alba90}, b) c) slope $\f_q$ as a function of order $q$ for
$y$, $\h$, $\vf$ or ($y$ and $\vf$) as variables for
NA22~\cite{Ajin89-90}.

\vs 2mm
Fig.~4.10 The log-log plot for TASSO data using a) 2 dimensional
 $y$--$\vf$ bins,\break b) one-dimensional $y$ bins, in comparison to
 a two-dimensional $\a$-model ~\cite{Brau89,Ochs}.

\vs 2mm
Fig.~4.11 Factorial moment of order $q=2$ for 1, 2 and 3 dimensional
analysis for a) DELPHI~\cite{Abreu90}, b) UA1~\cite{Alba90} and c)
NA22~\cite{Ajin89-90} as a function of $(\log_2 M)/d$ and $(\ln M)/d$,
respectively, where $M$ denotes the total number of boxes in a
$d$-dimensional analysis.

\vs 2mm
Fig.~4.12 a) Factorial moment $\ln F_2$ as a function of $(\ln M)/d$
from a 3 dimensional analysis of negative particles in pAu and central
OAu and SAu collisions~\cite{Dera90Sing};\break b) Factorial cumulant
$K_2$ from the same analysis in central OAu
collisions~\cite{Dera90Sing}.

\vs 2mm
Fig.~4.13 The second scaled factorial moment $F_2$ as a function of
the number of bins $M^3$ in the log-log scale~\cite{FiMPI91} for (a)
$\m$p data~\cite{Dera90}, (b) $\p/\PK$p data~\cite{Ajin89-90}, (c)
pAu, (d) OAu and (e) SAu data~\cite{Dera90Sing}.  Solid lines
represent (\ref{eq:3.15}) with parameter values: (a) $c$=0.025,
$\f_2$=0.45, $c'$=0; (b) $c$=0.02, $\f_2$=0.45, $c'$=0.16; (c)
$c$=0.02, $\f_2$=0.45, $c'$=0.2; (d) and (e) $c$=0.01, $\f_2$=0.5,
$c'$=0.

\vs2mm
Fig.~4.14 The factorial moments $\ln F_q$ for NA22~\cite{Ajin89-90} in
various variables and dimensions indicated, as compared to ECCO
{}~\cite{Hwa92}.

\vs 2mm
Fig.~4.15 a) Illustration of the modified power-law behaviour. The
lines indicate an 'eye-ball' fit to the data. Only the data in bins
where $F_2$ varies strongly are used; b) Test of the universal scaling
law (\ref{ochs:rel1}) for $\ln F_3$ and $\ln F_2$ in $\E$, lh, hh and
$AA$ collisions as indicated. The straight line is adjusted to the
$\E$ data and reproduced on the other data sets~\cite{Ochs}.

\vs 2mm
Fig.~4.16 Ratio of anomalous dimensions as a function of the order $q$
for a) NA22 results~\cite{Ajin89-90} and b) earlier results on all
types of collisions~\cite{Ochs}, c) and d) $d_q$ as a function of the
order $q$.  Continuous lines in c) and d) show the best fits using
(\ref{BB:134}) (c) $\bu -$ $Z^0$ decay, DELPHI, $r=1.6(8)$,
$\blacksquare -$ p-Ag/Br, KLM, $r=1.1(5)$, $\blacklozenge -$ S-Ag/Br,
KLM, $r=0.0(2)$; (d) $\bu -$ O+Ag/Br, KLM, $r=0.7(2)$, $\blacksquare
-$ $\Pp\Pap$, UA1, $r=0.5(3)$
\cite{chek94}.

\vs 2mm
Fig.~4.17 $\la_q$ as a function of $q$ for a) KLM, CMS and NA22
results \cite{BiaZa,Ajin89-90} and b) central C-Cu collisions at 4.5
$A$ GeV/$c$ \cite{Sark93}.

\vs 2mm
Fig.~4.18 Bose-Einstein correlation of order 2 to 5, as indicated. The
dashed lines represent fits by Gaussian terms, the full lines by
exponential terms. All data are corrected for Coulomb interaction
{}~\cite{Neum91}.

\vs 2mm
Fig.~4.19 The dependence of $R_2=N_2/N_2^{BG}$ on $Q^2_{q\p}$ for all
pairs of charged particles (full line), for opposite-charged pairs
(dotted) and for same-charged pairs (dashed)~\cite{Alba90}.

\vs 2mm
Fig.~4.20 a) $\ln F_q$ as a function of $-\ln \d y$ for various
$p_\rT$ cuts as indicated~\cite{Ajin89-90},\break b) Anomalous
dimensions $d_q$ as a function of the order $q$, for various $p_\rT$
cuts as indicated (lines are to guide the eye)~\cite{Ajin89-90}, c)
slope $\f_2$ as a function of the average transverse momentum $\bar
p_\rT$ in an UA1 event compared to PYTHIA 5.6~\cite{Wu94}.

\vs 2mm
Fig.~4.21 Factorial moment $F_2$ and $F_3$ as a function of resolution
for three $\E$ data sets with $p_\rT$ cuts as
indicated~\cite{Abreu90}. The lines correspond to the models as
indicated. Correction factors are given above the corresponding
sub-figures.

\vs 2mm
Fig.~4.22 $F_2$ and $F_3$ of the first, second and third jet ordered
by their energy. DELPHI corrected data (open symbols) are compared
with JETSET 7.2 PS Monte-Carlo predictions with re-tuned
settings~\cite{Abreu90}.

\vs 2mm
Fig.~4.23 a) Multiplicity dependence of the slope $\f_3$, compared to
that expected from a number of models, the crosses correspond to a
combination of independent events~\cite{Alba90}, b) slope $\f_2$ as a
function of particle density for NA22 (hp at 250 GeV) and heavy-ion
collisions as indicated~\cite{Adamo90}. The full line corresponds to
an extrapolation from the NA22 point to higher densities using
$\f_2\propto1/\rho$. The dotted lines show fits for the $^{16}$O and
$^{28}$Si samples and for the $^{32}$S samples.

\vs 2mm
Fig.~4.24 a) UA1 factorial moments (crosses) decomposed into cumulant
contributions: lower squares indicate 2-particle, upper squares
2+3-particle contributions~\cite{Egg91}, b) Upper row: normalized
factorial moments $F_q$ ($\blok$) from NA22, together with
contributions from 2-particle correlations ($\triangle$), the sum of
2- and 3- particle correlations ($\circ$), and the sum of 2-, 3- and
4-particle correlations ($\diamondsuit$); lower part: the normalized
cumulant moments $K_q$~\cite{Ajin89-90}. c) $K_3$ for nuclear
collisions~\cite{Egg91}.

\vs 2mm
Fig.~4.25 $\ln F_{pq}$ as a function of $-\ln D$ for four values of
$\d y$, as indicated~\cite{Aiv91}.

\vs 2mm
Fig.~4.26 Dependence of $\ln F_{pq}$ on the bin size $\d y$ for a
correlation distance $D=0.4$, a) for NA22 data, b) for a sample of
60~000 FRITIOF Monte-Carlo events. The dashed lines correspond to
horizontal-line fits through the points~\cite{Aiv91}.

\vs 2mm
Fig.~4.27 a) b) The increase of the slopes $\f_{pq}$ with increasing
order $pq$ compared to the expectation from FRITIOF, for two values of
$\d y$, respectively; c) d) the increase of $\f_{pq}/\f_2$ with
increasing order $pq$, compared to that expected from the $\a$ model
(dashed line), for two values of $\d y$, respectively~\cite{Aiv91}.

\vs 2mm
Fig.~4.28 a) $F_{11}$ versus $D$ ($\delta y=0.2$) and $F_2$ versus
$\delta y$ for a Gaussian-shaped two-particle correlation
function~\cite{dewolf92} compared to NA22 data~\cite{Aiv91}; b)
$F_{12}$ versus $D$ ($\delta y=0.2$) as in a).

\vs 2mm
Fig.~4.29 a) $\lan\ln G_q\ran$ as a function of resolution $(M=2^\m)$
for UA1 compared to the expectations from the Monte-Carlo models GENCL
and PYTHIA, b) $\lan\a_q\ran$ and $\lan \t_q\ran$ as a function of
$q$, c) the multifractal spectral function $\lan f(\a_q)\ran$ as a
function of $\lan \a_q\ran$~\cite{UA1G}.

\vs 2mm
Fig.~4.30 a) The universality function $\G_q$ as a function of
$\xi=\m-\n$ for $q=\pm5$ for UA1, b) the spectral function $\lan
f(\a_q)\ran$ as a function of $\lan\a_q\ran$~\cite{UA1G}.

\vs 2mm
Fig.~4.31 a)--c)~numerical results based on the Bernoulli-trial model
(\ref{3:eq:6}) averaged with a Negative Binomial distribution $P(n)$
with $\lan n\ran=6.26$ and dispersion $2.94$~\cite{EMCG}. The line in
(c) is a fit to $\tau_q=-C\,(q-1)$.

\vs 2mm
Fig.~4.32 Comparison of the exponents $\f_q$ of $F_q$ and
$1-q-\t_q^{\dyn}$ (respectively $1-q-\t_q$) of $G_q$ in a) ECCO
Monte-Carlo simulations \cite{HwaPan} and b) $\p^-$ AgBr data at 350
GeV/$c$ \cite{Ghosh92}.

\vs 2mm
Fig.~4.33 a) Plot of log $P$ vs $-\log \d y$ for $n=6$ and 14.  The
data correspond, downwards, to thresholds $n_{\th}\equiv
M^\g$=1,2,3,4,5; b) Codimension functions $c(\g)$ vs degree of
singularity $\g$ for $n=6$ and 14 \cite{Ratti92}.

\vs 2mm
Fig.~4.34 a) The integration domain $\W_{\PB}=\S_m\W_m$ of
$\r_2(y_1,y_2)$ for the bin-averaged factorial moments, b) the
corresponding integration domain $\W_\rS$ for the density
integral~\cite{Lipa91-53}, c) illustration of a $q$-tuple in snake
topology, d) GHP topology, e) star topology.

\vs 2mm
Fig.~4.35 a) The fourth factorial moment of the NA22 spike event
{}~\cite{AdamPL185-87}, b) the density strip integrals for $q=2-4$
{}~\cite{Lipa91-53}.

\vs 2mm
Fig.~4.36 Density strip integrals $C_q$ for $q=2-5$ for a) all-charged
and negatives in $\p^+\Pp$ and $\PK^+\Pp$ collisions at $\sqrt s$=22
GeV (the integral $F^\rS_2$ is also given for $(+-)$
combinations)~\cite{Ajin89-90}, b) all-charge and same-charge
combinations in $\Pp\Pap$ collisions at 630 GeV~\cite{Alba90}.

\vs 2mm
Fig.~4.37 ln $F^\rS_2$ in the NA22 data (full circles) compared to
FRITIOF2.0, FRITIOF2.0 with Dalitz decay and FRITIOF2.0 with Dalitz
decay and $\gamma$-conversion (open symbols), for (cc), (-- --) and (+
--) combinations, as indicated~\cite{Ajin89-90}.

\vs 2mm
Fig.~4.38 The slope $\f_2$ as a function of average transverse
momentum $\bar p_\rT$ and multiplicity $n$ for UA1 data and PYTHIA
\cite{Wu94}.

\vs2mm
Fig.~4.39 Comparison of density integrals for $q=2$ in their
differential form (in intervals $Q^2, Q^2+\rd Q^2)$ as a function of
$_2\log(1/Q^2)$ for $\E$ (DELPHI) and hadron-hadron
(UA1)~\cite{Mandl92}.

\vs2mm
Fig.~4.40 Density integrals $F^\rS_2$ (in their differential form) as
a function of $Q^2$ for like-charged pairs in UA1~\cite{Alba90} and
NA22~\cite{Ajin89-90}, compared to power-law, exponential,
double-exponential and Gaussian fits, as indicated.

\vs2mm
Fig.~4.41 $K_2(M_{\inv})$ for $cc, +-, --$ and $++$ pairs of tracks
with c.m. rapidity $-2<y<2$, in $\PK^+/\p^+\Pp$ collisions at 250
GeV/$c$~\cite{XX1}. The solid line is a power-law fit (see text).

\vs 2mm
Fig.~4.42 $K_2(M_{\inv})$ for a) $+-$, b) $--$ pairs of tracks with
c.m. rapidity $-2<y<2$, in $\PK^+/\p^+\Pp$ collisions at 250
GeV/$c$~\cite{XX1}, compared to FRITIOF2.

\vs 2mm
Fig.~4.43 $R=K_2(M_{\inv})+1$ for $\E$ annihilation compared to the
JETSET7.3 prediction with parameters tuned to the DELPHI data for a)
unlike-sign and b) like-sign combinations \cite{XX2}.

\vs 2mm
Fig.~4.44 $\ln K^*_q(Q^2)$ as a function of $-\ln Q^2$ for all charged
particles as well as for like-charged particles \cite{genuine}.

\end{document}